\documentclass[apj]{emulateapj}
\slugcomment{{\sc Accepted to ApJ:} November 11, 2008}
\def\gtorder{\mathrel{\raise.3ex\hbox{$>$}\mkern-14mu
                \lower0.6ex\hbox{$\sim$}}}
\def\ltorder{\mathrel{\raise.3ex\hbox{$<$}\mkern-14mu
                \lower0.6ex\hbox{$\sim$}}}

\shorttitle{Metal-Columns in Fast Shocks}
\shortauthors{Gnat \& Sternberg}

\begin{document}
\title{Metal-Absorption Column Densities in Fast Radiative Shocks}
\vspace{1cm}
\author{Orly Gnat\altaffilmark{1,2,3} and Amiel Sternberg\altaffilmark{1}}
\altaffiltext{1}{School of Physics and Astronomy and the Wise Observatory,
        The Beverly and Raymond Sackler Faculty of Exact Sciences,
        Tel Aviv University, Tel Aviv 69978, Israel}
\altaffiltext{2}{Theoretical Astrophysics, California Institute of Technology, 
        MC 130-33, Pasadena, CA 91125}
\altaffiltext{3}{Chandra Fellow}
\email{orlyg@tapir.caltech.edu}

\begin{abstract}
In this paper we present computations of the integrated metal-ion column
densities produced
in the post-shock cooling layers
behind fast,
radiative shock-waves.
For this purpose,
we have constructed a 
new
shock code
that calculates the non-equilibrium
ionization and 
cooling; 
follows the radiative transfer of the 
shock self-radiation through the post-shock cooling layers;  
takes into account the resulting photoionization and heating
rates; follows the dynamics of the cooling gas; and self-consistently
computes the initial photoionization state of the precursor gas.
We discuss the shock structure and emitted radiation, and 
study
the dependence on the 
shock velocity, magnetic field, and gas metallicity.
We present a complete set of integrated post-shock
and precursor metal-ion
column densities of all ionization stages of the 
elements  H, He, C, N, O, Ne, Mg, Si, S, and Fe,
for shocks with velocities of $600$ and $\sim2000$~km~s$^{-1}$,
corresponding to initial post-shock temperatures of
$5\times 10^6$ and $5\times 10^7$~K,
cooling down to 1000~K.
We consider shocks in which the magnetic field is negligible 
($B=0$)
so that the cooling occurs at approximately constant pressure (``isobaric''),
and shocks in which the magnetic 
pressure dominates everywhere
such that the cooling occurs at constant density (isochoric).
We present results for gas metallicities $Z$
ranging from $10^{-3}$ to twice the
solar abundance of heavy elements, and 
we study how the
observational signatures of fast radiative shocks
depend on $Z$.
We present our numerical results in convenient online figures and tables.

\keywords{ISM:general -- atomic processes -- plasmas --
absorption lines -- intergalactic medium -- shock waves}

\end{abstract}

\section{Introduction}
\label{introduction}

Shock waves are a common phenomenon in astrophysics,
and have a profound impact on the energetics and
distribution of gas in the interstellar and
intergalactic medium (McKee \& Hollenbach~1980; 
Draine \& McKee~1993 and references therein).
A detailed understanding
of the physical properties and observational signatures
of shock-heated gas is needed, for example, in the study of   
diffuse gas in supernova remnants 
(e.g. Shull \& McKee~1979; Dopita et al.~1984);
young stellar objects (e.g.~Dopita 1978);
high velocity clouds (e.g.~Fox et al.~2005);
active galactic nuclei (e.g.~Dopita \& Sutherland~1995) --
mostly in the context of  
narrow emission line regions, low ionization
emission line regions, and radio galaxies;
the shock-heated ``warm/hot'' intergalactic gas 
(e.g.~Savage et al.~2005; Furlanetto et al.~2005;
Cen \& Ostriker~2006; Bertone et al.~2008);
and in the virial 
and accretion shocks that may form in hot and
cold-flow accretion into dark-matter halos and
proto-disks in high redshift galaxies 
(e.g.~Birnboim \& Dekel 2003; Keres et al. 2005; Dekel et al.~2008).

In this paper we present new computations of the 
non-equilibrium
metal
ionization states and integrated metal column densities that
are produced behind 
fast,
steady, radiative shock waves,
and of the equilibrium columns produced in their radiative precursors.
Our work is motivated by recent {\it Hubble Space Telescope (HST)},
{\it Far Ultraviolet Spectroscopic Explorer (FUSE)},
{\it Chandra}, 
and {\it XMM-Newton}, detections and observations of hot gas
around the Galaxy (Collins et al.~2005; Fang et al.~2006), and of
gas that may be a part of a 10$^5$-10$^7$ K 
``warm/hot intergalactic medium'' 
(WHIM; Tripp et al.~2000, 2006, 2007; Shull et al.~2003; 
Richter et al.~2004; Sembach et 
al.~2004; 
Soltan et al.~2005;
Nicastro et al.~2005; Savage et al.~2005; Rasmussen et al.~2006;
Williams et al.~2006).
An intergalactic shock heated WHIM is expected to be a major 
reservoir of baryons in the low-redshift universe
(Cen \& Ostriker~1999, 2006; Dave et al.~2001; 
Furlanetto \& Loeb 2004;
Bertone et al.~2008;
Danforth \& Shull 2008).
Important ions for the detection of such gas include
\ion{C}{4}, \ion{N}{5},
\ion{O}{6}, \ion{O}{7}, \ion{O}{8}, \ion{Si}{3}, \ion{Si}{4},
\ion{Ne}{8}, and \ion{Ne}{9}.

Recent observations have confirmed the existence of warm/hot gas,
both in the local Universe, and in more distant environments 
(e.g.~Tripp et al.~2007; Savage et al.~2005). 
For example, Savage et al.~(2005) reported on an absorption-line
system at a redshift of 
$z=0.207$, containing \ion{C}{3},
\ion{O}{3}, \ion{O}{4}, \ion{O}{6}, \ion{N}{3}, \ion{Ne}{8}, \ion{Si}{3},
and \ion{Si}{6}.
They invoked an equilibrium two-phase model, in which most of the ions are
created in a warm ($\sim2\times10^4$~K) photoionized cloud, while \ion{O}{6}
and \ion{Ne}{8} are 
produced in a hotter ($\sim5\times10^5$~K), 
collisionally ionized
phase. The inferred temperature for the collisional phase is within
the range of temperatures where departures from equilibrium are expected
to be important
(Gnat \& Sternberg 2007; hereafter GS07).
While observations have confirmed the existence of the warm/hot clouds,
detailed theoretical
modeling of the physical properties in the gas is needed to 
determine what fraction of baryons they harbor, and whether they
confirm the theoretical predictions regarding the shock-heated 
WHIM (Furlanetto et al.~2005).
Investigating how the absorption line properties of fast radiative
shock waves depend on gas metallicity is particularly interesting in
the context of intergalactic shocks, as some of the WHIM absorbers are
expected to have significantly subsolar metallicities
(e.g.~Cen \& Ostriker~2006).

We focus on fast shocks, with velocities, $v_s$, of 
$600$ 
and
$2000$~km~s$^{-1}$, 
corresponding to post-shock temperatures, $T_s$, of 
$5\times10^6$ to $5\times10^7$~K.
After being heated by the shock, the gas
emits energetic radiation which may later be absorbed
by cooler gas further downstream, and by the unperturbed gas
that is approaching the shock front.
The shock self-radiation 
significantly affects 
the ionization states and thermal properties of the gas.

In following the time-dependent ion fractions, we rely 
on the code and work 
presented in GS07
where we focused on pure 
radiative cooling and ionization of gas clouds,
with no external source of heating or photoionization.
Below $\sim10^6$~K, cooling can become
rapid compared to ion-electron recombination. 
When recombination lags behind cooling
the gas at
any temperature tends to remain over-ionized compared to gas
in collisional ionization equilibrium (Kafatos et al.~1973;
Shapiro \& Moore~1976; Edgar \& Chevalier~1986; Schmutzler
\& Tscharnuter~1993; Sutherland \& Dopita~1993).
In GS07 we presented results for metal ionization states
assuming collisional ionization equilibrium,
as well as for the non-equilibrium ionization states
that occur in radiatively cooling gas in which recombination
lags behind cooling.
Here we extend 
our work
to compute the integrated metal-ion
column-densities produced in steady flows of
gas, 
in the post-shock cooling layers behind fast shocks.
The computations we present in this paper build upon the results we
presented in GS07. In addition
to calculating the non-equilibrium ionization and cooling,
we follow the dynamics of the cooling gas and 
the radiative transfer of the shock 
self-radiation through the post-shock cooling layers.  
We include the effects of photoionization and heating
by the shock radiation, and also self-consistently
compute the initial photoionization state of the precursor gas.

The diagnostic characteristics of gas cooling behind
steady shock waves have been studied by many authors dating back to
Cox~(1972), Dopita~(1976; 1977) and Raymond~(1979).
These works computed the time-dependent ionization and cooling
in the post-shock cooling layers.
Shull \& McKee (1979) added a self-consistent computation of the ionization
states in the radiative precursor produced by shock radiation
emitted in the up-stream direction.
All of these investigations considered intermediate or
low-velocity shocks, with $v_s\lesssim200$~km~s$^{-1}$.
Kang \& Shapiro (1992) considered shocks
with $v_s$ up to 400 km s$^{-1}$,
for gas with primordial compositions.
For slow shocks, the neutral fraction in the gas entering the
shock may be considerable, and 
this significantly affects the
thermal and dynamical evolution of the post-shock gas.
The metal-line signatures of slow shocks differ significantly
from those of faster shocks, both because collisional ionization
and excitation behind cooler shocks are less efficient, and
because the photoionizing self-radiation produced by the
shocked gas is much less intense and less energetic.

The observational signatures of fast shock waves 
($v_s>400$~km~s$^{-1}$) were 
initially studied by Daltabuit, MacAlpine \& Cox (1978), and
Binette et al.~(1985) assuming collisional ionization equilibrium,
later by  Dopita \& Sutherland~(1995; 1996), and more recently by 
Allen et al.~(2008), who explicitly followed the time-dependent ion 
fractions, and took into account the photoionization and heating
of both the ``downstream'' gas and radiative precursor by the
shock self-radiation.
Allen et al. present results for
shocks with velocities between $100$ and 
$1000$~km~s$^{-1}$, and with magnetic field
parameters $B/\sqrt{n}$ between $10^{-4}$ and $10$~$\mu G$~cm$^{3/2}$.
They considered five different heavy element abundance-sets,
all with a roughly solar metals-to-hydrogen ratios.
They presented results for the integrated 
metal column densities and
resulting emission line properties.

None of the previous discussions investigated how the
cooling column densities depend on the overall metallicity of
the shocked gas.
For the $600$
and $\sim2000$~km~s$^{-1}$ shocks that we consider, we 
present a complete set
of results for
gas metallicities $Z$
ranging from $10^{-3}$ to twice the
solar abundance of heavy elements.
We investigate
how the
observational signatures of fast radiative shocks
depend on $Z$. Our emphasis is on the
absorbing metal column densities that are produced
in the post-shock cooling layers.
We present computations for shocks in which the magnetic field
is negligible ($B=0$) so that the cooling occurs at
approximately constant pressure (isobaric), and for shocks
in which the magnetic field 
dominates the pressure everywhere
so that the gas cools at constant density (isochoric).
There are two key differences between isobaric and isochoric shocks
that we discuss in detail in this paper. First, in isobaric flows
the cooling times are shortened in the downstream gas due to the
higher densities that obtain as the gas cools and compresses.
This leads to a large suppression of the integrated columns densities
of species that are produced mainly in the cooler parts of the flows,
in isobaric versus isochoric shocks. Second, the effects of 
photoionization by the shock radiation are much more important in
isochoric shocks where the ``ionization parameters'' remain large
because the gas densities do not grow as the radiation is absorbed.
Photoionization in isochoric shocks leads to increased 
ionization in the cooler downstream absorbing layers, over and
above what is produced by recombination lags in the rapidly cooling gas.
The magnitude of these effects is sensitive to the metallicity,
via its control of the cooling rates and the associated shock self-radiation. 

In our computations we make several simplifying assumptions.
First, we assume steady state, one dimensional,
plane-parallel shocks.
Fast radiative shocks are known to be subject to thermal
instabilities, both locally -- where small inhomogeneities 
in the gas grow to produce clumps and create secondary shocks,
and globally -- where the shock velocity oscillates, thus affecting the
dynamics of the flowing gas (e.g.~Draine \& McKee~1993).
Global shock instabilities are expected to occur for shock
velocities in excess of $140$~km~s$^{-1}$, based on both
analytical (e.g.~Chevalier \& Imamura 1982; Innes et al.~1987;
Innes~1992; Toth \& Draine~1993)
and numerical work (e.g.~Langer et al.~1981; Gaetz et al.~1988; 
Toth \& Draine~1993; Strickland \& Blondin~1995; 
Sutherland et al.~2003; Pittard et al.~2005).
Global oscillations of the shock velocity may affect the
absorption and emission line signatures of the post-shock
cooling gas, and it is unclear to what extent predictions
based of steady-state shock models represent the complexity
of real physical shocks.
However, as noted by Allen et al.~(2008),
full three-dimensional models of unsteady and dynamically evolving
shocks, including radiative
transfer and
time-dependent evolution of the ion fractions in
the post-shock cooling layers is computationally
very complex. This is beyond the scope of this work.
Here we construct steady-state one-dimensional models as
approximations.
In our steady state, one-dimensional models, we 
self-consistently take into account the time dependent 
ionization, cooling, radiative transfer, and dynamics
of the post-shock gas.

Second, we assume that the electron and 
ion temperatures are equal throughout the flow.
Recent observational evidence has suggested that in
``Balmer-dominated'' shocks with $v_s\gtrsim400$~km~s$^{-1}$
the electrons and ions emerge from the shock with different
post-shock temperatures (e.g.~Ghavamian et al.~2007).
In $600$ and $2000$~km~s$^{-1}$ shocks, Ghavamian et al.~estimate 
that the post-shock electron
temperatures are $\sim0.5$ and $0.05$ the proton
post-shock temperatures. The validity of the
standard electron-ion equipartition assumption has therefore
been questioned (Chevalier et al.~1980;
Ghavamian et al.~2001, 2007; Yoshida et al.~2005;
Heng \& McCray~2007; Heng et al.~2007; 
Rakowski et al.~2008; Raymond et al.~2008).
However, the electron-proton equipartition time
$t_{ep}\simeq8.4\times10^3 T_{e7}^{3/2}/n_e$~years 
(where $n_e$ is the electron density, and $T_{e7} = T_e/10^7$~K,
Spitzer~1962; Yoshida et al.~2005),
is much shorter than the cooling times,
$t_c\approx 3.9\times 10^6 T_7^{3/2}/(n_eZ)$ (GS07), 
for the physical conditions that we consider, 
and any initial temperature difference will be rapidly
removed. The electron and ion velocity distributions are
therefore expected to remain closely coupled during the
cooling process. For example, for solar metallicity ($Z=1$), we
estimate, conservatively, that for $600$ and $2000$~km~s$^{-1}$ shocks
the initial ratios of the equipartition to cooling
times are $4.3\times 10^{-3}$ and $7.2\times 10^{-2}$ respectively\footnote{In 
estimating $t_{ep}$ we set $T=T_s$. This gives an
upper bound on the equipartition time. In estimating $t_{c}$ we set
$T$ equal to the initial electron temperature $T_e$. This gives
a lower bound on the cooling time. Thus, for example, for a 600 km s$^{-1}$ shock,
with $T_s=5\times 10^6$~K, and
for which initially $T_e=0.5T_p$ (where $T_p$ is the proton temperature),
and with $T_s=(T_e+T_p)/2$, it follows that $T_e=(2/3)T_s=3.3\times 10^6$~K.
For these values of $T_s$ and $T_e$ our conservative estimates for the
equipartition and cooling times are 
$t_{ep}=3.0\times 10^3/n_e$ and $t_c=7.1\times 10^5/(n_eZ)$ years. 
This gives 
$t_{ep}/t_c=4.3\times 10^{-3}$
for $Z=1$.
}.
The ratio
$t_{ep}/t_c$ is smaller for lower $Z$.
We therefore assume
$T_e=T_{\rm ion}$ everywhere.

Third, we assume that any initial
dust is rapidly destroyed (e.g.~by thermal sputtering) 
after passing 
through
the shock front, on a time scale that is
much shorter than the cooling time.
If a high dust mass can be maintained, cooling by
gas-grain collisions may dominate the total cooling at
high temperatures (Ostriker \& Silk~1973; Draine~1981).
Draine \& McKee~(1993) estimate that as much as $15\%$ of
the initial thermal energy may be radiated away before 
the grains are completely destroyed. Even in this case,
the integrated cooling column densities are only affected
at a level of $\lesssim10\%$ 
(except for the highest
ionization state of each element which may be affected
by $\lesssim30\%$).
More recently, the work of Smith et al.~(1996) has suggested that the 
dust sputtering destruction time-scale is shorter than 
the cooling time when $T\gtrsim3\times10^6$~K.
We consider higher shock temperatures, 
and assume that any dust is instantaneously 
destroyed.
The post-shock gas then evolves with constant gas-phase
elemental abundances as specified by the metallicity, $Z$.

The outline of our paper is as follows.
In \S 2 we write down the equations of ionization, dynamics,
cooling and heating, and radiative transfer
that we solve in our computations. 
We also discuss  
our treatment of 
the initial ionization states in the gas approaching the shock front, 
and describe our numerical method.
In \S 3 we discuss the shock structure and emitted radiation,
and investigate how they depend on the controlling parameters, including
the gas metallicity, shock velocity, magnetic field, and gas
density.
In \S 4 we present our computations of the non-equilibrium
ionization states in the post-shock cooling layers. We 
describe how photoionization by the shock self-radiation
and departures from ionization equilibrium affect the
ion fractions in the cooling gas.
In \S 5 
we discuss the radiative cooling 
and heating efficiencies.
Finally, in \S 6 we present the integrated metal-ion 
column-densities that are produced in 
steady flows 
of cooling
gas behind fast radiative shocks, for gas metallicities between
$10^{-3}$ and $2$ times solar abundances.
We discuss how the column densities depend on the gas metallicity,
shock velocity, and magnetic field.
In \S~7 we compute the 
equilibrium metal-ion column densities that are produced in the 
photoionized
precursors of these fast radiative shocks.
We summarize in \S 8.

In this work we have generated a large number of tables
containing our numerical results. These are
available as online data files at
http://wise-obs.tau.ac.il/$\sim$orlyg/shocks/.

\section{Basic Equations and Processes}
\label{physics}

We are interested in studying the evolving ionization states
in steady flows of cooling gas.
We focus on the ionization states in post-shock cooling
layers behind steady fast radiative shock waves.
The
gas is heated to some high temperature $T_s \gtrsim 5\times 10^6$ K
by the shock, and then cools and recombines as it flows away from the shock front.
In our models we follow a gas element as it advances through the post-shock flow.
If the gas cools faster than it recombines,
non-equilibrium effects become significant, and the gas remains over-ionized.
Non-equilibrium ionization leads to a suppression of the cooling rates. 
As in GS07, we compute this coupled time-dependent evolution.
We describe the ionization and cooling in \S~\ref{physics-ion}
and \S~\ref{phys-cooling} below.

The dynamics of a gas parcel that moves through the flow is determined
by the continuity equations for the mass-, momentum-, and energy-fluxes (see \S~\ref{flow-eqn}).
We explicitly follow these equations to derive the local gas velocity, density, temperature,
and pressure.

As gas enters the shock, it is heated to a high temperature $T_s$.
The gas then flows away from the shock front, and gradually radiates its
thermal energy. For fast shock waves, this radiation is intense and energetic.
The shock self-radiation is later absorbed by cooler gas further downstream,
providing a source of heating and photoionization, which greatly affects the
shock structure and the ion fractions in the cooling gas.
To compute the local ionization and heating rates everywhere in
the flow, we follow the transfer of the shock self-radiation as it
advances downstream (see \S~\ref{rad-transf}).

The radiation field that is emitted by the shock penetrates the gas approaching
the shock front, and creates a radiative precursor
that is photoionized by
this radiation (see \S~\ref{precursor-phys}).
The ionization states in the radiative precursor are in photoionization 
equilibrium with the shock self-radiation, and set the ion fractions in
the gas entering the shock front. As we describe in
\S~\ref{numeric}, we iterate to find a self-consistent solution for the initial ionization
state and the shock self-radiation.

As we describe in \S~\ref{introduction}, 
we assume that the
shocks are steady, and that after passing the
shock fronts the electron and ion temperatures are equal.
In computing the cooling rates and abundances, we neglect cooling by molecules and dust.

\subsection{Ionization}
\label{physics-ion}

As 
in GS07, we consider all ionization stages of the 
elements H, He, C, N, O, Ne, Mg, Si, S, and Fe. We include collisional
ionization by thermal electrons, radiative recombination, dielectronic 
recombination, and neutralization and ionization by charge transfer
reactions with hydrogen and helium atoms and ions (GS07, and references therein).
Here we add
statistical charge transfer rate coefficients for
high-ions with charges greater than $+4$ (Kingdon \& Ferland~1996).
In this work we also include photoionization (Verner et al.~1996) and 
multi-electron Auger
ionization processes (Kaastra \& Mewe~1993) induced by the X-ray photons
emitted by the cooling gas. 
(As we discuss in \S 4.6, the effects of Auger processes are small.)

The time-dependent equations for the ion abundance
fractions, $x_i$, of element $m$ in ionization stage $i$,
at every point in the flow, are
\begin{equation}
\label{ion-neq}
\begin{array}{l}
\displaystyle{\frac{dx_i}{dt}} = x_{i-1}~~[q_{i-1}n_{\rm e} + k^{\rm H}_{\uparrow i-1}n_{\rm H^+}
+ k^{\rm He}_{\uparrow i-1}n_{\rm He^+}]
+\displaystyle{\sum_{j<i}}x_j\Gamma_{j\rightarrow i}\\
\;\;\;\;\;\;\;\;\;\;\; + x_{i+1}~~[\alpha_{i+1}n_{\rm e} +
k^{\rm H}_{\downarrow i+1}n_{\rm H^0}
+ k^{\rm He}_{\downarrow i+1}n_{\rm He^0}] \\
\;\;\;\;\;\;\;\;\;\;\; - x_{i}~~[(q_{i} + \alpha_{i})n_{\rm e} + \Gamma_i +
k^{\rm H}_{\downarrow i}n_{\rm H^0}
+ k^{\rm He}_{\downarrow i}n_{\rm He^0}\\
\;\;\;\;\;\;\;\;\;\;\;\;\;\;\;\;\;\;\;\;\; + k^{\rm H}_{\uparrow i}n_{\rm H^+}
+ k^{\rm He}_{\uparrow i}n_{\rm He^+}] \ \ \ .
\end{array}
\end{equation}
In this expression, $q_i$ and $\alpha_i$ are the 
temperature-dependent
rate coefficients
for collisional ionization and recombination (radiative plus
dielectronic), and
$k^{\rm H}_{\downarrow i}$, $k^{\rm He}_{\downarrow i}$,
$k^{\rm H}_{\uparrow i}$, and $k^{\rm He}_{\uparrow i}$
are the rate coefficients for charge transfer reactions
with hydrogen and helium that lead to ionization or neutralization.
The quantities
$n_{\rm H^0}$, $n_{\rm H^+}$, $n_{\rm He^0}$,
$n_{\rm He^+}$, and $n_{\rm e}$ are the particle densities (cm$^{-3}$)
for neutral hydrogen, ionized hydrogen, neutral helium, singly ionized helium, 
and electrons, respectively. $\Gamma_{j\rightarrow i}$ are the 
local
rates of photoionization 
of ions $j$ which result in the ejection of $i-j$ electrons.
$\Gamma_{i}$ are the total photoionization rates of ions $i$ due
to externally incident radiation.

For each element $m$, the ion fractions $x_i = n_{i,m} / (n_HA_m)$ must
at all times satisfy
\begin{equation}
\sum x_i=1 \ \ \ ,
\end{equation}
where $n_{i,m}$ is the density of ions $i$ of element $m$,
$n_H = n_{\rm H^0} + n_{\rm H^+}$,
and $A_m$ is the abundance of element $m$ relative to hydrogen.

\subsection{Dynamics}
\label{flow-eqn}

As we discussed 
in \S~\ref{introduction}, 
we assume
one-dimensional, steady-state radiative shocks.
We neglect any instabilities that may form,
even though fast radiative shocks are knows to be subject to a
global oscillatory instability for shock velocities above
$\sim140$~km~s$^{-1}$ (e.g.~Ghavamian et al.~2007).
We also assume that the electron and ion temperatures are equal everywhere.
As discussed in \S 1, the electron-proton equipartition times
are much shorter than the cooling times for the shock velocities we consider.
The electron- and ion-velocity distributions are
therefore expected to remain closely coupled during the
cooling process.

Under these assumptions, we follow the standard Rankine-Hugoniot conditions (e.g. Cox~1972):
\begin{equation}
\label{conservations}
\begin{array}{l}
\rho v = \rm{const} \\
B / \rho = \rm{const}\\
\rho v^2 + P + \frac{1}{8\pi}B^2 = \rm{const} \\
\frac{1}{2} \rho v^3 + \frac{1}{\gamma-1} P v + P v + 
\frac{1}{4\pi}B^2v + \int_{0}^{x} (n_e n_H \Lambda - n \Upsilon) dx = \rm{const}
\end{array}
\end{equation}
where $\rho$ is the mass density, $v$ the velocity in the direction of the 
flow, $P$ is the gas pressure, $B$ is the magnetic field perpendicular
to the direction of the flow, $n$ is the particle density, $\Lambda$
is the cooling efficiency (erg~s$^{-1}$~cm$^3$), and $\Upsilon$ the heating 
rate (erg~s$^{-1}$).
These equations represent the conservation of the mass-, momentum-, and energy-flux
in the flow, and the condition of ``frozen in'' magnetic field.

The assumption of a strong shock implies the familiar jump conditions 
(e.g.~Draine \& McKee~1993)
relating  the pre-shock and post-shock physical properties: $n_{0} = 4 n_{pre}$,
$v_{0} = 1/4 ~v_s$. In these expressions $n_0$ and $v_0$ are the post-shock
particle density and velocity, $n_{pre}$ is the pre-shock density,
and $v_s$ is the shock velocity
in the frame of the pre-shocked gas.
The shock temperature and velocity are
related by $T_s = 3\mu v_s^2 / 16k_{\rm B}$ (e.g.~McKee \& Hollenbach~1980),
where $\mu$ is the mean mass per particle,
and $k_{\rm B}$ is the Boltzmann constant.

We consider two limiting cases.
First, we consider flows in which $B=0$ everywhere.
The gas then cools at approximately constant pressure (see \S~\ref{phys-cooling}).
Second, 
we study the case where the magnetic field is
``dynamically dominant'' everywhere, 
with
$B/\sqrt{\rho}\gg v_s$, so
that it dominates the pressure throughout the flow
(see equation~[3]).
In this case, the condition of ``frozen-in'' magnetic field
implies constant density
evolution.

Realistic shocks may have some intermediate value 
for the magnetic field,
and the observed signatures of such shocks will lie between the
two limiting cases presented in this paper (e.g. Dopita \& Sutherland~1996;
Allen et al.~2008).

\subsection{Cooling}
\label{phys-cooling}
The ionization equations are coupled to an energy
equation for the time-dependent heating and 
cooling, and resulting temperature variation.

We follow the electron cooling
efficiency, $\Lambda(T,x_i,Z)$ 
(erg~s$^{-1}$~cm$^3$), 
which depends on the gas temperature, the 
ionization state, and the total abundances of
the heavy elements specified by the metallicity $Z$.
As in GS07, we adopt the elemental abundances reported by
Asplund et al. (2005) for the photosphere of the Sun, and the 
enhanced
Ne abundance
recommended by Drake \& Testa (2005). We list our assumed solar abundances 
in Table~\ref{solar}. In all computations we assume a primordial
helium abundance $A_{\rm He}=1/12$ (Ballantyne et al.~2000),
independent of $Z$. 

\begin{deluxetable}{lr}
\tablewidth{0pt}
\tablecaption{Solar Elemental Abundances}
\tablehead{
\colhead{Element} & 
\colhead{Abundance}\\
\colhead{} &
\colhead{(X/H)$_{\odot}$} }
\startdata
Carbon   & $2.45\times10^{-4}$ \\
Nitrogen & $6.03\times10^{-5}$ \\
Oxygen   & $4.57\times10^{-4}$ \\
Neon     & $1.95\times10^{-4}$ \\
Magnesium& $3.39\times10^{-5}$ \\
Silicon  & $3.24\times10^{-5}$ \\
Sulfur   & $1.38\times10^{-5}$ \\
Iron     & $2.82\times10^{-5}$ \\
\enddata
\label{solar}
\end{deluxetable}

The electron cooling efficiency
includes the removal of electron {\it kinetic} energy
via recombinations with ions, collisional 
ionizations, collisional excitations followed by
prompt line emissions, and thermal bremsstrahlung.
We do not include the ionization potential
energies as part of the internal energy 
but instead follow the loss and gain
of the electron kinetic energy only (see GS07).

We also follow the heating
rate, $\Upsilon(x_i,Z,J_\nu)$ (erg~s$^{-1}$)
due to absorption of the shock self-radiation,
$J_\nu$,
by gas further downstream.
The net local cooling rate per volume is given
by $n_e n_H \Lambda - n\Upsilon$, where $n$ is the
total gas density.

For an ideal gas, the pressure $P=nk_{\rm B}T$,
and the thermal energy density $u=3/2nk_{\rm B}T$,
where $T$ is the gas temperature, and $n$ is the 
total particle density.
We follow the gas pressure, density,
temperature and velocity as is flows away from
the shock front.

When the magnetic field is set to zero,
the Rankine-Hugoniot conditions relate the net local
cooling to the gas deceleration:
\begin{equation}
\label{dv}
\frac{3}{2} \frac{c^2-v^2}{v} \frac{dv}{dt} = - \frac{n_e n_H \Lambda - n\Upsilon}{\rho},
\end{equation}
(e.g. Shu~1992), where $c^2 = (5/3)~P/\rho$ is the sound speed.
We use equation 
(\ref{dv})
to follow the evolution of
the velocity along the flow.
We later use the mass and momentum continuity
conditions to derive the gas density and pressure
via,
\begin{equation}
\begin{array}{l}
\rho = \rho_0 v_0 / v \rm{~~~~and} \\
P = \rho_0 v_0 (4v_0 - v),
\end{array}
\end{equation}
where $\rho = n\mu$ and $v$ are the local gas mass
density and velocity, $\rho_0$ and $v_0$ are the
post-shock mass-density and velocity, and $\mu$ is the
mean mass per particle.
Right after passing the shock front, the gas density
is determined by the jump conditions 
($\rho_0 = 4\rho_{pre};~~ v_0 = v_s/4$). The gas
pressure is then given by $P_0 = 3\rho_0 v_0^2$
(see Equation~[5]).
As the gas radiates all of its internal energy, its 
velocity becomes very small, and the final
pressure is then $P_{\infty}=4\rho_0 v_0^2 = 4/3~P_0$.
Thus, $B=0$ corresponds to approximately isobaric dynamics.

In the limit of a dynamically dominant magnetic field
($B/\sqrt{\rho}\gg v_s$), the 
Rankine-Hugoniot conditions imply constant density and 
velocity throughout the flow, so that $\rho=\rho_0$
and $v=v_0$ everywhere. 
This is an isochoric flow.
We refer to this as the ``strong-$B$'' limit.
Energy conservation then implies that the radiated
energy equals the loss of thermal energy, and
relates the net local cooling to the 
pressure change by
\begin{equation}
\label{dP}
\frac{3}{2}\frac{dP}{dt}=-(n_e n_H \Lambda - n\Upsilon)
\end{equation}
(see equation [7] in GS07).

Given a set of nonequilibrium ion abundances, 
$x_i(T)$, and the gas metallicity $Z$, we use the
cooling and heating functions included in 
Cloudy (ver. 07.02.00; Ferland et al. 1998)
to calculate $\Lambda(T, x_i, Z )$ and
$\Upsilon(x_i, Z, J_\nu)$.

\subsection{Radiative Transfer}
\label{rad-transf}

We follow the radiative transfer equation,
\begin{equation}
\label{rad_transf}
\frac{dI_\nu(\mu)}{d\tau_\nu} = I_\nu(\mu)-S_\nu,
\end{equation}
to evaluate the intensity of the radiation along the flow.
In equation (\ref{rad_transf}), $I_\nu(\mu)$ is the 
direction-dependent specific intensity
(erg~s$^{-1}$~cm$^{-1}$~Hz$^{-1}$~sr$^{-1}$),
$S_\nu=\epsilon_\nu/\alpha_\nu$ is the source function, 
$\epsilon_\nu$ is the specific emissivity (or emission coefficient, 
erg~s$^{-1}$~cm$^{-3}$~Hz$^{-1}$~sr$^{-1}$),
$\alpha_\nu$ is the absorption coefficient 
(cm$^{-1}$),
$\tau_\nu$ is the optical depth, 
and $\mu=cos(\theta)$.

We use a Gaussian quadrature scheme to evaluate the local
intensity of the radiation as a function of distance from the shock front
(Chandrasekhar 1960).
We follow the specific intensity $I_\nu(\mu)$ along ten downstream directions
between $\mu=1$ (parallel to the shock velocity)
and $\mu=0$ (perpendicular to the velocity).
We divide the flow into
thin slabs of thickness $l(T)$ (see \S~\ref{numeric}).
For each slab,
we use Cloudy to evaluate the absorption ($\alpha_\nu$)
and emission ($\epsilon_\nu$) appropriate for the local conditions
that we calculate.
We use our local non-equilibrium ionization states and temperature
at the 
upstream edge of each slab as input for the Cloudy models.
The 
Cloudy
output includes the continuum specific emissivities,
a list of line emissivities, and the continuum absorption 
coefficients as a function of energy\footnote{
It does not include line opacities. In the Cloudy computations,
we therefore do not include continuum pumping of the lines,
as these would not be consistently absorbed.
}.

Once the opacities and emissivities for a slab are known,
we compute the radiative transfer for each of the $10$ 
values of $\mu$. The optical depth in the downstream 
direction is given by $\tau^0_\nu = l~\alpha_\nu$,
where $l$ is the thickness of the slab in the downstream direction. 
The optical depth along the different directions
is then given by $\tau_\nu(\mu) = \tau^0_\nu / \mu$.
The source function $S_\nu = \epsilon_\nu / \alpha_\nu$.
Given an input specific intensity, $I^{in}_\nu(\mu)$, 
the specific intensity at the downstream edge of a slab is
\begin{equation}
I^{out}_\nu(\mu) = e^{-\tau_\nu(\mu)}(I^{in}_\nu(\mu)-S_\nu) + S_\nu \ \ \ .
\end{equation}
$I^{out}_\nu$ is then used as input radiation for the next slab.
Finally, the mean intensity, $J_\nu = \frac{1}{4\pi}\int{I_\nu(\mu) d\Omega}$, 
is computed locally, and used in evaluating the 
local
heating 
and photoionization rates.


\subsection{Radiative Precursor}
\label{precursor-phys}

The initial conditions for the shock depend on the ionization states
in the radiative precursor.
The precursor is photoionized by the intense radiation field created
by the shocked gas (Shull \& McKee~1979).
If the velocity of the ionization front in the gas approaching the shock, 
\begin{equation}
\label{vion}
v_{ion} \equiv \frac{1}{n_{pre}} \int^\infty_{\nu_0} \frac{J_\nu}{h\nu} d\nu,
\end{equation}
(where $n_{pre}$ is the preshock density)
is larger than the shock velocity
$v_s$,
a stable photoionization equilibrium
radiative precursor will form. 
This condition is met for shock velocities  $\gtrsim175$~km~s$^{-1}$ 
(Dopita \& Sutherland 1996), and a stable radiative precursor will therefore
form for the shock velocities that we consider here.

We self consistently calculate the ionization states
in the radiative precursor 
(see \S~\ref{prec-cols}).
Since the photoionizing radiation is emitted by the shocked gas, iterations
are required to obtain a self consistent solution.
In the first iteration, we assume that the gas starts out in collisional
ionization equilibrium (CIE) at the shock temperature, $T_s$.
We then compute the resulting shock model, and follow the radiative transfer in the
downstream direction (see \S~\ref{rad-transf}).

The 
self-radiation builds up with distance from the shock front,
as more and more emitting gas contributes to the intensity.
However, at some distance from the shock front,  the gas becomes optically
thick, and the intensity begins to decrease.
We label the distance at which the gas becomes thick at the Lyman
limit $l_{thick}$.
At $l_{thick}$, most of the initial energy has been emitted, but it
has not yet been absorbed.
We assume 
that the mean intensity at this point,
$J_\nu(l_{thick})$, is similar to the intensity entering
the radiative precursor, and use $J_\nu(l_{thick})$ to compute the
ionization states in the radiative precursor.

Since our shock velocities are high enough to
ensure the formation of a steady equilibrium photoionized precursor,
we use the photoionization code Cloudy 
to compute
the ion fractions in the radiative precursor.

We then use these ionization states as initial conditions for 
a second iteration.
This process can be repeated
until the resulting photoionizing radiation $J_\nu(l_{thick})$ has converged.
We find that the second iteration is sufficient, and that
although
the ionization states in the precursor are significantly underionized
relative to CIE at the shock temperature, the
radiation field created by the shocked gas is not substantially altered.
The shock structure (e.g.~the temperature
as a function of time, and the integrated column densities) is also
similar in the first and second iterations.

The assumption that the upstream radiation field equals $J_\nu(l_{thick})$
is based on the fact that the flowing gas is optically thin between 
the shock front and $l_{thick}$.
Nevertheless, geometrical effects as well as scatterings in the downstream
gas, may increase the intensity of the radiation in upstream directions.
Our assumption may only lead to an 
underestimate of
the intensity of radiation photoionizing the precursor.
The true ionization states in the precursor may therefore be higher
than we compute, but still lower than the CIE values used in the
first iteration.
Since the two iterations lead to similar shock structures and mean
intensities, we conclude that this assumption does not
significantly affect the solution.
It may have a minor impact on the ionization states immediately following
the shock-front before the gas adjusts to CIE at the shock temperature.
This will have a negligible impact on the integrated column densities
(see \S~\ref{columns}).

\subsection{Numerical Method}
\label{numeric}

The abundance equations (\ref{ion-neq}) and the flow equation
(eq. [\ref{dv}] when $B=0$ or eq. [\ref{dP}] in the 
strong-$B$ limit) 
are a set of 103 coupled ordinary 
differential equations.
When $B=0$, we advance the numerical solution in small
velocity steps $\Delta v=\varepsilon v$, where $\varepsilon\lesssim0.005$,
and $v$ is the 
velocity associated with a temperature $T(v)$.
For strong-$B$
isochoric conditions 
we advance the solution in small pressure steps
$\Delta P = \varepsilon P$, where $P$ is the gas pressure associated 
with a temperature $T(P)$.

For any temperature $T$, we compute the total cooling and heating rates
by passing the current non-equilibrium ion fractions $x_i(T)$,
temperature $T$, and mean intensity $J_\nu$ to the 
Cloudy
cooling
and heating functions.
We then compute the time interval $\Delta t$
associated with the velocity change $\Delta v=\varepsilon v$
(when $B=0$) or pressure change $\Delta P=\varepsilon P$ 
(for strong-$B$).
In some cases, additional constraints were set when determining
the time step, 
to ensure numerical accuracy and computational
efficiency. As the gas approaches thermal equilibrium, the time steps 
$\Delta t$ become very large, and may exceed the recombination time.
We therefore apply an upper limit on the time steps,
$\Delta t_{\rm max} = {\rm{max}} (0.05\times t, 10^{11}/n_0/Z)$, 
that depends
on the current time, and on the gas density and metallicity.  
We further demand that each time step will be at most $3$ times 
the previous step. 
We find that this choice prevents significant 
numerical ``noise'', but allows the computation to proceed efficiently.

We integrate equations (\ref{ion-neq}) over the interval
$\Delta t$ using a Livermore ODE solver (Hindmarsh 1983, see GS07).
We assume that over the time step $\Delta t$ the velocity 
(or pressure) evolves linearly with time. 
In the integration, the estimated local errors on the fractional
ion abundances are controlled so as to be smaller than $10^{-6}$, 
$10^{-5}$, and $10^{-4}$, for hydrogen, helium and metals, 
respectively.

The radiation intensities
in the different directions are stored
as vectors indicating the intensities at different energies.
The energy grid contains $565$ data point between an
energy of $0.001$~Ryd and $\sim850$~Ryd. The energy grid 
is ``dense'' near 
ionization edges 
to ensure an accurate computation
of the photo-absorption rates.

When computing the ionization and heating rates, 
we assume that the intensity of the radiation field 
is constant within a slab of thickness $l=v\times \Delta t$, 
and is equal to the intensity of the radiation 
entering that slab.
For each time step, we use the ion fractions at
the beginning of the slab to compute the emissivities
and opacities of the gas in the slab. 
In following the
radiative transfer, we assume that the emissivities
and opacities are constant within the slab. 

In the first iteration, we assume that the gas starts at collisional ionization equilibrium
at the shock temperature.
We then compute the self radiation emitted by the cooling gas.
We use this radiation field to compute the initial photoionization equilibrium 
ion fractions for the second iteration (see \S~\ref{precursor-phys}).
The equilibrium photoionization solution is significantly underionized 
relative to CIE at the shock temperature.
Very rapid changes therefore occur as the gas adjusts to CIE just below
the shock temperature. In order to follow this rapid evolution 
accurately, we set  $\varepsilon = 0.0001 (\frac{5\times10^6}{T_s})^{1.5}$ 
during this rapidly-evolving period.
We find that the mean intensity emitted by the cooling gas converges by the
second iteration to a level of $10\%$, and no further iterations are
required.

\section{Shock Structure and Scaling Relations}
\label{struct}

In this section we describe the shock structure, and how it depends on 
the controlling parameters: shock temperature (or velocity), gas metallicity, 
and magnetic field.
We show results for two values of shock temperatures: 
$5\times10^6$~K ($v_s\simeq600$~km~s$^{1}$),
and $5\times10^7$~K ($v_s\simeq1920$~km~s$^{-1}$).
We explore five different values of the gas metallicity $Z$,
from $10^{-3}$ to $2$ times
the metal abundance of the Sun.
For each shock temperature and gas metallicity we study the shock structure
and ion fractions for the $B=0$,
and strong-$B$ 
limits.

\begin{figure*}
\epsscale{1}
\plotone{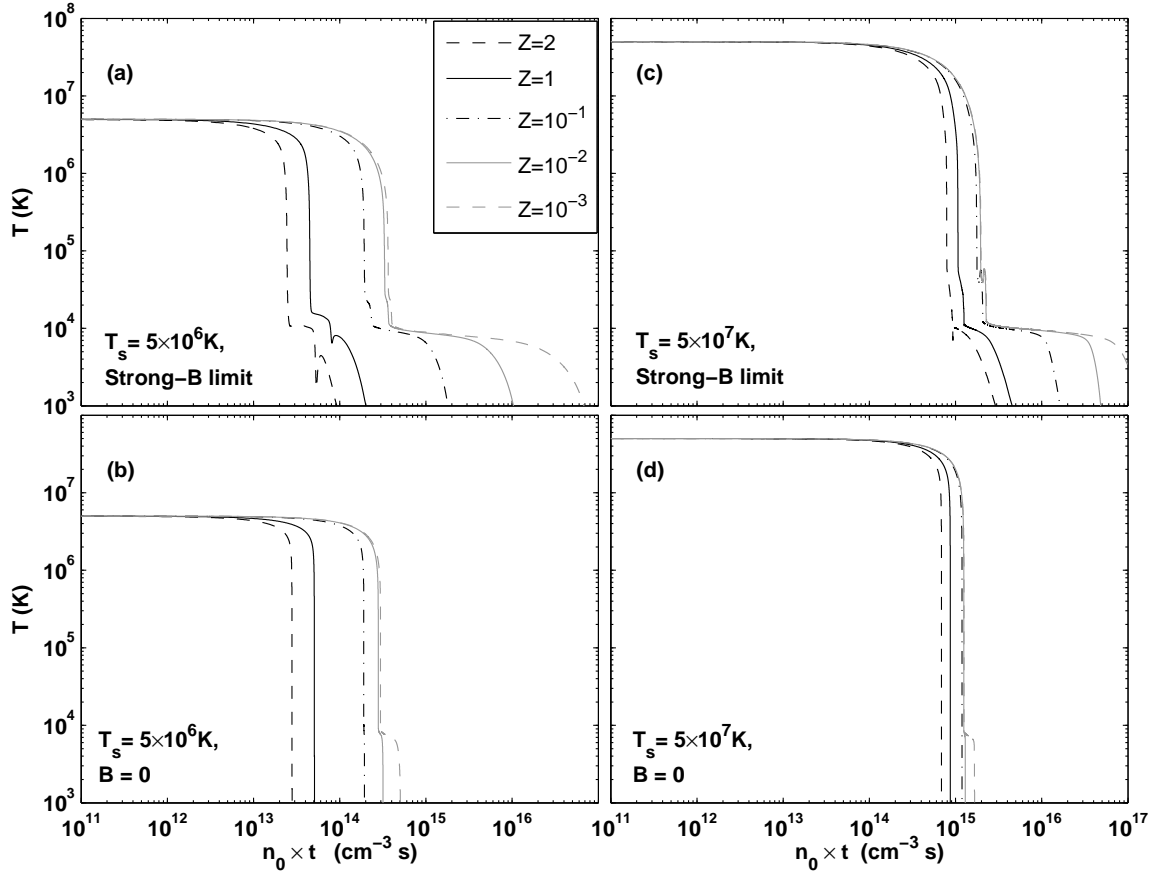}
\caption{ Temperature versus post-shock hydrogen
density times time (cm$^{-3}$~s), 
for gas metallicities ranging from $10^{-3}$ to $2$ times solar.
The upper panels show results in the
strong-$B$ limit (isochoric)
and the lower
panels show results for $B=0$ (approximately isobaric).
The left hand panels show results for $T_s = 5 \times 10^6$~K, 
and the right hand panels for $T_s = 5 \times 10^7$~K.
The results presented in this figure were computed assuming a
post-shock hygrogen density, $n_{\rm H}=0.1$~cm$^{-3}$,
but are nearly independent of density (see \S~3.5).
}
\label{Tt}
\end{figure*}

Figure~\ref{Tt} shows the temperature profiles of the post shock cooling
layers for the different cases that we study.
Panel~(a) shows results for $T_s=5\times10^6$~K, 
for a 
``strong-$B$'' isochoric shock.
The different curves show results for different gas metallicities.
The horizontal axis shows $n_0\times t$ - the initial post-shock hydrogen
density times time. This scheme makes the results nearly independent of
density, as we discuss in section 3.5 below.

\subsection{A strong-$B$ (isochoric), $T_s=5\times10^6$~K, shock at a metallicity of $Z=2$.}
\label{shock-details}

To illustrate the behavior of the post-shock cooling gas in the 
strong-$B$ limit, 
we first focus on the results for $T_s=5\times10^6$~K and 
$Z=2$. 

The initial ionization states of the gas that enters the shock,
are set by photoionization equilibrium of the precursor gas with
the shock self radiation.
This gas has an equilibrium temperature 
which is
significantly lower than the shock temperature.

As the precursor gas passes through the shock front, its temperature
abruptly rises to $T_s$, leaving the gas under-ionized relative to
CIE at the shock temperature.
The gas then very rapidly adjusts to CIE at a temperature close to 
$T_s$. 
During this phase, the gas 
radiates
very efficiently, 
as the hot thermal electrons efficiently excite the low-energy
transitions of the under-ionized gas.
Dopita and Sutherland~(1996) 
refer to this phase as the ``ionization zone''.
The cooling efficiencies are 
about 
$60$ times higher than at CIE.
The evolution 
to CIE is very rapid, and occurs
within $n_0\times t = 10^{11}$~cm$^{-3}$~s.
After this time, the ion fractions and cooling efficiencies reach CIE,
at a temperature very close to $T_s$ ($>0.95T_s$).

For the shock temperatures that we consider ($\gtrsim5\times10^6$~K),
the CIE cooling efficiency at $\sim T_s$ is low. The gas therefore stays
hot for a long time. 
This 
phase is the ``hot radiative zone''
(e.g.~Draine \& McKee~1993; Dopita \& Sutherland~1996).
Metal line emissions 
(and bremsstrahlung emission for low gas metallicity) dominate the 
cooling in this zone (Sutherland \& Dopita~1993; GS07; see \S~\ref{cooling}). 

\begin{figure}[!h]
\epsscale{1}
\plotone{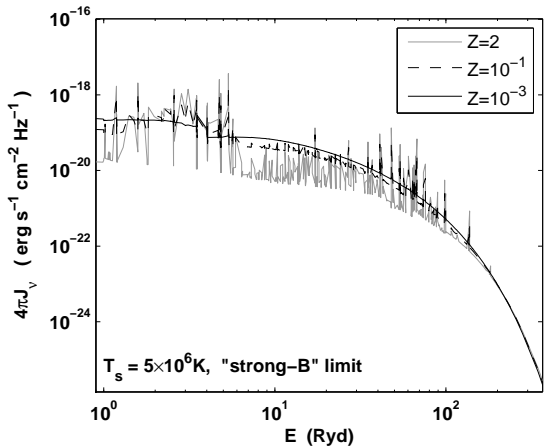}
\caption{ Emitted mean intensity~(erg s$^{-1}$~cm$^{-2}$~Hz$^{-1}$) 
versus energy (Ryd) for $T_s = 5 \times 10^6$~K in the 
strong-$B$ (isochoric)
limit.
The gray curve shows the emitted spectrum for $Z=2$, the dark
dashed curve for $Z=10^{-1}$, and the dark solid curve for $Z=10^{-3}$.}
\label{specs-Z}
\end{figure}

The hot radiative
phase ends once the gas cools to a temperature
for which the cooling efficiency is high, $T\lesssim10^6$~K.
The shocked gas therefore emits most of its initial thermal energy
while in the hot radiative 
phase.
Figure~\ref{specs-Z} shows the mean intensity of the radiation field 
emitted by the shocked gas. The gray curve shows results for $Z=2$ 
times solar metallicity gas.
For $Z=2$, the relative contribution of 
resonance
lines is very large, and the
line to continuum contrast 
is
high. The spectrum shows a very prominent
``UV-bump'' created by numerous UV emission lines,
with excess radiation between $\sim1-5$~Ryd. The line
contribution is also significant at far-UV and even X-ray energies.
This radiation is absorbed by cooler gas further downstream,
providing a source of heating and photoionization.

Once the temperature drops sufficiently to bring the gas closer to the
cooling-efficiency peak, the temperature decline becomes very rapid,
and the gas cools to a temperature of a few~$\times10^4$~K.
If cooling becomes faster than recombination, departures from 
equilibrium occur, and the gas tends to stay over-ionized (Sutherland
\& Dopita~1993; GS07).
We refer to this phase as the ``non-equilibrium cooling zone''
(e.g.~Dopita \& Sutherland~1996).
During this rapid cooling stage, metal line 
emissions 
(and Hydrogen-Helium line emission 
for $Z\lesssim10^{-2}$) dominate the cooling.

\begin{figure}[!h]
\epsscale{1}
\plotone{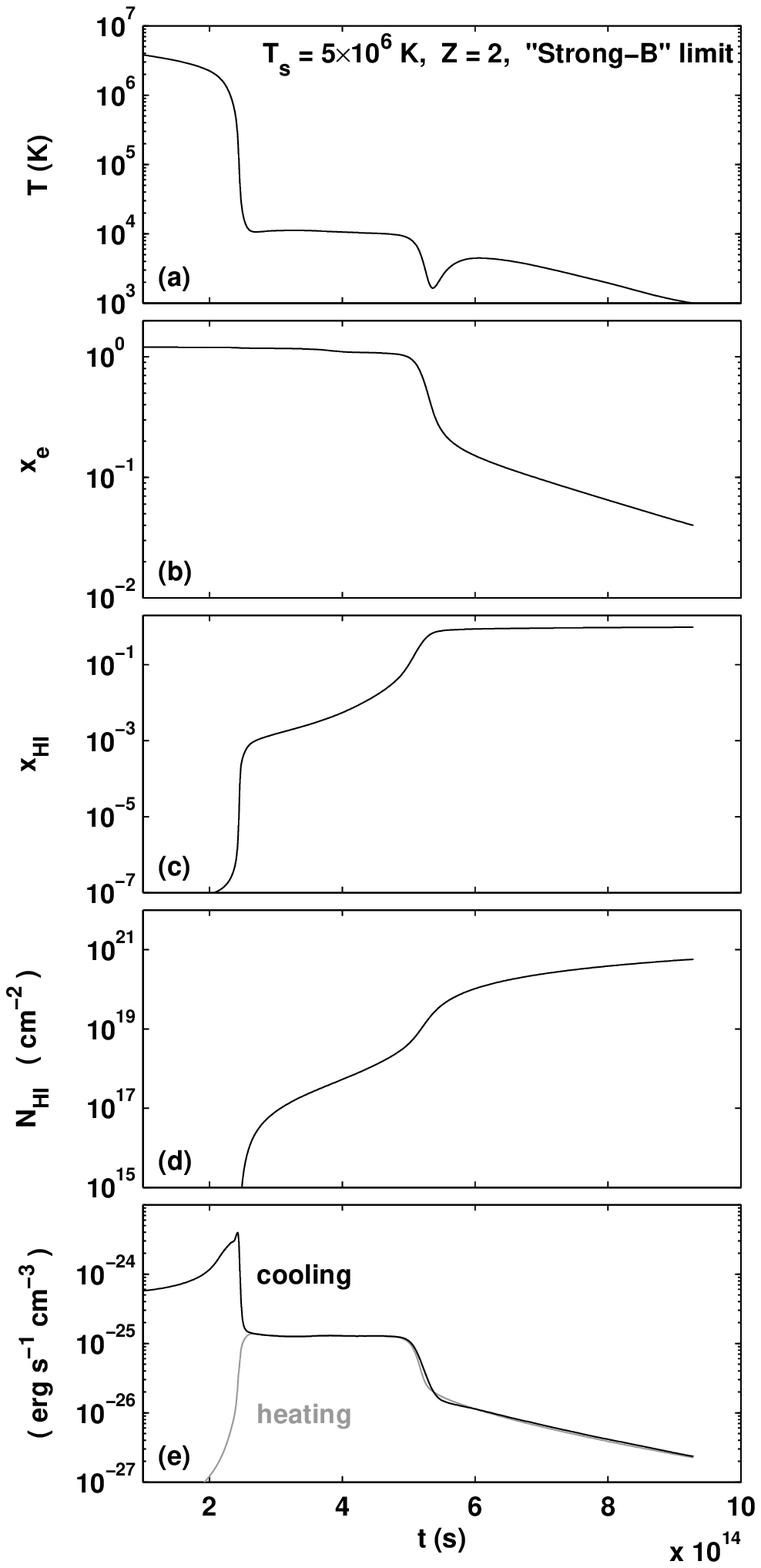}
\caption{ Shock structure versus time for $T_s=5\times10^6$~K,
for $Z=2$ solar, and for a post-shock hydrogen density $n_0=0.1$~cm$^{-3}$,
in the 
strong-$B$ limit.
(a)~Temperature~(K). (b)~Electron fraction. (c)~Neutral hydrogen fraction.
(d)~Integrated \ion{H}{1} column density~(cm$^{-2}$) from the shock front.
(e)~Local cooling (dark) and heating (gray) rates per volume~(erg~s$^{-1}$~cm$^{-3}$).}
\label{isoc-ex}
\end{figure}

In Figure~\ref{isoc-ex} we display 
various 
physical parameters in
the cooling gas. The upper panel shows the temperature as a 
function of (linear) time, starting at the non-equilibrium
cooling zone.
As the gas cools, hydrogen starts to recombine (see panel~c).
Eventually, the neutral hydrogen fraction is high enough that
it allows for efficient absorption of the shock 
radiation.
This occurs
when $x_{\rm H~I}\sim10^{-3}$. 
This significantly raises the heating rate, as is shown by
the gray line in panel~(e). The integrated \ion{H}{1} column
density measured from the shock front is shown in panel~(d).

Once the heating rate reaches the cooling rate (see panel~e), 
the rapid temperature decrease stops, and the gas enters a
``plateau'' in which the temperature remains roughly constant, 
and the radiation is gradually 
absorbed.
This is the ``photoabsorption plateau''
(e.g.~Dopita \& Sutherland~1996; also called the ``recombination zone''
[Shull \& McKee~1979] 
or the ``thermalization zone''
[Draine \& McKee~1993]).
As the radiation is 
removed by absorption, the neutral
fraction of the gas rises. 
In this optically thin and ionized part of the  plateau,
the heating rate changes very slowly\footnote{ The heating rate 
$\Upsilon \simeq x_{\rm H\;II}^2 \alpha <E>$, where $<E>$ is 
the mean photoelectron energy,
and $\alpha$ the recombination coefficient.
While the gas is optically thin and $x_{\rm H\;II}^{} \sim 1$,
the heating rate is close to constant.}.
Once the neutral fraction is of order unity ($\gtrsim0.1$),
the optical depth increases and
the heating rate declines more rapidly.
We refer to the part of the plateau in which hydrogen is 
still
ionized
as the warm ionized medium (WIM) plateau.

The depth of the ``WIM plateau'' depends on the 
flux, $F_\nu$, of ionizing
radiation entering the plateau. The thickness of the
ionized region will be
\begin{equation}
\label{WIM}
l_{\rm WIM} = \frac{1}{n_Hn_e\alpha}\int\frac{F_\nu}{h\nu}~d\nu,
\end{equation}
where $\alpha$ is the recombination coefficient.
The time spent in the WIM plateau is then 
$t_{\rm WIM} = l_{\rm WIM}/v_0\simeq2.5\times10^{14}$~s,
following a $5\times10^6$~K,
$v_0=v_s/4 = 150$~km~s$^{-1}$, $n_0=0.1$~cm$^{-3}$ isochoric shock.

For the set of parameters considered here ($T_s=5\times10^6$~K,
$Z=2$ solar, 
strong-$B$), after
the
hydrogen recombines
the gas rapidly cools down to a temperature $T_{\rm low}=1000$~K,
at which we 
terminate
the computation. 
For other 
(lower)
gas metallicities,
an additional warm neutral medium (WNM) plateau may form,
with a temperature between $8000$ and $10,000$~K.
The depth of the ``WNM plateau'' depends on the gas 
X-ray opacity,
which is set by the gas metallicity.
The WNM plateau is 
more extended
for lower metallicity gas (see Figure~\ref{Tt}).

\begin{figure}[!h]
\epsscale{1}
\plotone{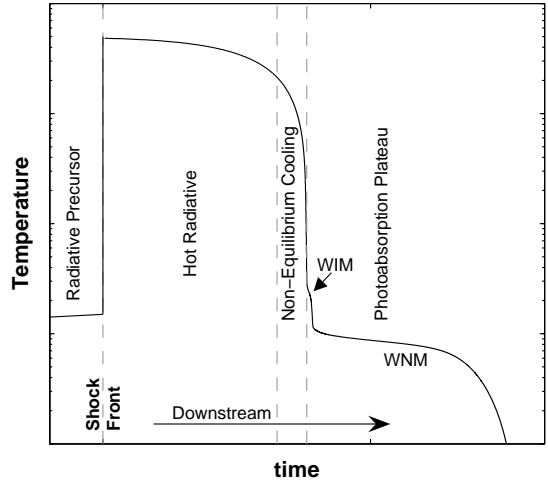}
\caption{ A schematic plot of shock temperature versus time.
The different zones are shown (from left to right):
The preshocked
upstream gas is initially in a 
photoionized
``radiative precursor'';
After passing the shock front, the
gas enters the ``Hot Radiative''
zone, in which cooling is slow.
It later passes through the ``Non-Equilibrium Cooling'' zone, in which
cooling is rapid, and departures from equilibrium ionization and cooling may occur.
After the neutral ion fraction rises, photoabsorption heating becomes efficient,
and the gas enters a ``Photoabsorption Plateau'',
consisting of WIM and WNM components,
in which the temperature
declines much more slowly. Finally the gas cools to the minimal computed
temperature. See text.}
\label{schem}
\end{figure}

The various components in post-shock cooling layers 
(e.g.~Draine \& McKee 1993; Dopita \& Sutherland 2003)
are illustrated schematically in Figure~\ref{schem}.
Here we have made the further distinction
between warm ionized (WIM) and warm neutral (WNM) plateaus
in the downstream absorbing layers.

Throughout most of the photoabsorption plateau and the final
cooling following it, the radiation field is absorbed slowly enough that
the gas has time to approach photoionization and thermal equilibrium.
The photoabsorption time scale is given by $t_{abs} = \frac{1}{<n\sigma>v}$,
and the recombination time scale by $t_{rec} = \frac{1}{n_e\alpha}$.
As long as $t_{rec}<<t_{abs}$, the gas stays at a state of near
photoionization equilibrium with the local radiation field.

Once the temperature drops below $\sim10^4$~K, Ly$\alpha$ 
emission and other allowed transitions become very inefficient, 
and the gas cools primarily via fine-structure (FS)
line emissions, including \ion{O}{3}~88.3$\mu$m,
\ion{O}{3}~51.8$\mu$m, \ion{Ne}{3}~15.5$\mu$m,
\ion{Si}{2}~34.8$\mu$m, \ion{S}{3}~33.5$\mu$m,
and \ion{C}{2}~157.7$\mu$m. (We do not include molecules or dust in
this computation).

As mentioned above, for $Z=2$ metal line cooling dominates throughout.
Metal lines are important even for $T\sim10^4$~K where Ly$\alpha$ cooling
dominates in CIE. 
The Ly$\alpha$ cooling efficiency 
is suppressed 
because the hydrogen is still mostly ionized,
while the existence of over-ionized metal species, such as
Ne$^{4+}$, O$^{3+}$, S$^{3+}$,
allows for efficient line
cooling that is not available in CIE.
Initially, the heating in the plateau is due to \ion{H}{1} photoionization
(for $n_0 \times t < 5.3 \times 10^{13}$~cm$^{-3}$~s), then
\ion{He}{1} photoionization
(for $5.3 \times10^{13} < n_0 \times t < 7.4 \times 10^{13}$~cm$^{-3}$~s),
and 
then by photoionization of metals, mostly by 
neutral oxygen 
(for $n_0 \times t > 7.4 \times 10^{13}$~cm$^{-3}$~s).
These transitions 
occur as the photo-absorptions remove lower energy photons
from the radiation field. For photon energies greater than $1.8$~Ryd,
helium absorbs more efficiently than hydrogen, and at yet higher energies oxygen
is more efficient than helium. 

Below $\sim10^4$~K, the cooling rate per volume 
generally decreases due to the decreasing
electron density. 
Below $\sim6000$~K, the cooling rate per electron
($n_H\Lambda$) is constant to within ~30\% (for $Z=2$),
with small variations due to recombinations that increases
the abundance of the dominant coolants 
(e.g. \ion{C}{2} 157.7$\mu$m and \ion{Si}{2} 34.8 $mu$m).

Because the radiation field controls both the heating rate, and the
electron fraction (which influences the cooling rates), 
heating and cooling remain strongly 
coupled for the remainder of the cooling process.
While both the heating and cooling rates decrease with time,
they remain nearly equal, 
leading to the slow thermal evolution in the plateau.
However, departures from exact thermal equilibrium enable the gas
to cool further, down to a temperature of $1650$~K at 
$n_0 \times t ~\sim 5.4 \times 10^{13}$~cm$^{-3}$~s.

As the gas cools, 
and
the heating radiation is 
absorbed, the 
H and He preferentially absorb low energy photons,
and the 
remaining 
ionizing radiation therefore becomes harder
with increasing depth in the photoabsorption plateau.
Figure~\ref{isoc-ex}e shows that after the UV 
photons 
are
absorbed (at $t\sim5\times10^{14}$~s), 
the heating rate decreases much more rapidly,
and the cooling rate lags 
slightly 
behind the heating. 
This leads to the brief temperature dip at 
$n_0\times t=5.4\times10^{13}$~s~cm$^{-3}$ in Fig.~3a.
After reaching a temperature minimum of $1650$~K at this time,
the gas heats up again to $\sim 4500$~K.
After this,
the cooling rate overcomes the heating rate again, 
and the gas then cools monotonically to
$1000$~K where we terminate the computation.

The shock structure and time-scales that are shown in Figure~\ref{Tt}
are in qualitative agreement with previous computations
(Dopita \& Sutherland~1996, Allen et al.~2008).
For example, our results for the temperature profiles in a solar
metallicity shock, are similar to those presented in Allen et 
al.~(2008) for their ``Dopita~2005'' abundance set (middle panel 
of their Figure~7), where gas stays in the hot radiative phase
for a few $\times10^{13}$~s, and then rapidly cools through the
non-equilibrium cooling zone to enter the photoabsorption
plateau.
Differences between our results and those of Allen et al.~are
likely due to differences in the assumed abundances, and in the
strength of the magnetic field, $B$.

\subsection{Dependence on Gas Metallicity}
\label{struct-Z}

The time-scale over which the gas cools from the initial
shock temperature down to $1000$~K, depends on the shock 
velocity
that sets the initial energy content of the shocked gas,
and on the cooling efficiency of the post-shock gas.
For $T_s=5\times10^6$~K, the cooling is initially dominated by metal
resonance line cooling for $Z\gtrsim0.1$, and by bremsstrahlung emission
for $Z\lesssim10^{-2}$.
For $T_s\gtrsim10^7$~K, bremsstrahlung cooling becomes important
even for high-metallicity gas.
Metal lines continue to dominate the cooling at lower temperature for
$Z\gtrsim0.1$, whereas for $Z\lesssim0.1$, hydrogen and helium
dominate the cooling at $T\lesssim3\times10^5$~K 
(see also GS07).

The cooling efficiency therefore strongly depends on the metal content 
within the gas. For $Z\gtrsim0.1$, the cooling efficiency is roughly
proportional to the gas metallicity.
However, at lower gas metallicity, as the metal contribution to the
cooling rate above $\sim10^4$~K becomes negligible, 
the cooling efficiency
approaches a 
limit set by the primordial helium abundance 
(see Boehringer \& Hensler~1989; GS07).
This can be seen in Figure~\ref{Tt}.
For high gas metallicities the
cooling time is proportional to $Z$, whereas the curves for $Z=10^{-2}$ and
for $Z=10^{-3}$ nearly overlap 
for temperatures
above $10^4$~K. 

For $Z>1$, even at $\sim10^4$~K, permitted metal transitions dominate
the cooling, 
and are
much more efficient than Ly$\alpha$.
For $Z=1$ Ly$\alpha$ provides $\sim15\%$ of the cooling at
its peak
efficiency.
For lower gas metallicities, Ly$\alpha$ provides most of the
cooling at $T\sim10^4$~K.
For example, for $Z=0.1$, Ly$\alpha$
provides as much as $80\%$ of the cooling.
Therefore, the cooling times at $\sim10^4$~K are nearly independent
of gas metallicity, except for $Z>1$.

As the gas cools below $\sim10^4$~K, metal fine structure 
emissions
start
to dominate the cooling, even for the lowest gas metallicities.
In addition, the gas opacity and 
associated
depth over which the
heating radiation is 
absorbed are functions of the gas
metallicity. The cooling times below $\sim10^4$~K are roughly
proportional to gas metallicity for all $Z$.
The overall {\it total} cooling times in isochoric flows 
is therefore
sensitive to
$Z$, as can be clearly seen
in panel~(a) of Figure~\ref{Tt}.

The different cooling times also set the degree of non-equilibrium
ionization in the gas (GS07), as affected by the ratio of the cooling-time
and the (metallicity-independent) recombination time.
When the cooling is faster than recombination, departures from
equilibrium may occur.
Since cooling is faster for high gas metallicities, non-equilibrium
effects, and the over-ionization in the gas, 
are more important for higher $Z$.
These effects occur at temperatures between $\sim10^6$K and $10^4$~K
where the cooling is rapid.
Below $10^4$~K, the heating becomes efficient, and the cooling times 
becomes long, so that the gas stays close to photoionization equilibrium.

A second factor that is strongly affected by the gas metallicity is
the radiation field emitted by the cooling gas.
As discussed before, the emitted radiation is composed mainly of
bremsstrahlung continuum, and of emission-lines and recombination continua.
The total flux of radiation emitted by the cooling gas
equals the input energy flux into the flow,
\begin{equation}
\label{flux}
F\propto n_0 k_{\rm B} T_s v_s \propto n_0 v_s^3 \ \ \ .
\end{equation}
However, the relative contribution of 
metal emission lines 
to the total flux depends on $Z$. 

For high-metallicity gas, a 
large fraction of the input energy
is radiated as line emission.
For low 
metallicity, the relative contribution of lines is 
small, and most of the initial energy flux is radiated 
as thermal bremsstrahlung.
This can be clearly seen in Figure~\ref{specs-Z}.
For $Z=2$ (the gray solid line), the lines-to-continuum contrast is
large, producing the ``UV-bump'' discussed above.
For $Z=0.1$ (dashed dark curve), lines are still important, but are
much less pronounced than for $Z=2$.
For $Z=10^{-3}$ (black solid curve) the spectrum is very smooth,
and consists almost entirely of thermal bremsstrahlung.

The differences in the spectral energy distribution
of the emitted radiation
field affect the thermal evolution of the cooling gas in
the photoabsorption region, and the resulting ion distributions.
We discuss the impact of the changing spectral energy distribution
on the ion fractions in detail in \S~\ref{ion-frac}.

Additional features in the shock profiles are affected by
the gas metallicity and associated cooling efficiency.
The 
start 
of the photoabsorption plateau is associated 
with the increase in photoabsorption efficiency that occurs
as 
the neutral hydrogen fraction becomes sufficiently large
(of order $10^{-3}$ for $5\times10^6$~K shocks).
Higher-metallicity gas is more
over-ionized, and
therefore
reaches this 
critical 
neutral fraction at a
lower temperature. The 
start of the WIM plateau therefore
takes place at lower temperature for higher metallicity gas. 
For $Z=2$, the plateau starts at $10^4$~K, for $Z=1$ at
$1.7\times10^4$~K, and for $Z\lesssim0.1$ at $2.5\times10^4$~K.
The depth of the WIM plateau is given by 
equation~(\ref{WIM}).
Since the energy flux emitted by models with the same 
initial shock temperature and magnetic field
is similar, the depth of the WIM plateau is
independent of gas metallicity.
For our $T_s=5\times 10^6$~K shocks
it is $\sim2.5\times10^{14}$~s 
(for $n_0=0.1$~cm$^{-3}$ and strong-$B$)
for $Z$ between $10^{-3}$ and $2$.

For $Z<1$, a second warm ionized medium plateau is apparent
in Figure~\ref{Tt}. This plateau has a temperature $T\sim10^4$~K.
The depth over which the heating 
self-radiation is absorbed and the
WNM plateau persists, depends on the gas opacity which is a 
function of $Z$. Lower metallicity gas produces longer WNM plateaus.

In the final cooling of the gas, fine-structure 
transitions dominate the cooling.
The intensity of fine-structure cooling is proportional to gas metallicity as discussed above.
At high gas metallicities ($Z\gtrsim1$) fine-structure cooling is efficient, and starts to
dominate the cooling at $\sim10^4$~K. At lower metallicities ($Z\lesssim 0.1$),
fine-structure cooling starts to dominate at $\sim7500$~K.

\subsection{Shock Temperature}

Panel~(c) of Figure~\ref{Tt} shows the temperature profiles in a shock
with 
an initial temperature $T_s=5\times10^7$~K,
in the strong-$B$ limit.
The overall characteristics of these temperature profiles are similar to
those of the lower-velocity ($T_s=5\times10^6$~K) shocks discussed 
in \S~\ref{shock-details}.
The gas is initially heated to $T_s$, then goes through a 
prolonged hot radiative phase during which
it emits most of its initial thermal energy as radiation. 
As $T$ drops, cooling becomes more efficient,
and a phase of rapid non-equilibrium cooling begins, 
bringing the
gas to a temperature of a few $\times10^4$~K.
At this point the neutral 
hydrogen
fraction becomes large enough that
photoionization
heating 
becomes
significant, and a temperature plateau
is 
formed.
The WIM plateau ends when the neutral fraction is of order
unity. A WNM plateau then follows.
The gas finally cools to 
our cut-off temperature, $T_{\rm low}=1000$~K.

Since the initial temperature is $10$ times higher in a 
$5\times10^7$~K shock than in a $5\times10^6$~K shock,
and the velocity 
is 
$10^{1/2}\sim3$ times higher, the total
energy flux input is $\sim30$ times higher. The overall cooling 
time is therefore longer, as the gas has to radiate more energy. 

\begin{figure}[!h]
\epsscale{1}
\plotone{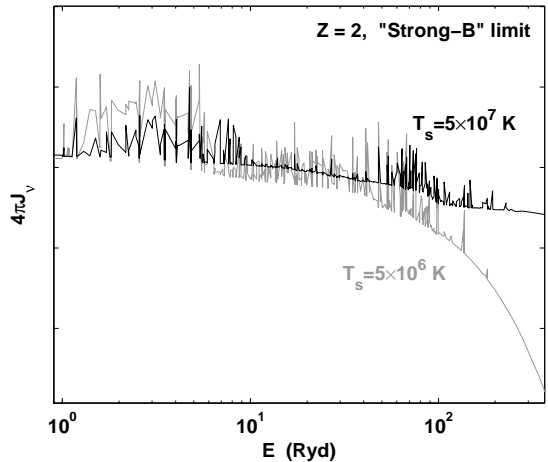}
\caption{ Shock self-radiation as a function of shock temperature for 
the strong-$B$ limit,
and for $Z=2$. The dark curve is the mean intensity 
(erg s$^{-1}$~cm$^{-2}$~Hz$^{-1}$) for $T_s = 5 \times10^6$~K. The gray curve
is the mean intensity for $T_s = 5\times 10^7$~K.
The spectra are normalized so that their UV continuum-intensities overlap,
to emphasize the different spectral energy distributions.}
\label{specs-T}
\end{figure}

The emitted radiation field is 
affected by the increased initial temperature.
The higher temperature produces a more intense bremsstrahlung 
continuum with more energetic photons
(the exponential bremsstrahlung cut-off
occurring at higher photon frequencies).
In addition, since at temperatures above $\sim2\times10^7$~K bremsstrahlung is
the dominant cooling process even at $Z=2$, a large fraction of the initial
energy is radiated as 
bremsstrahlung continuum 
rather than in lines.
Lines still dominate the cooling at lower temperatures, but their total
contribution to the integrated spectrum is smaller.
This is shown in Figure~\ref{specs-T} 
that
compares the 
(normalized)
spectral
energy distributions 
for $5\times10^6$~K and $5\times10^7$~K shocks.
The hotter shock provides a flatter spectrum extending to higher energies,
and its line-to-continuum contrast is smaller due to the larger
fraction of radiation emitted at high temperatures dominated by 
bremsstrahlung emission.

Because much of the initial energy in the $5\times 10^7$~K shock
gas is radiated as 
bremsstrahlung, the cooling times above
$\sim10^4$~K are much less sensitive to the gas metallicity than
for $5\times10^6$~K-shocks. In fact, the cooling time from $5\times10^7$~K
to $\sim10^4$~K increases only by a factor of $1.6$ when 
the metallicity is reduced from
$Z=1$ 
to 
$Z=0.1$.

\subsection{Magnetic Field}
\label{struct-B}

Panel~(b) of Figure~\ref{Tt} shows 
the temperature profiles
for $T_s=5\times10^6$~K,
assuming
$B=0$. As discussed in \S~\ref{flow-eqn}, the hydrodynamics 
follow 
equations~(\ref{conservations}), 
and the evolution is nearly isobaric,
with $P_\infty = 4/3\;P_0$.
The 
most important
difference between shocks in which $B=0$ and the previously discussed
strong-$B$ limit, is that
for $B=0$ the density increases as the gas cools.
Since the cooling time is inversely proportional to the gas density, this implies
a rapidly decreasing cooling time within the flow. 
The overall cooling times are therefore
much shorter, and the evolution below $10^4$~K, while qualitatively similar
to that of the previously discussed isochoric shocks, 
is 
compressed into a very short interval.
This can be seen in Figure~\ref{Tt}b.
Once the gas starts to cool, the decline
down to a temperature of $1000$~K is very rapid.

\begin{figure}[!h]
\epsscale{1}
\plotone{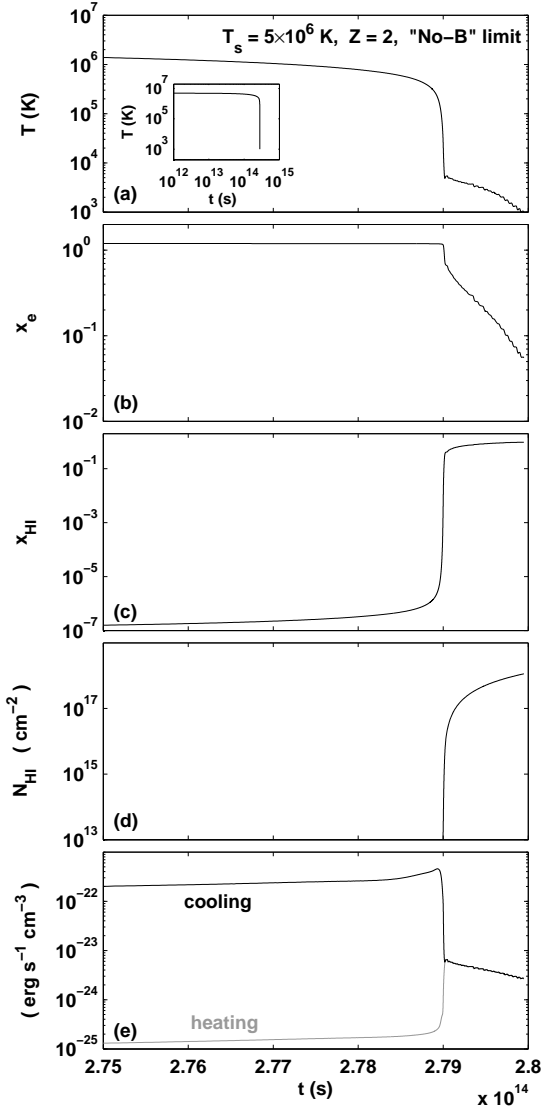}
\caption{ Shock structure versus time for $T_s=5\times10^6$~K,
$Z=2$ solar, for $B=0$. The plots focus on the final evolution,
between $2.75\times10^{14}$~s and $2.8\times10^{14}$~s for $n_0=0.1$~cm$^{-3}$.
(a)~Temperature~(K). (b)~Electron fraction. (c)~Neutral hydrogen fraction.
(d)~Integrated \ion{H}{1} column density~(cm$^{-2}$) from the shock front.
(e)~Local cooling (dark) and heating (gray) rates per volume~(erg~s$^{-1}$~cm$^{-3}$).
}
\label{hydro-ex}
\end{figure}

Figure~\ref{hydro-ex} shows a ``zoomed-in'' snapshot of the final stages
of the evolution for $Z=2$. All times are shown assuming a post shock hydrogen
density of $0.1$~cm$^{-3}$.
The insert in panel~(a) shows the full
temperature profile in the flow. The panels focus on the final evolution, between
$t=2.75\times10^{14}$~s and $2.8\times10^{14}$~s.
The final evolution shows similar features 
to those discussed for the strong-$B$ isochoric shocks.
The gas cools rapidly to a point where photoabsorption becomes
significant enough that the heating balances the cooling.
It then enters a photoabsorption region in which the shock self-radiation
is gradually absorbed, and the temperature decline becomes slower.
This happens on much shorter time scales due to the increasing density
(c.f.~Figure~5 in Allen et al.~2008).

Equation~(\ref{flux}) 
states that the total flux created by the cooling gas is
proportional to the input energy flux, $F \propto n_0 v_s^3$.
Almost all of this flux ($>99.8\%$) is emitted before the plateau starts,
and most of it ($>80\%$) is emitted within the hot radiative zone.
The level of photoionization in the 
gas is determined by the
ionization parameter, which is proportional to $F/n$.
For strong-$B$ shocks, the density in the flow is constant,
and the ionization parameter 
is therefore $\propto v_s^3$.
For $B=0$, the density increases as the gas cools, and the ionization
parameter is therefore $\propto v_s^3 n_0 / n$, which is smaller by a factor
$n_0/n$ (or $\sim T/T_s$).
Photoionization is therefore much less important in $B=0$ models,
and its contribution to the creation of intermediate- and high-ions 
is diminished.

\begin{figure}[!h]
\epsscale{1}
\plotone{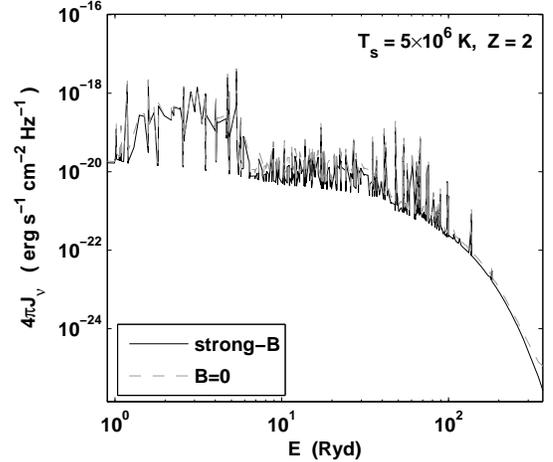}
\caption{ Shock self radiation for $T_s = 5 \times10^6$~K shocks and for $Z=2$ solar metallicity gas.
The solid curve shows the mean intensity (erg s$^{-1}$~cm$^{-2}$~Hz$^{-1}$) in the
strong-$B$ limit. 
The gray dashed curve shows the mean intensity for $B=0$, divided
by $5/3$ to remove the $PdV$ contribution. As expected, the spectra are similar.}
\label{specs-B}
\end{figure}

A more subtle effect is related to the $PdV$ work that appears in
$B=0$ shock models, due to the gas compression.
The work done on the cooling gas implies that the overall emitted
radiation field is $5/3$ times larger for $B=0$ shocks than for isochoric shocks
(Edgar \& Chevalier~1986; GS07).
The final total flux in strong-$B$ flows is $3/2~n_0 k_{\rm B} T_s v_s$,
while for $B=0$ it is $5/2~n_0 k_{\rm B} T_s v_s$. This is shown in Figure~\ref{specs-B}.
However,
the 
effect 
of the $5/3$-factor on 
the
downstream ionization
parameter is much smaller than the impact of the increasing density
in $B=0$ shocks.

\begin{figure}[!h]
\epsscale{1}
\plotone{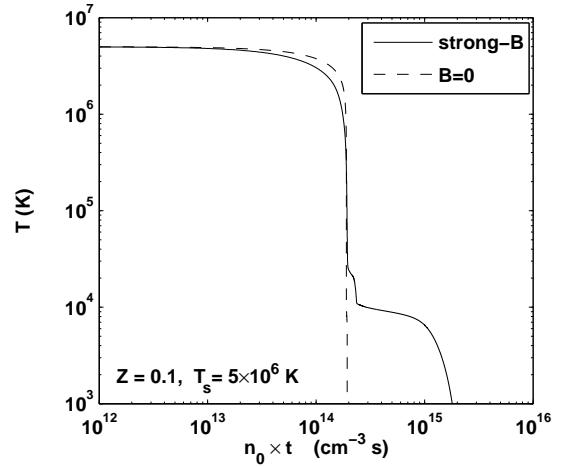}
\caption{ Temperature profiles for $T_s=5\times10^6$~K and $Z=0.1$ solar metallicity.
The solid curve shows results in the strong-$B$ limit. The dashed curve shows results for
$B=0$.}
\label{Tt-comp}
\end{figure}

Figure~\ref{Tt-comp} compares the temperature profiles for
strong-$B$ and $B=0$ models.
Initially, due to the $PdV$ work, the cooling in the isochoric, 
strong-$B$,
shock is faster
than in the $B=0$ nearly isobaric flow.
However, 
for $B=0$
the density and
cooling rates quickly grow, and the final $B=0$ cooling is much more rapid.

\begin{figure}[!!ht]
\epsscale{1}
\plotone{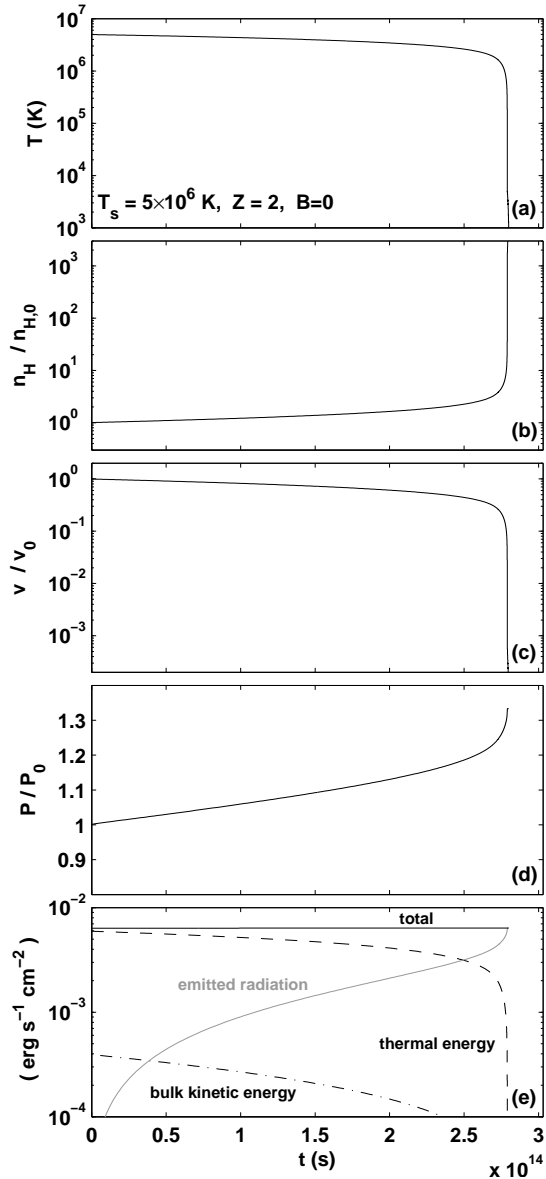}
\caption{ Dynamics in a $5\times10^6$~K, $Z=2$, $B=0$ shock.
(a)~Temperature~(K). (b)~Density profile. (c)~Velocity profile.
(d)~Pressure profile. (e)~Energy flux~(erg~s$^{-1}$~cm$^{-2}$).
The dark solid curve shows the total
energy flux in the flow, which is conserved as it should.
The dashed line shows the thermal energy flux ($\frac{5}{2}Pv$),
which is initially the dominant
component of the energy flux, but drops as the gas cools.
The dash-dotted line shows the bulk kinetic energy flux ($\frac{1}{2}\rho v^3$).
The gray solid curve is the integrated cooling radiation, which is accumulated in
the flow as the gas cools, until it reaches the total input energy flux
when all thermal energy is lost.}
\label{hydro-dyn}
\end{figure}

Figure~\ref{hydro-dyn} shows the hydrodynamic evolution of a 
$B=0$, $T_s=5\times10^6$~K, $Z=2$ flow.
Panel~(a) shows the gas temperature.
Panel~(d) shows the gas pressure. As discussed above 
(see equation~[5]),
the flow is nearly isobaric, with a total pressure change of $4/3$.
This implies that the gas density is roughly inversely proportional to 
the temperature, as
can be seen in panel~(b).
The mass flux conservation equation implies that the velocity is inversely proportional
to the gas density as can be seen in panel~(c).

For $B=0$, the pressure is dominated by the thermal gas pressure through the flow.
We can therefore 
verify
mass, momentum and energy conservation,
and we
indeed find that they are conserved to better than $1\%$.
Panel~(e) shows the energy flux components in the flow.
The total energy flux, shown by the upper solid line,  
remains constant as it should.
The thermal energy component ($5Pv/2$) is shown by the dark dashed line.
Initially it dominates the energy flux, but as the gas cools, the thermal component
gradually decreases.
The bulk kinetic energy ($\rho v^3/2$) is shown by the dash-dotted line.
Initially, 
this is only a small fraction of the total energy flux, 
and 
it
decreases further
as the gas cools and decelerates.
The gray line shows the integrated cooling radiation. As expected, this component
grows with depth into the flow, and
finally reaches
a value
equal to the total initial energy flux when all 
internal
thermal energy is lost.

The impact of gas metallicity on $B=0$ 
isobaric
shocks is similar to
that
which occurs in strong-$B$ isochoric shocks discussed 
in \S~\ref{struct-Z}.
Again,
as the gas metallicity increases, the metal-line cooling efficiency
increases, and cooling becomes more rapid.
This is clearly seen in Figure~\ref{Tt}b, where the initial cooling
time is proportional to gas metallicity, except at $Z\lesssim 10^{-2}$ where
the metals contribution to the cooling is negligible.

The cooling below $\sim10^4$~K is 
very rapid 
due to the increased gas density.
While the cooling times below $\sim10^4$~K are proportional 
to gas metallicity, this cannot
be seen in Figure~\ref{Tt}b since the total cooling times below $\sim10^4$~K
are 
so
much shorter than the cooling times from $5\times10^6$~K to $10^4$~K.
The low-temperature cooling time only becomes long enough
to be seen in Figure~\ref{Tt}b for $Z\lesssim10^{-2}$, 
for which
the cooling time 
during the plateau phase is of order
the cooling time 
to reach the plateau.

\subsection{Gas Density}
\label{struct-n}

The results discussed above for 
$T$ versus $n_0 \times t$ are independent of density for 
strong-$B$ 
isochoric
models. 
When $B=0$, the gas is compressed as it cools, and the thermal
evolution below $\sim10^4$~K may depend on the initial density.
In the high-density compressed gas, 
the cooling efficiencies may be suppressed by collisional
de-excitations of the cooling transitions if the densities
become sufficiently high.
The temperature at which 
such collisional quenching occurs depends on the shock velocity 
as well as on the initial post-shock density, $n_0$. 

\begin{figure}[!h]
\epsscale{1}
\plotone{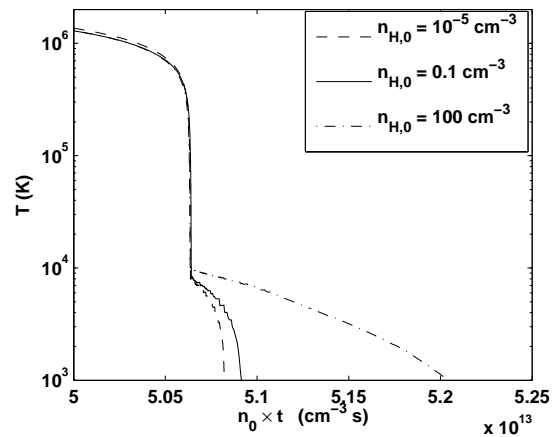}
\caption{ Temperature profiles for different post-shock hydrogen
densities, for a $5\times10^6$~K shock 
and $B=0$.
The plot focuses on the final evolution.
Higher density models cool 
less efficiently because collisional quenching occurs sooner
as the gas is compressed. 
}
\label{n-isob}
\end{figure}

Figure~\ref{n-isob} shows the temperature as a function of $n_0\times t$
in $T_s=5\times10^6$~K shocks,
assuming post-shock hydrogen densities of $10^{-5}$~cm$^{-3}$ (dashed line),
$0.1$~cm$^{-3}$ (solid line), and $100$~cm$^{-3}$ (dash-dotted line).
For 
these initial densities, the gas density never becomes
high enough 
for significant collisional de-excitation
of the permitted transitions.
However, 
once the temperature drops 
below
$\sim10^4$~K, 
some of the fine-structure cooling is quenched,
and the thermal evolution is altered.
As shown by Figure~\ref{n-isob}, 
for higher initial post-shock densities this collisional
quenching occurs sooner, and the effective cooling times, $n_0\times t$,
are therefore longer for higher $n_0$
(c.f.~Figure~6 in Allen et al.~2008, in which the 
cooling times from $T_s$ to the photoabsorption plateau scale
as $n^{-1}$, whereas the cooling below $\sim10^4$~K 
depends more strongly on $n$).

\subsection{Steady-State Conditions}
\label{sizes}

The results presented in this paper rely on the assumption that
the shocks have reached a steady-state structure.
Attaining steady-state requires that the shock 
exists 
for
a time-scale
that is longer than the cooling time. Otherwise, the shocked gas
only cools partially, and the shock structure, self-radiation,
and integrated column densities are time-dependent.

In Table~\ref{size-scales} we list the total cooling times (yr), from $T_s$
down to $1000$~K, assuming a post-shock hydrogen density, $n_0=1$~cm$^{-3}$.
We also list the associated length-scales (kpc), and total hydrogen column
densities, $N_{\rm H,tot}$ (cm$^{-2}$).
These are then the hydrogen columns and length scales required for the
formation of steady-state shocks.
For example, Table~\ref{size-scales} implies that for a $T_s=5\times10^6$~K
shock in solar metallicity gas, the time required to reach steady-state is 
$6.4\times10^6$~yrs, in the strong-$B$ isochoric limit. The cooling length is
then $1$~kpc, and the associated hydrogen column density is 
$3\times10^{21}$~cm$^{-2}$.
The time- and size-scales are generally proportional to $n_0^{-1}$,
while the total hydrogen columns are generally independent of $n_0$
(but see \S~\ref{struct-n}).

\begin{deluxetable*}{llllllll}
\tablewidth{0pt}
\tablecaption{Shock Parameters (for $n_0=1$~cm$^{-3}$)}
\tablehead{
\colhead{} &
\multicolumn{3}{c}{$T_s=5\times10^6$~K}&&
\multicolumn{3}{c}{$T_s=5\times10^7$~K}\\
\colhead{Metallicity} &
\colhead{Time} & 
\colhead{Size} & 
\colhead{$N_{\rm H,tot}$} && 
\colhead{Time} & 
\colhead{Size} & 
\colhead{$N_{\rm H,tot}$} \\
\colhead{} &
\colhead{(yr)} &
\colhead{(kpc)} &
\colhead{(cm$^{-2}$)} &&
\colhead{(yr)} &
\colhead{(kpc)} &
\colhead{(cm$^{-2}$)} }
\startdata
\underline{strong-$B$ (isochoric)} & & & & & & \\ 
$Z=2$       & $2.9\times10^{6}$ & $0.45$ & $1.4\times10^{21}$ && $9.2\times10^{7}$ & $45$   & $1.4\times10^{23}$ \\
$Z=1$       & $6.4\times10^{6}$ & $1.0$  & $3.0\times10^{21}$ && $1.4\times10^{8}$ & $71$   & $2.2\times10^{23}$ \\
$Z=10^{-1}$ & $5.7\times10^{7}$ & $8.8$  & $2.7\times10^{22}$ && $5.2\times10^{8}$ & $250$  & $7.8\times10^{23}$ \\
$Z=10^{-2}$ & $3.3\times10^{8}$ & $50$   & $1.6\times10^{23}$ && $1.5\times10^{9}$ & $750$  & $2.3\times10^{24}$ \\
$Z=10^{-3}$ & $2.1\times10^{9}$ & $330$  & $1.0\times10^{24}$ && $4.1\times10^{9}$ & $2000$ & $6.3\times10^{24}$ \\
\hline\\
\underline{$B=0$ (``isobaric'')}      & & & && & & \\ 
$Z=2$       & $8.9\times10^{5}$ & $0.10$ & $4.3\times10^{20}$ && $2.2\times10^{7}$ & $7.0$  & $3.3\times10^{22}$ \\
$Z=1$       & $1.6\times10^{6}$ & $0.18$ & $7.8\times10^{20}$ && $2.8\times10^{7}$ & $8.6$  & $4.2\times10^{22}$ \\
$Z=10^{-1}$ & $6.1\times10^{6}$ & $0.63$ & $2.9\times10^{21}$ && $3.8\times10^{7}$ & $11.0$ & $5.7\times10^{22}$ \\
$Z=10^{-2}$ & $1.0\times10^{7}$ & $0.87$ & $4.8\times10^{21}$ && $4.1\times10^{7}$ & $11.4$ & $6.3\times10^{22}$ \\
$Z=10^{-3}$ & $1.6\times10^{7}$ & $0.91$ & $7.7\times10^{21}$ && $5.3\times10^{7}$ & $11.5$ & $8.0\times10^{22}$ \\
\enddata
\tablecomments{Cooling-times (yr), size-scales (kpc), and total hydrogen column
densities (cm$^{-2}$) for post-shock gas cooling from $T_s$ to $1000$~K, assuming
a post-shock hydrogen density, $n_0=1$~cm$^{-3}$. 
}
\label{size-scales}
\end{deluxetable*}

\section{Ion Fractions}
\label{ion-frac}

We have 
computed
the ionization states of H, He, C, N, O,
Ne, Mg, Si, S, and Fe in the post-shock cooling layers.
When the photoionized precursor-gas enters the shock, its ionization state
rapidly reaches CIE at a temperature very close to $T_s$.
As the hot gas flows away from the shock front, it recombines,  cools, and
radiates away its thermal energy.
This radiation is later absorbed by the cooler gas further downstream,
providing a source of heating and photoionization.
We follow the time dependent ion fractions in the flow, taking into account
photoionization by the shock self-radiation. As we discuss below, the
photoionizing radiation significantly affects the ion fractions in the
gas.

When
the cooling time becomes short compared to the recombination time,
departures from equilibrium may occur, keeping the gas over-ionized
compared to 
ionization equilibrium at the local conditions
(as specified by the mean intensity of the photoionizing radiation, 
gas density, and temperature). 
We consider the non-equilibrium ionization states as a function of
the time dependent temperature and mean 
radiation intensity.
We present results 
for gas cooling behind
shocks 
with initial post-shock temperatures
$5\times10^6$~K and $5\times10^7$~K, with metallicities
$Z=10^{-3}$, $10^{-2}$, $10^{-1}$, $1$, and $2$ times the 
solar metal abundances,
in the 
$B=0$ (nearly isobaric) and 
``strong-$B$''
(isochoric) limits.

Table~\ref{ion-frac-table} lists the ion fractions as a function of time and
temperature for the various models that we consider, as outlined in Table~\ref{guide}. 
All the times given in Table~\ref{ion-frac-table} were
computed assuming a post-shock hydrogen density $n_{{\rm H},0}=0.1$~cm$^{-3}$.
As we discussed in \S~\ref{struct}, 
the shock structures, as functions of $n_0\times t$, 
are independent of the gas density in
the strong-$B$ limit, 
and are density-dependent
only at low temperatures ($\lesssim10^4$~K) when $B=0$.

\begin{deluxetable*}{llcccc}
\tablewidth{0pt}
\tablecaption{Ion Fractions in a $5\times10^6$~K, $2$ times solar, strong-$B$-shock}
\tablehead{
\colhead{Time} &
\colhead{Temperature} &
\colhead{H$^0$/H} & 
\colhead{H$^+$/H} & 
\colhead{He$^0$/He} & 
\colhead{\ldots} \\
\colhead{s}&
\colhead{K}&
\colhead{}&
\colhead{}&
\colhead{}&
\colhead{}
}
\startdata
$0.00$          &$5.000\times10^6$&$1.63\times10^{-4}$&$0.00$&$4.93\times10^{-3}$&\ldots\\
$7.25\times10^8$&$4.998\times10^6$&$1.52\times10^{-5}$&$0.00$&$3.59\times10^{-3}$&\ldots\\
$1.57\times10^9$&$4.997\times10^6$&$1.01\times10^{-6}$&$0.00$&$2.48\times10^{-3}$&\ldots\\
\enddata
\tablecomments{The complete version of this table is in 
the electronic edition of the Journal. The printed edition contains only a sample. 
The full table lists ion fractions for
the $B=0$ (isobaric) and strong-B (isochoric)
magnetic field limits,
for shock temperatures of $5\times10^6$~K and $5\times10^7$~K, and for
$Z=10^{-3}$, $10^{-2}$, $10^{-1}$, $1$, and $2$ times solar metallicity
gas (for a guide, see Table~\ref{guide}). The times in the first column 
are 
fo an assumed 
post-shock hydrogen density of $0.1$~cm$^{-3}$.}
\label{ion-frac-table}
\end{deluxetable*}

\begin{deluxetable*}{lllcccc}
\tablewidth{0pt}
\tablecaption{Ionization and Cooling Tables}
\tablehead{
\colhead{} & \colhead{} & \colhead{} &
\colhead{} &\colhead{} &\colhead{Post-Shock} &\colhead{Precursor}\\
\colhead{Data} & \colhead{} & \colhead{} &
\colhead{Ion Fractions} &\colhead{Cooling} &\colhead{Columns} &\colhead{Columns}}
\startdata
$5\times10^6$~K & strong-$B$ & $Z=2$       & 3A & 5A & 6A & 8A \\
                &            & $Z=1$       & 3B & 5B & 6B & 8B \\
                &            & $Z=10^{-1}$ & 3C & 5C & 6C & 8C \\
                &            & $Z=10^{-2}$ & 3D & 5D & 6D & 8D \\
                &            & $Z=10^{-3}$ & 3E & 5E & 6E & 8E \\
                & $B=0$      & $Z=2$       & 3F & 5F & 6F & 8F \\
                &            & $Z=1$       & 3G & 5G & 6G & 8G \\
                &            & $Z=10^{-1}$ & 3H & 5H & 6H & 8H \\
                &            & $Z=10^{-2}$ & 3I & 5I & 6I & 8I \\
                &            & $Z=10^{-3}$ & 3J & 5J & 6J & 8J \\
$5\times10^7$~K & strong-$B$ & $Z=2$       & 3K & 5K & 6K & 8K \\
                &            & $Z=1$       & 3L & 5L & 6L & 8L \\
                &            & $Z=10^{-1}$ & 3M & 5M & 6M & 8M \\
                &            & $Z=10^{-2}$ & 3N & 5N & 6N & 8N \\
                &            & $Z=10^{-3}$ & 3O & 5O & 6O & 8O \\
                & $B=0$      & $Z=2$       & 3P & 5P & 6P & 8P \\
                &            & $Z=1$       & 3Q & 5Q & 6Q & 8Q \\
                &            & $Z=10^{-1}$ & 3R & 5R & 6R & 8R \\
                &            & $Z=10^{-2}$ & 3S & 5S & 6S & 8S \\
                &            & $Z=10^{-3}$ & 3T & 5T & 6T & 8T \\
\enddata
\label{guide}
\end{deluxetable*}

\begin{figure}[!th]
\epsscale{1}
\plotone{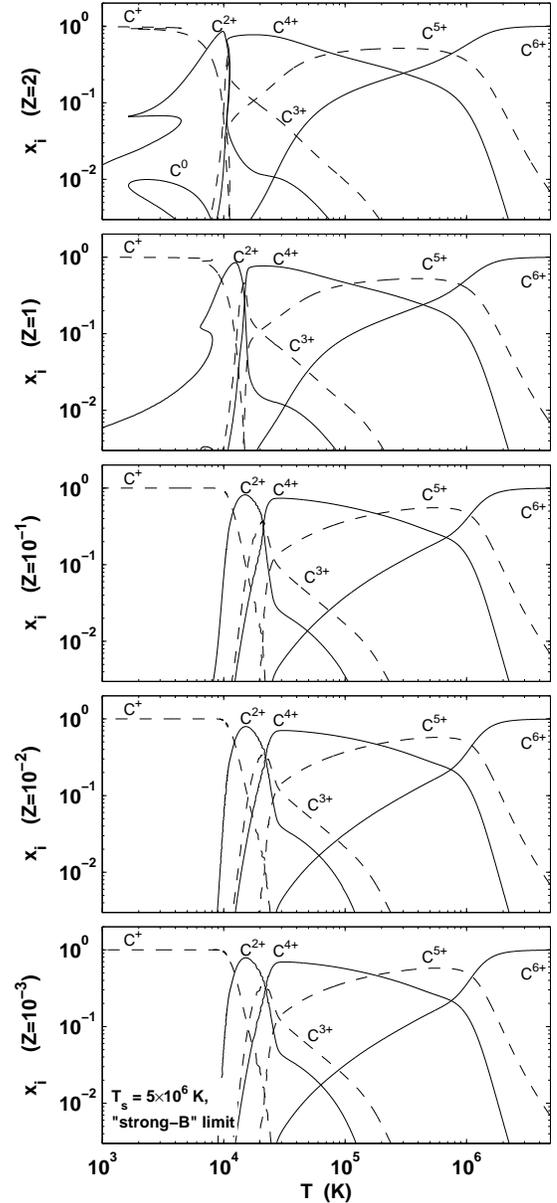}
\caption{ Carbon ion fractions 
versus
gas temperature in a $T_s=5\times10^6$~K shock in
the strong-$B$ limit.
The various
panels show results for 
gas metallicities 
ranging from $10^{-3}$ to $2$
times solar.}
\label{frac-Z}
\end{figure}

As a detailed example, in Figure~\ref{frac-Z} we show the carbon ion fractions 
for $T_s=5\times 10^6$~K, in the strong-$B$ limit, for the various values of $Z$.
The second panel 
(Fig~\ref{frac-Z}b)
shows the results for 
$Z=1$. 
We first focus on these results, and 
then consider how the results 
depend on $Z$.

\begin{figure}[!h]
\epsscale{1}
\plotone{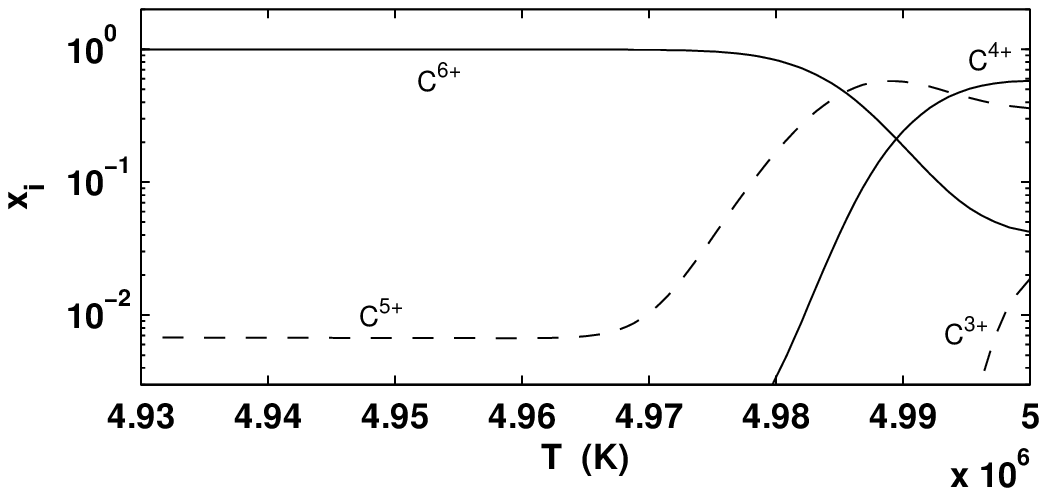}
\caption{ Carbon ion fraction 
versus gas temperature just after 
crossing the shock front, for
$T_s=5\times10^6$~K, strong-$B$, and $Z=1$ gas.
The plot shows how the initial ionization states quickly adjust to CIE at the
initial
shock temperature.}
\label{rise}
\end{figure}

As the photoionized precursor gas enters the shock, the carbon ion fractions
approach CIE at the shock temperature. This initial rise is very rapid, and cannot
be seen in Figure~\ref{frac-Z}.
Figure~\ref{rise} focuses on 
this initial evolution.
The abundant carbon ions in the precursor gas are C$^{4+}$ and C$^{5+}$.
As can be seen from Figure~\ref{rise}, these ionization states
are quickly replaced by C$^{6+}$, which is the abundant ion at $T_s=5\times10^6$~K
(see GS07).
By the time the temperature drops to $4.97\times10^6$~K,
the carbon ion fractions reach CIE. 
Later on, as carbon recombines in the cooling flow, 
the lower ionization
states reappear at lower temperatures, as can be seen in 
the second panel of Figure~\ref{frac-Z}.
For example, C$^{4+}$ is again the most abundant carbon ion between 
$T\sim10^5$~K, and $T\sim2\times10^4$~K.
Many ionic species show such a double peak abundance pattern, where the abundance
peaks 
first immediately after passing through the shock front, 
and then a second time as the gas recombines further downstream.

The time scales of the two peaks are very different. For the example
shown is Figure~\ref{rise},
the width of the first peak in the C$^{4+}$ abundance is of order
$10^{12}$~s, while the second peak lasts of order $10^{14}$~s.
The total  contribution of the first abundance peak to the total
ionic column density is therefore  very small.
However, since the gas is underionized as it goes through this first peak,
the hot ambient electrons efficiently excite the transitions of 
the underionized ions (McCray~1987), creating an enhanced emissivity.
This enhanced emissivity also implies that the cooling rates during this
adjustment phase are enhanced. 
In lower velocity shocks, the first peaks can contribute significantly to
the integrated column denisties (Krolik \& Raymond~1985).

Figure~\ref{frac-Z} shows that
in the hot radiative zone, the gas remains
close to CIE, with C$^{6+}$ being the most dominant carbon ion.
During the non-equilibrium cooling 
phase ($T\lesssim10^6$~K), 
the gas cools and recombines, and
the most abundant ionization state drops to C$^{5+}$ and later to C$^{4+}$.
The ionization state stays higher than at CIE due to photoionization
by the shock self-radiation as we discuss below.
The photoabsorption plateau starts at $T\sim2\times10^4$~K, and as the
shock self radiation is absorbed within the plateau, the dominant carbon 
ion gradually drops from C$^{3+}$ to C$^{2+}$, and finally to C$^+$.
For the parameters considered here, C$^+$ 
remains
the dominant species
until the gas reaches 
our termination temperature
$T_{\rm low}=1000$~K.
Our results are in qualitative agreement with those of Allen et al.~(2008;
see their Figure~9),
in which C$^{6+}$ is the most dominant carbon ion in the hot radiative phase,
C$^{5+}$-C$^{3+}$ dominate in the non-equilibrium cooling zone, and 
in the plateau C$^{2+}$ recombines to from C$^+$, which remains dominant
down to $1000$~K.

\subsection{Photoionization}

\begin{figure}[!h]
\epsscale{1}
\plotone{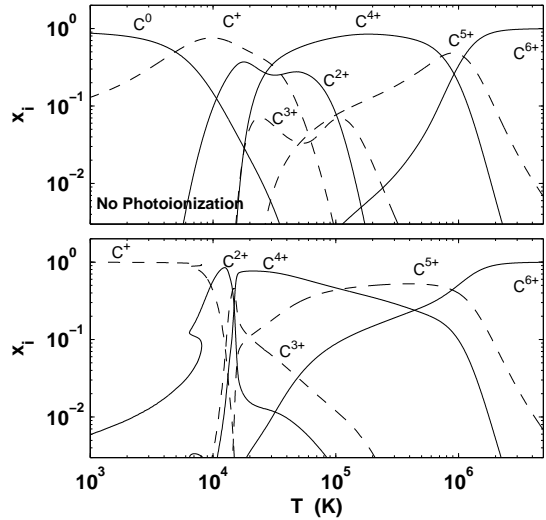}
\caption{ Carbon ion fractions versus gas temperature in a $T_s=5\times10^6$~K 
strong-$B$ shock for $Z=1$.
The upper panel shows a non-physical model in which photoionization is 
not included in the computation.
The lower panel shows results when photoionization is included.}
\label{photo}
\end{figure}

The radiation emitted by the hot post-shock gas has a profound 
effect
on the ion fractions in the cooling gas.
To illustrate the importance of photoionization by the gas self-radiation
on the evolution and ion fractions, we have computed a 
comparison-model 
in which the radiation field at
any point in the flow 
is 
artificially set to zero, 
so that there is no photoionization and associated heating anywhere in the flow.
The results (for $T_s=5\times10^6$~K, strong-$B$, and
$Z=1$)
with and without photoionization
are presented in Figure~\ref{photo}.

In the hottest parts of the flow where $T\gtrsim10^6$~K, the ionizing
radiation has 
a negligible effect 
on the ion fractions, because the
typical ionization potentials are too high to be efficiently affected
by the radiation. 
At lower temperatures, 
photoionization becomes
important.
For example, in the post-shock gas the C$^{5+}$ ion fraction remains
higher than $10\%$ down to a temperature of $\sim2\times10^4$~K,
whereas without photoionization it drops below $10\%$ at $T\sim10^5$~K.
As expected, photoionization 
becomes dominant
at lower temperatures,
maintaining C$^{2+}$ 
as an abundant species below $10^4$~K, and C$^+$ the
most abundant species even at $T\sim10^3$~K.
Without radiation, the ionization level is much lower, and neutral carbon is
the most dominant species below $5000$~K.
Photoionization 
thus
strongly affects the resulting integrated column densities
in the cooling layer.
The column density of 
the high-ion
C$^{5+}$ is 
enhanced by a factor of $\sim1.5$ 
due to photoionization.
The column densities of the mid- and low-ions C$^{3+}$, C$^{2+}$,
and C$^+$ are enhanced by factors of more than $100$.

\subsection{Shock Temperature}

The shock self-radiation depends on the shock velocity and
associated initial post-shock temperature $T_s$.
As discussed in \S~\ref{struct-Z} 
(equation~[\ref{flux}]),
the mean intensity of the 
radiation field
is proportional 
to 
$T_s^{1.5}$. 
In addition to the overall intensity dependence,
the spectral-energy distribution hardens with increasing $T_s$.
Figure~\ref{specs-T} shows the 
spectral energy
distributions 
for $T_s$ equal to $5\times10^6$~K and $5\times10^7$~K.
Even high ions, which are 
only
collisionally ionized in a $5\times10^6$~K
shock, are photoionized by the harder photon emitted in a 
$5\times10^7$~K shock. The gas in the hotter shock is more
highly ionized, 
both due to the higher intensity, and 
the harder spectral shape
of the shock radiation.

\begin{figure}[!h]
\epsscale{1}
\plotone{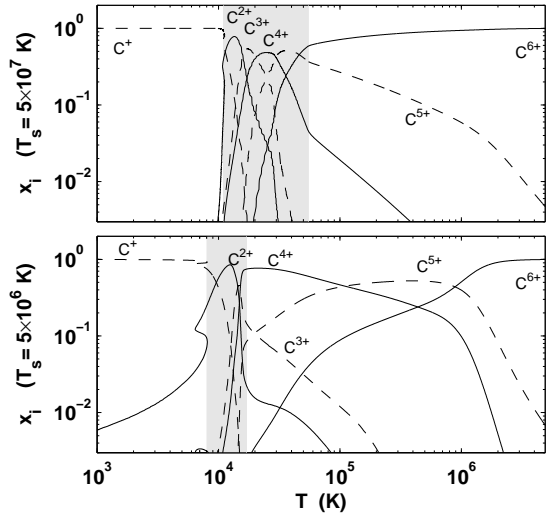}
\caption{Carbon ion fraction versus
gas temperature in a solar metallicity gas,
for a strong-$B$ (isochoric) shock.
The upper panel shows results for a shock temperature of $5\times10^7$~K,
and the lower panel shows results for $5\times10^6$~K.}
\label{frac-T}
\end{figure}

In Figure~\ref{frac-T} we show the carbon ion fractions as a
function of gas temperature.
The upper panel 
is for
$T_s=5\times10^7$~K, and the lower panel 
is
for $T_s=5\times10^6$~K.
It is clear that the gas is more 
highly
ionized for the hotter shock.
This is evident even at very high temperatures, $>10^6$~K, where
the more energetic photons created by the hotter bremsstrahlung
continuum of the $5\times10^7$~K-shock efficiently ionize C$^{5+}$
to C$^{6+}$.

Some of the differences between the ion distributions
shown in Figure~\ref{frac-T} for the two shock velocities, result from the 
fact that the various ``zones'' occur at different gas temperatures
depending on $T_s$.
For example, the ranges of temperatures at which the gas is in the
WIM plateau (see \S~\ref{struct}) are marked in Figure~\ref{frac-T} by the 
gray shaded zones.
The WIM plateau starts at a temperature of $\sim5.5\times10^4$~K
for $T_s=5\times10^7$~K, and at a temperature of $\sim1.6\times10^4$~K
for  $T_s=5\times10^6$~K.
As the radiation is 
absorbed in the plateau, the carbon ionization
states drops to C$^{3+}$,  then C$^{2+}$, and eventually to C$^+$ towards
the end of the WIM plateau where hydrogen and helium become neutral.
While this recombination process takes place within the WIM plateau for
both values of $T_s$, the ion distributions versus temperature, $x_i(T)$,
are different.

In the absence of photoionization,
the ion fractions,
for ions that are produced collisionally at temperatures
less than $T_s$, are independent of the shock velocity.
Thus, without photoionization the ion fractions versus temperature
shown in Fig.~17 would be identical for the two shock temperatures. 
The integrated column densities through the flow are then
proportional to the shock velocity 
(e.g.~Heckman et al.~2002). 
However, as is clearly seen
from Figures~\ref{photo} and \ref{frac-T}, photoionization plays
a major role in setting the ion fractions in the flow, and the
ion distributions are affected by the metallicity, 
shock temperature, and magnetic field, 
through their control of the
ionization parameter.
We present detailed results for
the ionic column densities in \S~\ref{columns}.

\subsection{Magnetic Field}
\label{results-B}

One of the parameters that determines the level of photoionization
in the post-shock gas is the 
strength of the magnetic field.
As discussed in \S~\ref{struct-B}, the magnetic field strongly affects
the ionization parameter in the downstream gas, due to the compression
that takes place when $B$ is small.
We therefore expect that at low temperatures, after significant compression
has taken
place, models with $B=0$ will be much less ionized than 
strong-$B$ isochoric
models.

\begin{figure}[!h]
\epsscale{1}
\plotone{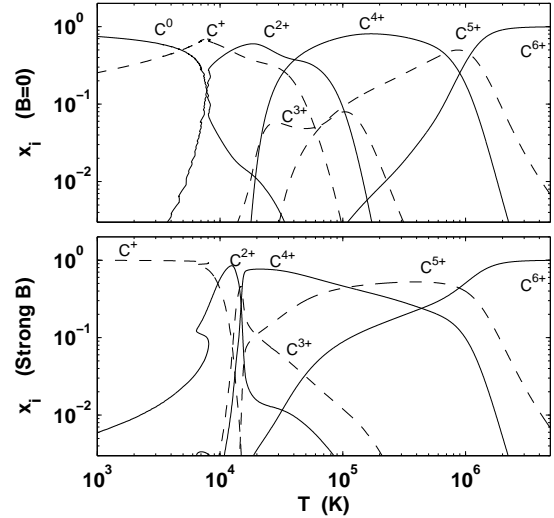}
\caption{ Carbon ion fractions versus gas temperature in a $T_s=5\times10^6$~K shock
for $Z=0$. The upper panel is for $B=0$, and the lower panel is for strong-$B$.
}
\label{frac-B}
\end{figure}

Figure~\ref{frac-B} shows the carbon ion fractions 
in the isobaric
$B=0$ (upper panel), and isochoric strong-B (lower panel) limits. 
The lower ionization parameter in
the downstream gas when $B=0$ results in lower ionization states
at all temperatures.
For example, at a temperature of $30,000$~K, $>98\%$ of the carbon 
is in the state of C$^{3+}$-C$^{5+}$ in the limit of strong-$B$, 
whereas for $B=0$ more than half of the carbon is in the state
of C$^0$-C$^{2+}$.

For $B=0$, the temperature versus time in the
post-shock cooling layers depends on the density, at temperatures
lower than $\sim10^4$~K (see \S~\ref{struct-n}) where collisional 
de-excitations
of fine-structure transitions
become significant. 
The 
physical conditions below $T\sim10^4$~K, therefore depend
on the density of the shocked medium. The computations 
presented
here assume
a post-shock hydrogen density of $0.1$~cm$^{-3}$.

\subsection{Departures from Equilibrium Ionization}

In a photoionized gas with a high ionization parameter, the ion fractions depend
only weakly on the gas temperature, as opposed to 
purely collisionally ionized gas. 
Since the gas does not have to significantly adjust its ionization
state to the time-dependent temperature during cooling, departures from 
equilibrium are expected to be smaller than 
for pure radiative cooling.

\begin{figure*}
\epsscale{1.0}
\plotone{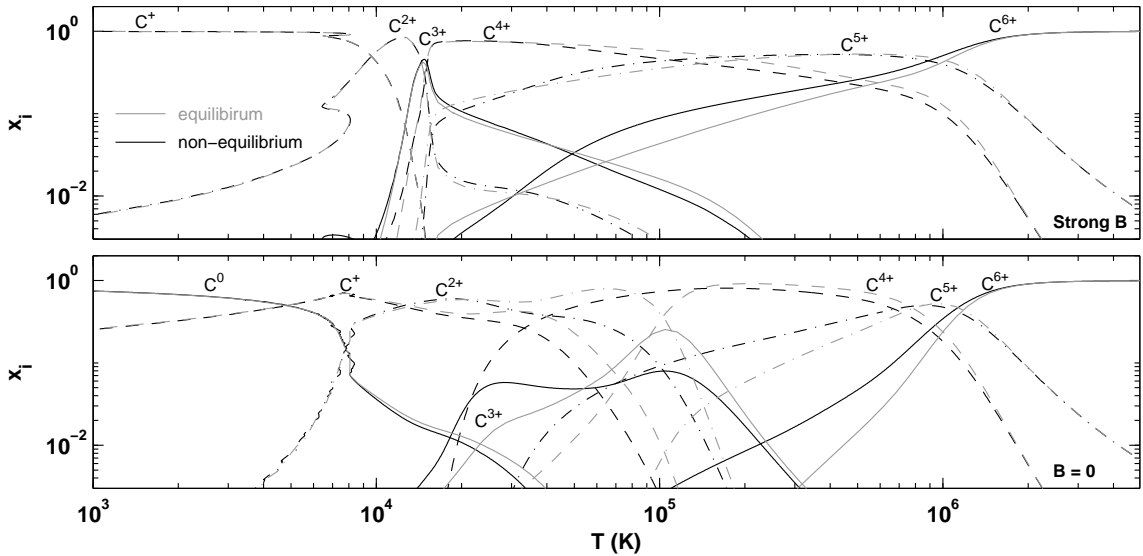}
\caption{ Carbon 
ion fractions
versus 
gas temperature in $T_s=5\times10^6$~K
shocks, for a solar metallicity gas. 
The upper 
panel is
for the strong-$B$ limit, 
and the lower
panel is
for $B=0$.
The gray curves show the local
photoionization equilibrium ion fractions. }
\label{eq}
\end{figure*}

To demonstrate how departures from equilibrium affect the ion fractions, we 
compare our time-dependent models to a computation in which 
local equilibrium ionization is imposed everywhere, given the
local mean intensities and temperatures that obtain for 
$5\times10^6$~K, $Z=1$ shocks.
In the comparison calculations we 
use the mean intensities and temperatures derived
from the full time-dependent computations. (These mean intensities are different
from those that would 
be derived in a ``self-consistent'' equilibrium model.)
Figure~\ref{eq} shows the time-dependent (dark) and equilibrium (gray) 
carbon 
ion fractions as a function of temperature.
The upper 
panel is for a strong-$B$ isochoric model,
and the lower panel is for $B=0$.

For strong-$B$ shocks, the upper
of Figure~\ref{eq} 
shows
that departures from photoionization equilibrium occur at temperatures 
between $\sim2\times10^4$~K and $\sim10^6$~K, but the ion fractions
differ by $\lesssim25\%$.
These differences are significantly smaller than the differences in the
absence of photoionization (GS07).
In strong-$B$ shocks, photoionization increases the abundances of
high-ions at low temperatures, beyond the enhancement of  
the ``collisional'' recombination lag (see Figure~\ref{photo}).
The time-dependent ionization states then remain close to
photoionization equilibrium (to within $25\%$) even in the
non-equilibrium cooling zone.

The
lower
panel shows that when $B=0$, departures from equilibrium
ionization are larger, due to the smaller effect of photoionization
resulting from the gas compression (see \S~\ref{results-B}).
For example, 
the non-equilibrium abundance of C$^{4+}$ is greater
than $0.05$ down to $\sim2.2\times10^4$~K, whereas the equilibrium
abundance vanishes below $\sim3.3\times10^4$~K.
Departures from equilibrium ionization tend to keep the gas at
any temperature over-ionized, as recombination lags behind 
cooling. The contribution of photoionization increases the 
equilibrium ion fractions relative to the case of pure radiative
cooling (GS07), especially at $T\lesssim10^5$K, where the shock
self-radiation is energetic enough to efficiently ionize
the abundant species, but the compression is still 
not too large
to maintain a significant ionization parameter.
At lower temperatures, the non-equilibrium ion-fractions
are due entirely to the recombination lags in the collisionally
ionized gas.
Below $\sim10^4$~K, efficient heating increases
the net cooling times significantly, and the gas approaches
photoionization equilibrium as can be seen by the near overlap
between the dark and gray curves.

\subsection{Gas Metallicity}

Our results for the non-equilibrium ion fractions, $x_i(T)$, for
metallicities $Z$ equal to $10^{-3}$, $10^{-2}$, $10^{-1}$, $1$
and $2$,
are presented in Table~\ref{ion-frac-table}.
Figure~\ref{frac-Z} shows, as an example,
the carbon ion fractions for the different
values of $Z$.
The 
assumed
metallicity affects the ion fractions in
several ways.
First, the cooling times depend on the metal abundance.
Higher $Z$ leads to enhanced metal lines cooling, and therefore shorter
cooling times. Departures from equilibrium 
and recombination lags 
are therefore larger for
higher metal abundances.
Higher metallicity
gas will tend to be more over-ionized.

Second, the spectral shape of the shock self radiation
depends on the gas metallicity.
Since metal line emission is enhanced for higher $Z$, a larger fraction
of the initial energy is radiated via line emission, and a smaller fraction
as bremsstrahlung continuum.
High-metallicity shocks therefore produce more photons with energies between
$1$ and $\sim5$~Rydbergs, and 
fewer
photons with $E\gtrsim6$~Rydbergs.
At low $Z$ the shock self-radiation is harder.
The changing spectral energy distribution as a function of
gas metallicity affects the photoionization rates, and therefore
the ion fractions as a function of temperature.

\begin{figure*}
\epsscale{1.0}
\plotone{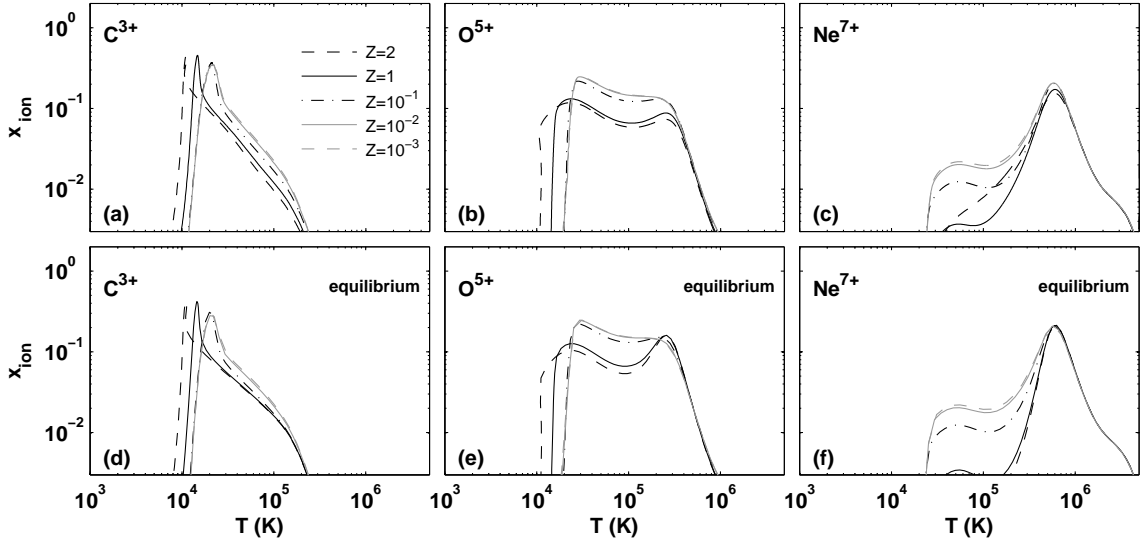}
\caption{ C$^{3+}$, O$^{5+}$, and Ne$^{7+}$ ion fraction versus
temperature for gas metallicities between $10^{-3}$ and $2$ times solar,
in a $5\times10^6$~K shock in the limit of strong-$B$  (panels [a]-[c]).
Panels (d)-(f) show, for comparison,  the local {\it equilibrium} ion
fractions, given the local mean intensities and temperatures that
obtain in the shock.}
\label{ionsZ}
\end{figure*}

As examples illustrating the various effects,
in Figure~\ref{ionsZ}a-c we display the C$^{3+}$, O$^{5+}$, 
and Ne$^{7+}$ distributions for the different 
values of $Z$ 
for a $5\times10^6$~K shock in the strong-B limit.
For comparison, in panels (d)-(f) we shows the local equilibrium
ion fractions, given the local mean intensities and temperatures that
obtain in the shock.

Panel (a) shows that carbon is more over-ionized for higher
metallicity gas, and that the C$^{3+}$ ion fraction peaks at lower
temperatures for higher $Z$. 
For example, for $Z=2$ the C$^{3+}$ distribution peaks
at $T\simeq10^4$~K. For smaller $Z$, the ion fraction peaks at higher 
temperatures. For $Z=10^{-3}$, it peaks at $T\simeq2\times10^4$~K.
The sharp decline in the C$^{3+}$ fraction at $\sim10^4$~K, is related
to the onset of the photoabsorption plateau, which occurs at a
metallicity-dependent temperature (see \S~\ref{struct-Z}).
Panels (a) and (d) show that the C$^{3+}$ fraction remains close
to its equilibrium distribution for any value of $Z$, and for the
entire temperature range over which it is abundant ($\sim10^4-2\times10^5$~K). 

O$^{5+}$ shows similar behavior.
The higher metallicity shocks are
more over-ionized, and the O$^{5+}$ fraction 
persists
to lower
temperatures where the photoabsorption plateau occurs.
A comparison of panels (b) and (e) shows that departures from equilibrium
ionization occur only for $Z\gtrsim1$, and are limited to $T\sim3\times10^5$~K
where 
the role of photoionization is still minor.

For Ne$^{7+}$, 
the behavior is a bit more complicated.
At high gas metallicities ($Z>1$), the recombination lags
enhance the Ne$^{7+}$ abundances above the enhancements
due to photoionization, for temperatures between $4\times10^4$ and $2\times10^5$~K.
Because the spectral energy distributions for $Z=2$ and for $Z=1$ are similar,
the equilibrium distributions are identical. They are also
narrower than the non-equilibrium distributions.
However, for lower values of $Z$ ($<0.1$), the varying spectral
energy distributions affects the Ne$^{7+}$ fractions.
Figure~\ref{specs-Z} 
shows that there are more photons capable of ionizing
Ne$^{6+}$
(with an ionization threshold of $15.2$~Ryd) for low $Z$, than for $Z=2$. 
Indeed, for decreasing $Z$
the Ne$^{7+}$ ion distribution becomes broader again, due
to enhanced photoionization by the shock radiation.
This can be seen in panel (f). A comparison with 
panel (c) then shows that for $Z<0.1$ the Ne$^{7+}$-fraction approaches
the equilibrium distribution controlled by photoionization.

At low gas metallicities ($\lesssim10^{-2}$), the total contributions of 
the metals to the gas cooling and to the emissivity becomes negligible
(GS07).
The cooling rates and spectral energy distributions therefore become
independent of $Z$. This can be clearly seen in Figure~\ref{ionsZ},
where the ion distributions for $Z=10^{-2}$ and for $Z=10^{-3}$ are
similar for C$^{3+}$, O$^{5+}$, and Ne$^{7+}$.

\subsection{Auger Effects}

\begin{figure}[!h]
\epsscale{1}
\plotone{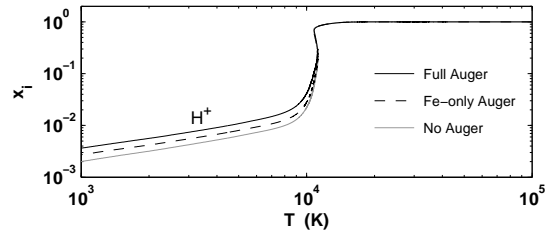}
\caption{ H$^+$ 
fraction as a function of temperature for a 
strong-$B$, $5\times10^7$~K shock at $Z=1$.
The dark solid line shows the H$^+$ fraction when multi-electron
Auger ionizations are included. 
The gray solid line shows a model in which Auger ionizations are
excluded.
The dashed curve shows the H$^+$ ion fraction when Auger ionizations 
are only included for iron ions.}
\label{auger}
\end{figure}

We have included multi-electron Auger ionizations using cross section and
yields from Kaastra \& Mewe (1993). Auger ionization is a potential source of multiply charged
metal species in largely neutral (and cool) hydrogen gas, due to the
large penetration depths of high-energy photons. However, we find that
rapid charge-transfer neutralization with atomic hydrogen
quickly removes such ions.
The metal abundances are therefore not significantly affected by the
Auger processes (cf.~Dopita and Sutherland 1996).
However, the charge transfer reactions do lead to a slight rise
in the ionized-hydrogen (proton) densities.
We demonstrate this effect in Figure~\ref{auger} for a 
strong-$B$, $Z=1$,
$5\times10^7$~K shock, in which intense X-rays are produced.
The dark and gray solid lines show the ionized hydrogen fractions with and
without the inclusion of Auger processes.
Major contributors to the increase in the ionized hydrogen fraction
are charge-transfer reactions with
Fe and Si ions. The dark
dashed curve shows the H$^+$ abundances if Auger ionizations 
are included for iron ions only.

\section{Heating and Cooling}
\label{cooling}

We have carried out 
computations
of the heating and cooling efficiencies
in the post-shock cooling layers for shock temperatures of
$5\times10^6$~K and $5\times10^7$~K, for gas metallicities
$Z$
between
$10^{-3}$ and $2$, 
for $B=0$ 
and strong-$B$
dynamics.
Our results for the heating rates and cooling efficiencies are listed in
Table~\ref{cool-table}, as outlined by Table~\ref{guide}.
We use the self-consistent non-equilibrium ion
fractions, $x_i$, obtained in \S~\ref{ion-frac} 
to compute $\Lambda(T,x_i,Z)$
and $\Upsilon(x_i,Z,J_\nu)$
for all temperatures in the downstream gas.
There is no external source of photoionization
and heating, and 
all of the heating radiation is produced by the shock itself.
In all of our shock models, 
photo-heating is negligible above temperatures of a few $\times10^4$~K
and the gas undergoes pure radiative cooling. However, as discussed
in \S~\ref{ion-frac}, photoionization of the metals can become a very important
process even in the hotter components, and this can alter the
radiative cooling efficiencies.

\begin{deluxetable}{llcccc}
\tablewidth{0pt}
\tablecaption{Cooling and Heating in a $5\times10^6$~K, $2$ times solar, strong-$B$-shock}
\tablehead{
\colhead{Time} &
\colhead{Temperature} &
\colhead{$\Lambda$} & 
\colhead{$\Upsilon$} \\
\colhead{(s)}&
\colhead{(T)}&
\colhead{(erg~s$^{-1}$~cm$^{3}$)}&
\colhead{(erg~s$^{-1}$)}
}
\startdata
$0.00$          &$5.000\times10^6$&$2.72\times10^{-21}$&$2.98\times10^{-27}$\\
$7.25\times10^8$&$4.998\times10^6$&$2.34\times10^{-21}$&$4.38\times10^{-28}$\\
$1.57\times10^9$&$4.997\times10^6$&$2.10\times10^{-21}$&$1.79\times10^{-28}$\\
\enddata
\tablecomments{The complete version of this table is in 
the electronic edition of the Journal. The printed edition contains only a sample. 
The full table lists the cooling and heating
for strong-$B$ and for $B=0$,
for shock temperatures of $5\times10^6$~K and $5\times10^7$~K, and for
$Z=10^{-3}$, $10^{-2}$, $10^{-1}$, $1$, and $2$ 
(for a guide, see Table~\ref{guide}). The times in the first column 
are given assuming a post-shock hydrogen density of $0.1$~cm$^{-3}$.
$\Lambda$ and $\Upsilon$ are defined such that the cooling rate per
volume is $n_en_{\rm H}\Lambda$, and the heating rate
per volume is $n_{\rm tot}\Upsilon$.}
\label{cool-table}
\end{deluxetable}

For $Z\gtrsim0.1$, the radiative cooling between a few $\times10^4$~K and
$\sim10^7$~K is dominated by electron impact excitation of resonance
line transitions of metal ions. Above $\sim10^7$~K, bremsstrahlung cooling
dominates for all gas metallicities.
At lower temperatures ($\lesssim10^4$~K), Ly$\alpha$ cooling and
fine-structure lines dominate the cooling, with relative
contributions depending on $T$ and $Z$, as we discuss below
(Boehringer \& Hensler~1989; Sutherland \& Dopita~1996; GS07).

\begin{figure*}
\epsscale{1.0}
\plotone{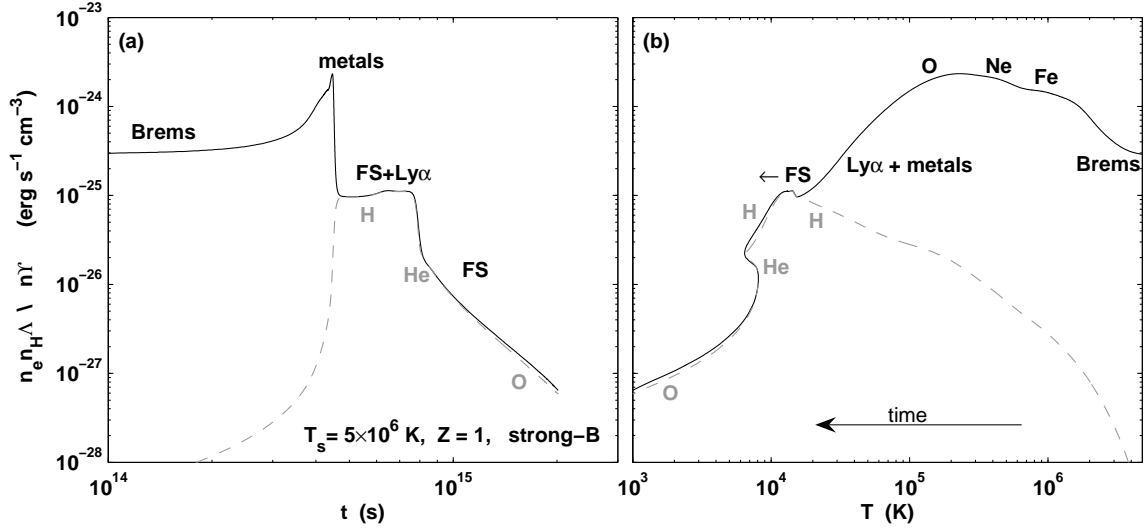}
\caption{ The cooling (dark) and heating (gray) rates per 
unit
volume
(erg~s$^{-1}$~cm$^{-3}$) 
for a strong-B $5\times10^6$~K at $Z=1$.
(a)~Cooling and heating rates per 
unit
volume as a function of time,
assuming a post-shock hydrogen density $n_0=0.1$~cm$^{-3}$.
(b)~Cooling and heating rates per 
unit
volume as a function of
temperature.
The dominant cooling and heating elements are indicated near
the curves.
}
\label{cool-case}
\end{figure*}

As an example, in Figure~\ref{cool-case} we 
show the cooling and heating rates per unit volume,
$n_en_H\Lambda$ and $n\Upsilon$,
as a function of time 
(in panel~[a]),
and as a function of temperature 
(in panel[~b]) for a 
strong-$B$, $T_s=5\times10^6$~K shock at $Z=1$.
In panel~(a), we have assumed a post-shock hydrogen density of 
0.1~cm$^{-3}$
for the cooling time-scales.
The dominant cooling and heating elements are indicated near the curves.

The gas starts out in the ``hot radiative'' phase.
In this model it remains in this phase for
$\sim3.5\times10^{14}$~s, 
and cools gradually, via metal resonance transitions and bremsstrahlung emission,
down to a temperature of $\sim2.5\times10^6$~K.
It then enters the ``non-equilibrium cooling'' phase, where it 
rapidly cools to a temperature of $\sim10^4$~K via very 
efficient metal line cooling. This cooling is due mainly to
iron, neon, and oxygen ions, as indicated in 
Figure~\ref{cool-case}b.
Throughout the ``hot radiative'' and ``non-equilibrium cooling'' 
phases, heating is
negligible, and the gas cools radiatively
from the initial hot shock temperature
down to $\sim10^4$~K.
However, since photoionization has a major influence on the ion fractions,
the non-equilibrium cooling rates differ significantly from those presented
in GS07 in which no sources of ionizing radiation were considered.
We discuss this point further below.

As discussed in \S~\ref{struct},
once the neutral hydrogen fraction becomes large 
enough, 
$\gtrsim 10^{-3}$,
to allow for efficient photoabsorption, 
the gas enters the 
``WIM 
photoabsorption plateau''.
In this stage, heating is mainly due to 
H and He$^+$ photoabsorption.
As previously discussed,
for $Z=1$, Ly$\alpha$ provides only $15\%$ of the cooling at its peak
efficiency.
The rest of the cooling is provided by 
metal
resonance lines 
for
$T\gtrsim2\times10^4$~K, and by 
fine-structure (FS) and resonance 
line
transitions at
lower temperatures.
For lower gas metallicities, Ly$\alpha$ constitutes a larger 
fraction
of the cooling at $T\sim10^4$~K. 
Below $\lesssim8000$~K, 
Ly$\alpha$ cooling vanishes
and the cooling is dominated 
entirely
by fine-structure transitions.

Heating is 
first
dominated by hydrogen
photoabsorption. As the
radiation is increasingly absorbed and the gas cools to
$\sim6500$~K, helium photoabsorption
dominates the heating.
Finally,
below $\sim2000$~K, 
and 
as the ionizing radiation 
becomes even harder, oxygen and other metals become significant absorbers.
In the 
WIM and WNM plateaus
the gas remains
close to thermal equilibrium, as is evident by the fact that the heating
and cooling curves nearly overlap.
The delicate balance between heating and cooling that depends 
on the temperature, electron density and ion fractions, determines
the temperatures profiles that were discussed in \S~\ref{struct}.

The way in which the cooling efficiency, $\Lambda$, depends 
on the gas metallicity determines the column-density ratios
in the post-shock cooling layers, as we discuss in 
\S~\ref{columns} below. For $T>10^7$~K, cooling is
dominated by bremsstrahlung emission for all gas
metallicities, and the cooling efficiency is therefore
only weakly dependent 
on $Z$.

For $2\times10^4$~K$~\lesssim T \lesssim 10^7$~K, 
resonance metal line transitions are efficiently excited.
For $Z\gtrsim0.1$, the dominant cooling process is
metal line cooling, and $\Lambda\propto1/Z$.
For $Z\lesssim0.01$, the contribution of metals
to the gas cooling is negligible. 
The cooling is dominated
by bremsstrahlung emission and by hydrogen and helium
Ly$\alpha$, and $\Lambda$ is therefore
independent of $Z$.

At $T\sim10^4$~K, a mixture of 
metal fine structure
and Ly$\alpha$ emissions provide the gas cooling.
For isochoric flows with $Z>1$, fine-structure 
emissions dominate 
the cooling, and $\Lambda$ is therefore proportional to
$1/Z$.
For lower gas metallicities, and in 
$B=0$
flows 
Ly$\alpha$ emission 
dominates
the cooling 
and $\Lambda$ is independent of $Z$.

Below $\sim7000$~K, fine structure line emissions are the
only available coolants we have included
and $\Lambda$ is therefore proportional to $Z$ for all gas
metallicities. The dependence of $\Lambda$ on $Z$ is
summarized in Table~\ref{LamZ} below.
These scalings will be useful in \S~\ref{columns}.

\begin{deluxetable}{lccc}
\tablewidth{0pt}
\tablecaption{The Dependence of $\Lambda(x_i,T)$ on $Z$}
\tablehead{
\colhead{Temperature} & 
\multicolumn{3}{c}{Cooling efficiency, $\Lambda(x_i,T)$}\\
\colhead{K}&
\colhead{$Z\lesssim10^{-2}$} &
\colhead{$10^{-2}< Z < 1$}&
\colhead{$Z\gtrsim1$}
}
\startdata
$\lesssim7000$     & $\propto Z$~$^{(1)}$ & $\propto Z$~$^{(1)}$ & $\propto Z$~$^{(1)}$\\
$\sim10^4$         & $\propto 1$~$^{(2)}$ & $\propto 1$~$^{(2)}$ & $\propto Z$~$^{(1)}$\\
$2\times10^4-10^7$ & $\propto 1$~$^{(3)}$ & $\propto Z$~$^{(4)}$ & $\propto Z$~$^{(4)}$\\
$\gtrsim10^7$      & $\propto 1$~$^{(5)}$ & $\propto 1$~$^{(5)}$ & $\propto 1$~$^{(5)}$           \\
\enddata
\tablecomments{Given a set of ion fractions $x_i$, and a 
temperature $T$, the table shows how the cooling efficiency
$\Lambda(x_i,T)$ depends on the gas metallicity $Z$.
The entry ``$\propto1$'' means that the cooling efficiency 
is nearly independent of $Z$ (but does depend on $x_i$
and $T$).\\
(1) FS line cooling.\\
(2) H Ly$\alpha$ cooling.\\
(3) H, He, and bremsstrahlung cooling.\\
(4) Metal resonance line cooling.\\
(5) Bremsstrahlung cooling.}
\label{LamZ}
\end{deluxetable}

\begin{figure*}
\epsscale{1}
\plotone{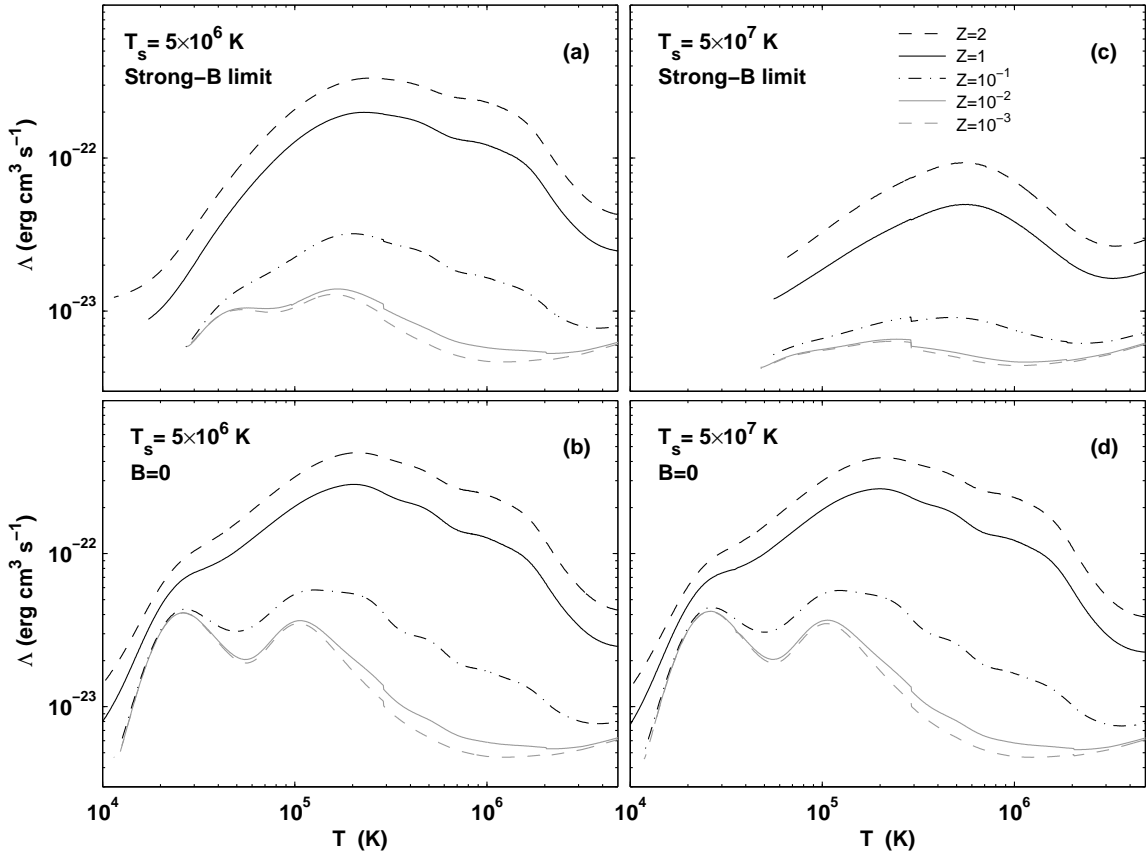}
\caption{ The cooling efficiencies, $\Lambda$~(erg~cm$^3$~s$^{-1}$),
as a function of temperature in the ``hot radiative'' and ``non-equilibrium
cooling'' zones. Each panel shows results for gas metallicities 
between $10^{-3}$ and $2$ times solar. For the range of temperatures
displayed in this figure, heating is 
negligible.
The curves end where the
photoabsorption plateaus begin.
(a)~$T_s = 5\times10^6$~K, for strong-$B$. 
(b)~$T_s = 5\times10^6$~K, for $B=0$.
(c)~$T_s = 5\times10^7$~K,  for strong-$B$. 
(d)~$T_s = 5\times10^7$~K, for $B=0$.
}
\label{highT-cool}
\end{figure*}

In Figure~\ref{highT-cool}, we display the cooling efficiencies 
for $Z$ between $10^{-3}$ and $2$, 
in the ``hot radiative''
and ``non-equilibrium cooling'' 
phases for the 
$T_s=5\times 10^6$~K (left panels) and $T_s=5\times 10^7$~K (right panels)
shocks that we have considered, in the $B=0$ (lower panels)
and
strong-B limits (upper panels).
We display results for temperatures between $5\times10^6$~K, 
and the
temperatures at which the photoabsorption plateau begins.

Consider first the $5\times 10^6$~K shocks.
For $B=0$ (isobaric) the gas compression reduces the 
photoionization parameter, and the non-equilibrium ion-fractions
are mostly due to the recombination lags in the collisionally
ionized gas. In this limit, therefore, the cooling curves
are almost identical to the non-equilibrium cooling curves
presented in GS07 in which photoionization processes were excluded. 
However, as discussed in \S~\ref{ion-frac}, 
in the strong-B (isochoric) limit, photoionization by the shock
radiation significantly increases the ionization state of the gas above
the non-equilibrium level that obtains in a purely radiatively cooling gas.
More highly ionized species generally 
have more energetic resonance line transitions, and these are less
efficiently excited by the thermal electrons at any given
temperature (McCray 1987).
The cooling curves are therefore suppressed,
compared to $B=0$ shocks.

For the hotter $5\times 10^7$~K shock, the shock 
radiation is more intense, and photoionization drives the ion
fractions to still higher ionization states. Again,
the more highly ionized gas cools less efficiently, and the cooling
rates at a given temperature are therefore 
lower compared to the $5\times10^6$~K shocks.
In the strong-$B$ limit
the intensity of the radiation is high enough
that even minute amounts of neutral hydrogen allow for efficient 
photoabsorption heating, and the photoabsorption plateaus begin
at a higher temperature 
of $\sim5\times10^4$~K.

\section{Cooling Columns in Shocked Gas}
\label{columns}

In this section we present 
computations of
the integrated metal-ion column
densities 
that are produced in the post-shock cooling layers.
In computing the cooling columns, we integrate from the shock front
to a distance where the gas 
has cooled down to our termination temperature $T_{\rm low}=1000$~K.

\begin{figure*}
\epsscale{1.0}
\plotone{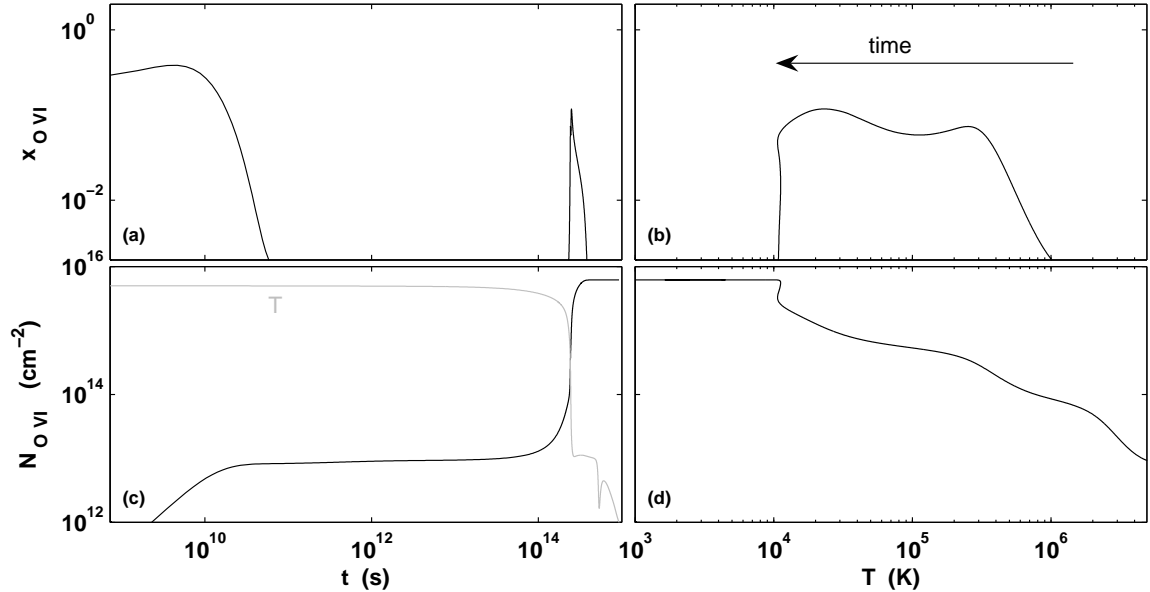}
\caption{ O$^{5+}$ ion fraction (upper panels) and accumulated
column density (lower panels) as a function of time (left hand
panels) and temperature (right hand 
panels),
for 
a strong-B, $T_s=5\times10^6$~K, $Z=1$ shock.
The gray curve in panel (c) shows the temperature 
(rescaled by $\times10^9$) 
as a function of time.}
\label{OVI-column}
\end{figure*}

As an example of what these integrations look like,
in Figure~\ref{OVI-column} we display the O$^{5+}$ ion fraction and
accumulated column density, in a 
strong-$B$,
$5\times10^6$~K shock, for $Z=1$.
Panel~(a) shows the O$^{5+}$ ion fraction as a function of time, for
an assumed post-shock hydrogen density of $0.1$~cm$^{-3}$.
Initially, the O$^{5+}$ abundance peaks as the 
gas
is ionized after passing the shock front. This first peak is short,
with a duration of less than $10^{11}$~s.
Later on, a second peak is 
formed as the gas 
cools behind the
shock front.
This second peak lasts 
for 
$\sim10^{14}$~s.
Panel~(b) shows the ion fraction as a function of temperature.
The first abundance peak 
occurs over such a narrow temperature
range ($T\sim T_s$) that it cannot be resolved in panel~(b).
The second abundance peak is seen to occur between a temperature
of $\sim10^6$~K, and $\sim10^4$~K.

Panel~(c) shows the accumulated O$^{5+}$ column density as a function
of time.
The initial abundance peak contributes an O$^{5+}$ column density of
$\sim10^{13}$~cm$^{-2}$. 
The column 
density
builds rapidly, and 
remains
unchanged until $\gtrsim10^{14}$~s.
The gray curve 
in panel (c)
shows a scaled temperature as a function of time.
During 
the hot radiative 
phase the accumulated column density
remains unaltered, as higher 
ions dominate the 
oxygen
abundance.
The column remains constant until the 
non-equilibrium cooling phase, 
during
which the second abundance 
peaks occurs. The column density then builds up rapidly to
$\sim10^{15}$~cm$^{-2}$.
The O$^{5+}$ column 
keeps increasing in the photoabsorption
plateau, until O$^{5+}$ finally recombines at $T\sim10^4$~K.
The accumulated column density as a function of temperature is shown
in panel~(d). 
Panels~(c) and (d) show that the contribution of the first abundance
peak to the total column density in the post-shock cooling layers is
negligible.

More generally,
the integrated metal ion column density, $N_i$, through the
post-shock cooling gas is given by,
\begin{equation}
\label{Ni_t}
N_i = \int^{0}_{t_{\rm end}} n_H Z A_{\rm el} x_i v dt
\ \ \ ,
\end{equation}
with
\begin{equation}
\label{dt}
dt \propto \frac{dT}{n\Lambda(x_i,T,Z)}
\end{equation}
In this expression, $n_H$ is the hydrogen density, $A_{\rm el}$ is
the abundance of the element relative to hydrogen
in a solar metallicity gas, $x_i$ is the ion
fraction, and $v$ is the gas velocity.

Since $n_H v$ is constant throughout the flow 
(eq.~[\ref{conservations}]),
equations~(\ref{Ni_t}) and (\ref{dt}) imply that 
\begin{equation}
\label{Ni_T}
N_i \propto Z\int^{T_s}_{T_{\rm low}} x_i \frac{dT}{n\Lambda}.
\end{equation}

For strong-$B$ isochoric flows 
$n=n_0$ everywhere, 
and $N_i \propto Z\int x_i dT/\Lambda$.
However, 
for
$B=0$ isobaric flows
the density increases 
and the cooling times are shortened
as the gas cools (see \S~\ref{struct}).
Therefore, the column densities of ions produced in low
temperature gas are substantially suppressed (by the
shorter cooling times) 
in isobaric versus isochoric cooling flows (e.g.~Edgar \& Chevalier 1986). 
For an ion that is produced at a characteristic
temperature $T$, the column density 
in a $B=0$ shock is quenched by a
factor approximately $n_0/n(T) \simeq T/T_s$ relative to 
a strong-$B$ shock.
Because
$T/T_s$ is smaller for higher $T_s$,
the quenching effect is more significant for higher shock velocities.

Given an ion distribution $x_i(T)$, 
equation~(\ref{Ni_T})
implies
that $N_i$ 
is independent of $Z$
if $\Lambda\propto Z$ (as occurs when metals dominate the cooling).
When $\Lambda\propto Z$, reduced elemental abundances are compensated
by longer cooling times, 
and $Z$ cancels out in equation~(14).
On the other hand, for ions that are abundant at temperatures
where the cooling is dominated by hydrogen, helium, or bremsstrahlung
emissions, 
equation~(\ref{Ni_T}) implies that the column densities
are proportional to $Z$.

The actual behavior is more complicated because
the ion distributions, $x_i(T)$, 
are themselves 
a function of gas 
metallicity and shock temperature as discussed in \S~\ref{ion-frac}.
For example, higher shock temperatures imply stronger
photoionizing radiation fields, which affect the ionization
states in the post-shock cooling layers. The gas metallicity
affects the spectral energy distribution of the shock
self-radiation, changing the ion fractions for different values of $Z$.
Departures from equilibrium ionization and the degree of
over-ionization in the gas are also a function of gas
metallicity. All these factors imply that the column densities 
depend on
the metallicity even for temperatures where
$\Lambda\propto Z$.

\begin{figure*}
\epsscale{1.0}
\plotone{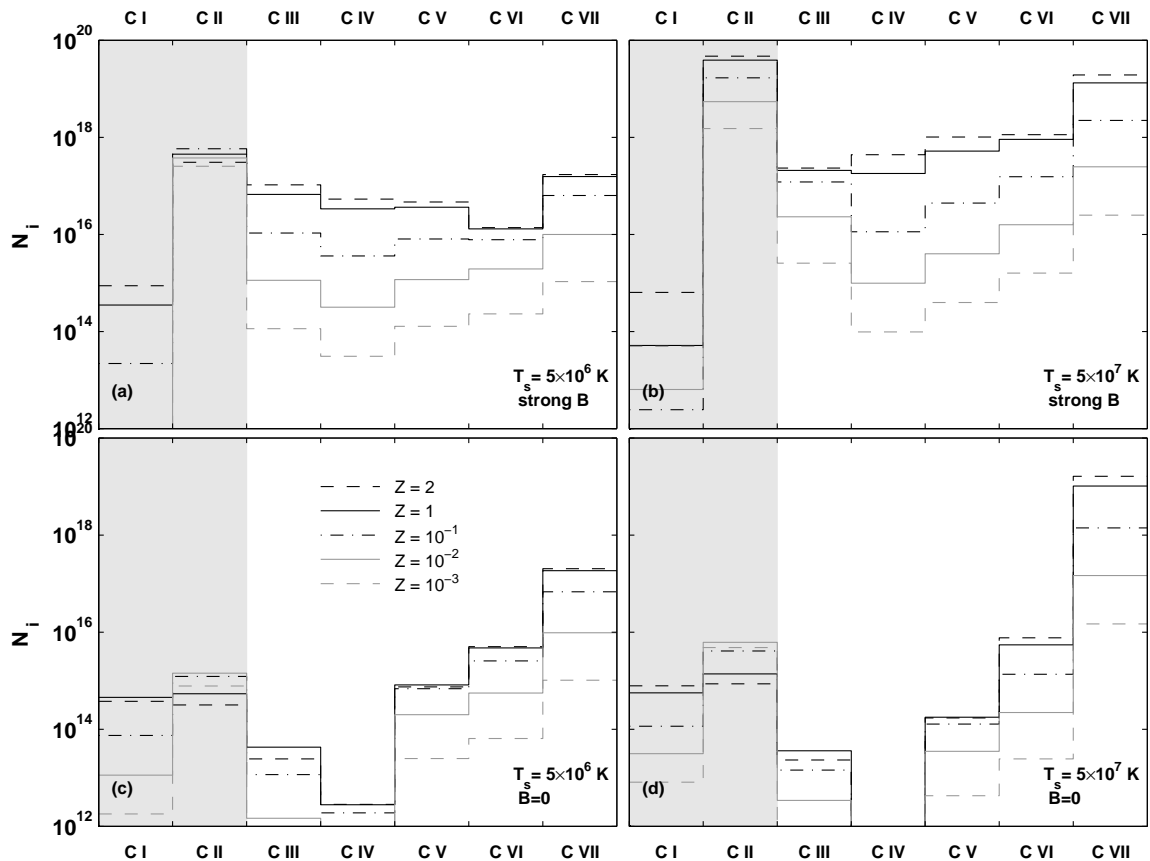}
\caption{ Carbon ion-column densities in post-shock cooling layers,
for $Z$ from $10^{-3}$ to $2$ times solar.
(a)~$T_s = 5\times10^6$~K, for strong-$B$.
(b)~$T_s = 5\times10^7$~K, for strong-$B$.
(c)~$T_s = 5\times10^6$~K, for $B=0$.
(d)~$T_s = 5\times10^7$~K, for $B=0$.
The shaded areas show ions that are still abundant at $T_{\rm low}=1000$~K
where we stop the computation. The column densities of these ions are
sensitive to the choice of $T_{\rm low}$.
}
\label{C-col}
\end{figure*}

\begin{figure*}
\epsscale{1.0}
\plotone{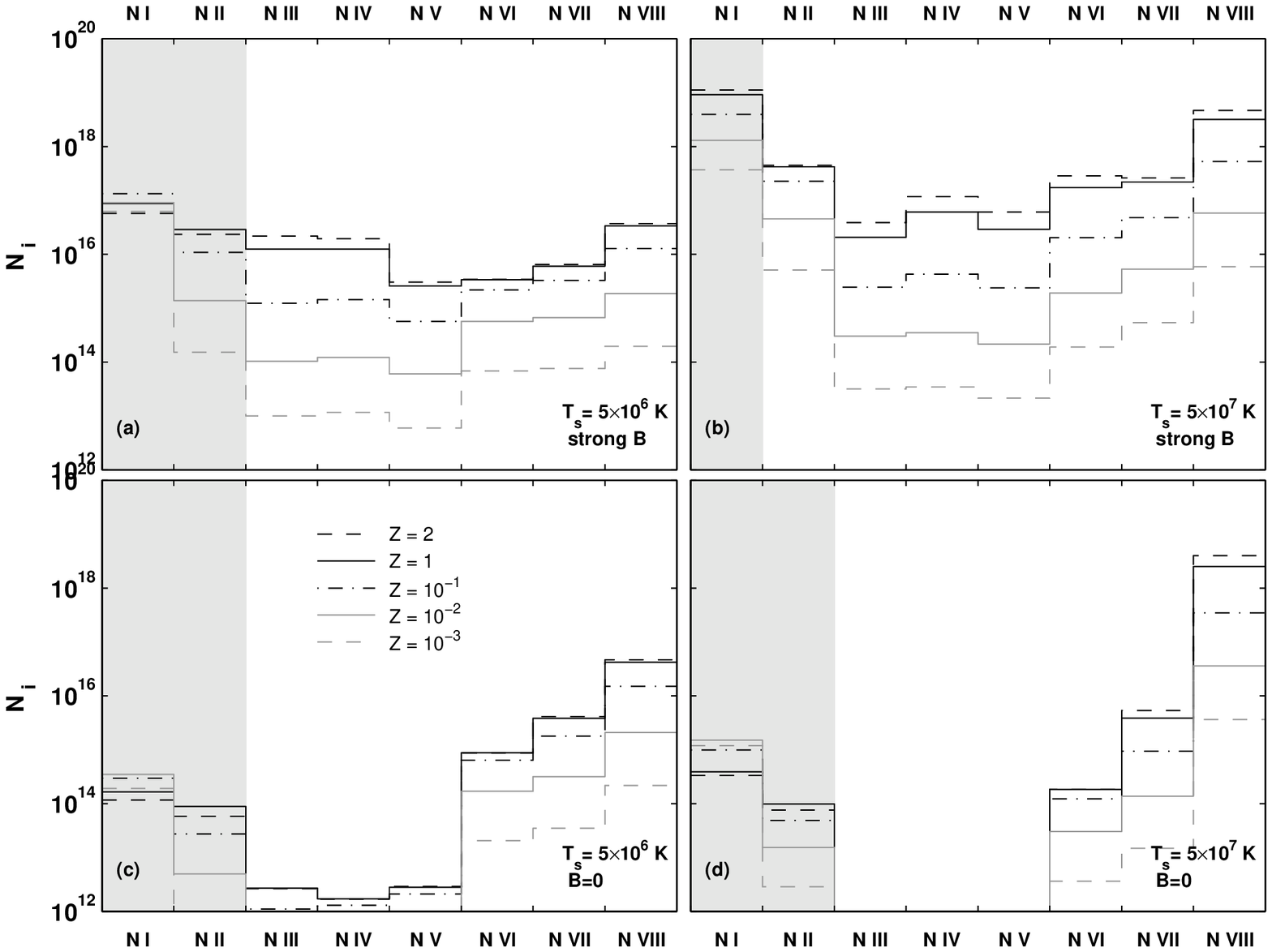}
\caption{ Same as Figure \ref{C-col} but for Nitrogen.}
\label{N-col}
\end{figure*}

\begin{figure*}
\epsscale{1.0}
\plotone{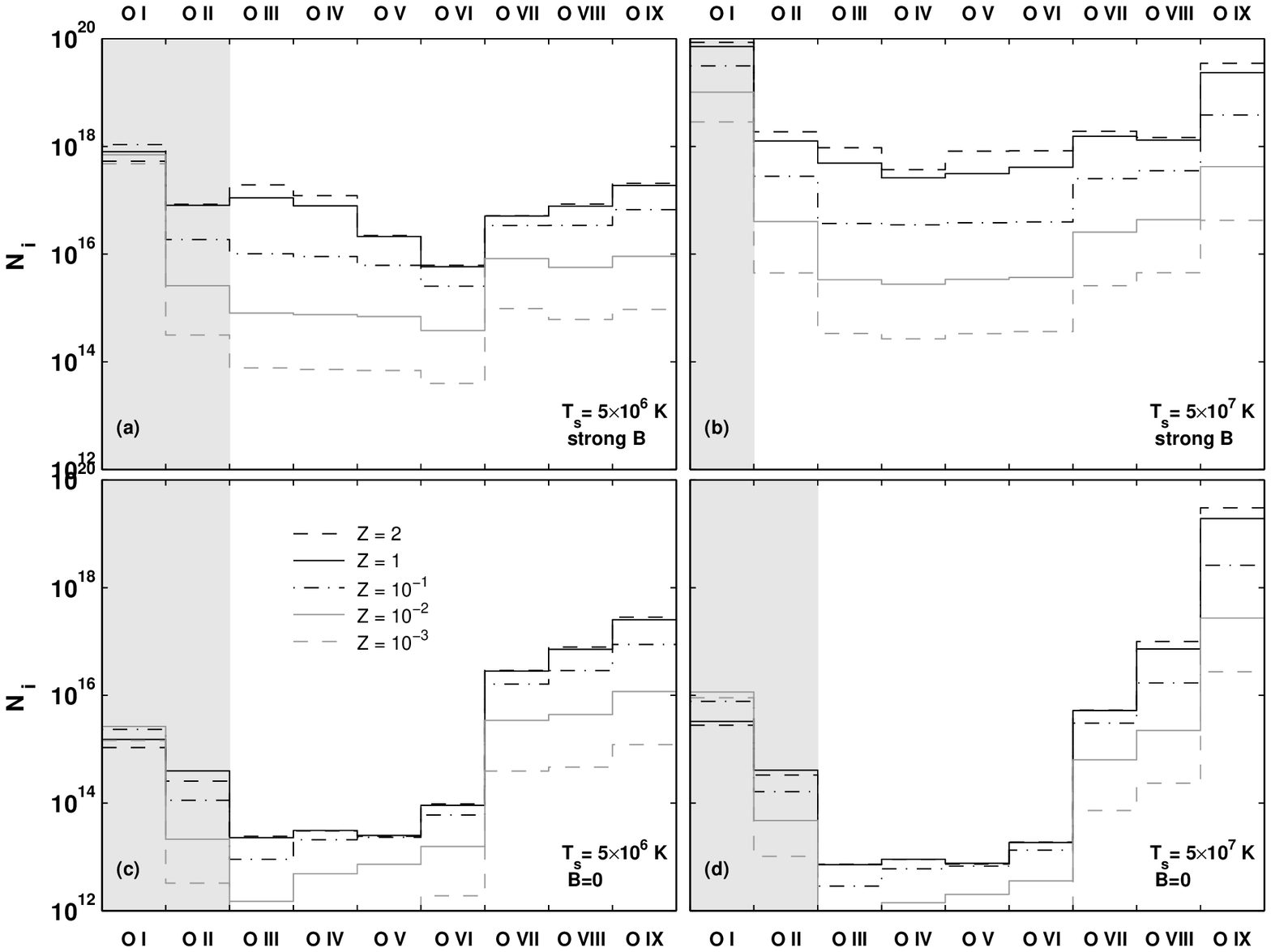}
\caption{ Same as Figure \ref{C-col} but for Oxygen.}
\label{O-col}
\end{figure*}

\begin{figure*}
\epsscale{1.0}
\plotone{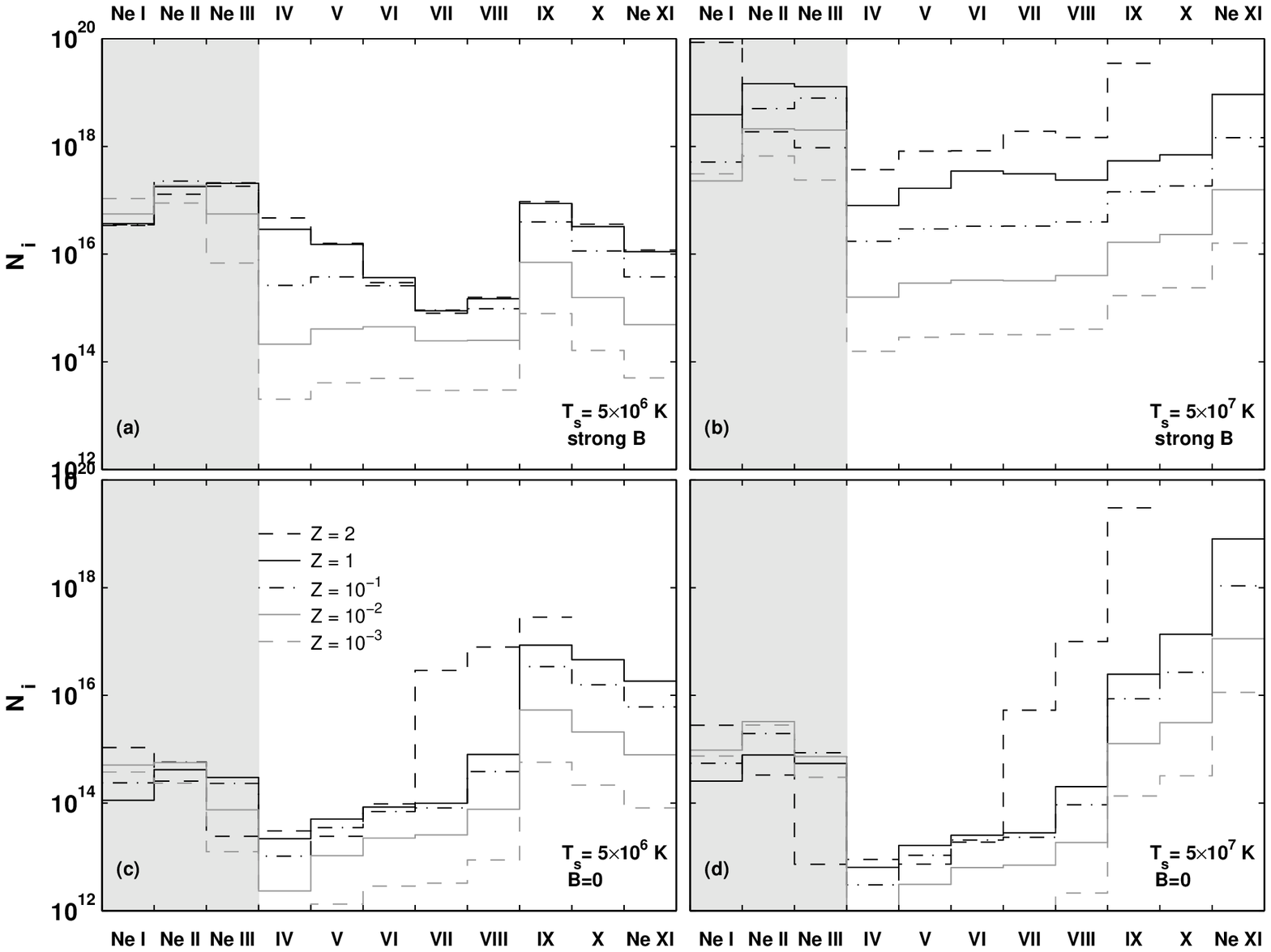}
\caption{ Same as Figure \ref{C-col} but for Neon.}
\label{Ne-col}
\end{figure*}

In Figures~\ref{C-col}-\ref{Ne-col} we 
display the full set of carbon-, nitrogen, oxygen, and neon-ion
column densities, for $Z$ between $10^{-3}$ and $2$,
for strong-B (upper panels) and B=0 (lower panels) post-shock
cooling layers.
Left hand panels 
are
for $T_s=5\times10^6$~K,
and the right hand panels for  $T_s=5\times10^7$~K.
The shaded areas show ions that are still abundant at
$T_{\rm low}=1000$~K. 
The column densities of these ions 
(e.g.~O and O$^+$)
are sensitive to the
choice of $T_{\rm low}$. 
The full set 
of
column densities for all the metal ions that we consider
are listed in Table~\ref{columns-table}, as described in 
Table~\ref{guide}.
\footnote{The columns presented in
Table~\ref{columns-table} and
Fsigures~\ref{C-col}-\ref{Ne-col} assume that the shock is
observed ``face-on'' (so that the line-of-sight is parallel
to the shock velocity). These columns should be multiplied
by a geometrical correction factor,
$1/cos\theta$,
for shocks that are viewed at an angle $\theta$
to the normal direction.}

\begin{deluxetable*}{llcccc}
\tablewidth{0pt}
\tablecaption{Metal-ion Column Densities in a $5\times10^6$~K, $2$ times solar, strong-$B$-shock}
\tablehead{
\colhead{ionization} &
\colhead{H} &
\colhead{He} & 
\colhead{C} & 
\colhead{N} & 
\colhead{\ldots} \\
\colhead{}&
\colhead{(cm$^{-2}$)}&
\colhead{(cm$^{-2}$)}&
\colhead{(cm$^{-2}$)}&
\colhead{(cm$^{-2}$)}&
\colhead{\ldots}
}
\startdata
I   & $5.7\times10^{20}$ & $4.4\times10^{19}$ & $8.8\times10^{14}$&$5.7\times10^{16}$&\ldots\\
II  & $8.5\times10^{20}$ & $2.7\times10^{19}$ & $3.1\times10^{17}$&$2.3\times10^{16}$&\ldots\\
III & --                 & $4.8\times10^{19}$ & $1.0\times10^{17}$&$2.2\times10^{16}$&\ldots\\
\enddata
\tablecomments{The complete version of this table is in the electronic
edition of the Journal. The printed edition contains only a sample.
The full table lists metal-ion column densities for $B=0$ and in the
strong-$B$ limit,
for shock temperatures of $5\times10^6$~K and $5\times10^7$~K, and for
$Z=10^{-3}$, $10^{-2}$, $10^{-1}$, $1$, and $2$ times solar metallicity
gas (for a guide, see Table~\ref{guide}). }
\label{columns-table}
\end{deluxetable*}

Our integrated metal-ion cooling column densities are in good
qualitative agreement with the recent results of Allen et al.~(2008)
for the case of a solar metallicity gas cooling behind
a $600$~km~s$^{-1}$ shock, in which $B\simeq0$ (their model ``J\_n1\_b0'').
Remaining differences are likely due to differences in the assumed
elemental abundances, and in the magnetic field strength.
The computations presented in Allen et al.~cover a different
region of the parameter space, and do not show results for the other
cases than we consider here (namely $v_s=2000$~km~s$^{-1}$, 
sub-solar metallicities, and the limit of strong-$B$).

We now focus on the oxygen ion fractions
shown in  Figure~\ref{O-col}
as an example.
Table~\ref{LamZ} predicts that the column densities in
low metallicity gas ($\lesssim10^{-2}$) are always
proportional to $Z$ except at very low temperatures.
As can be seen from Figure~\ref{O-col} the ratio
between the oxygen ion column densities for $Z=10^{-2}$ 
and the columns for $Z=10^{-3}$ is $\sim10$ as expected.
For example, in panel~(a) the columns of O$^{3+}$,
O$^{4+}$, and O$^{5+}$ are $\sim8\times10^{14}$, 
$\sim7.5\times10^{14}$, and $\sim7\times10^{14}$~cm$^{-2}$
for $Z=10^{-2}$, and $\sim8\times10^{13}$, $\sim7.5\times10^{14}$,
and $\sim7\times10^{13}$ for $Z=10^{-3}$.
The same applies for other shock temperatures and 
magnetic field limits, as can be seen in panels~(b)-(d).

For $Z>1$, Table~\ref{LamZ} predicts that for
a given ion distribution $x_i$, the column densities are
independent of $Z$. Figures \ref{O-col}c and \ref{O-col}d
show that 
for
$B=0$ where photoionization is
less significant, the column densities for $Z=1$ and for 
$Z=2$ are indeed similar. Departures from equilibrium 
have only a very minor impact on the column densities.
For example, in Figure~\ref{O-col}c, the O$^{5+}$ 
column is
$9.6\times10^{13}$~cm$^{-2}$ for $Z=2$, and 
$9.1\times10^{13}$~cm$^{-2}$ for $Z=1$, due to the 
longer recombination lags for  higher $Z$.
However, in strong-$B$ isochoric shocks, photoionization introduces a 
$Z$ dependence.
For example, Figure~\ref{O-col}b shows that the O$^{3+}$ column
density is $8.1\times10^{17}$~cm$^{-2}$ for $Z=2$,
but is $3.1\times10^{17}$~cm$^{-2}$ for $Z=1$.
Although $\Lambda\sim Z$, 
the O$^{3+}$ column density ratio of $2.6$
is in this case even larger than the metallicity-ratio of $2.0$.

Figures~\ref{C-col}-\ref{Ne-col} clearly show the differences between
strong-$B$ (isochoric) shocks and $B=0$ (approximately 
isobaric) shocks. For $B=0$,
the gas compression and shorter cooling times
reduce the column densities of low and intermediate ions
that are formed at $T<T_s$.
A comparison of Figures~\ref{O-col}a and \ref{O-col}c shows that while
the column densities of the high ions (O$^{6+}$, O$^{7+}$) is
similar in both cases, the column densities of O$^{2+}$-O$^{5+}$
are greatly suppressed in $B=0$ models.
For example, in solar metallicity gas, the O$^{6+}$ column is,
as expected, similar for the two cases.
It is $5\times10^{16}$~cm$^{-2}$
for a strong-$B$ shock, 
and $3\times10^{16}$~cm$^{-2}$ for $B=0$.
However, for O$^{3+}$ the column density is $8\times10^{16}$~cm$^{-2}$
for strong-$B$, and $3\times10^{13}$~cm$^{-2}$ for $B=0$,
$3.5$ orders of magnitude lower.
The magnetic field therefore has a very strong impact on the
column density ratios between high- and intermediate-ions,
and X-ray and UV observations will be able to easily 
distinguish between the cases of strong versus weak magnetic fields.

The shock velocity affects the column densities profoundly.
Equation \ref{Ni_t} implies that for ions that only exist below
$T_s$, and given an ion distribution $x_i(T)$, the integrated
column density will be proportional to the shock velocity.
However, since the shock self-radiation and the degree of
photoionization in the gas depends on the shock
temperature, this scaling is not exact.
For $B=0$, photoionization is less important, and the
ion distributions $x_i(T)$ are therefore less sensitive to $T_s$.
For strong-$B$ shocks, photoionization plays a major role in
setting the ion fractions in the cooling gas, and the ion 
distributions are more affected by $T_s$.

For example, panels~(a) and (b) show that the column density
of O$^{4+}$ is $8.1\times10^{17}$~cm$^{-2}$ in a $5\times10^7$~K
shock of a two solar metallicity gas, compared to 
$2.2\times10^{16}$~cm$^{-2}$ in a $5\times10^6$~K shock.
This is a factor of $\sim37$ in the column density,
compared to a factor of $10^{1/2}\sim3$ in the shock velocities.
For $B=0$ an additional factor $T/T_s$ suppresses the ion column
densities as discussed above, and the exact range of
temperatures over which the ion is abundant will determine the
column density ratio.

\section{Metal Columns in the Radiative Precursor}
\label{prec-cols}

In this section we present equilibrium photoionization 
computations of the integrated metal-ion column densities 
that are produced in the radiative precursors.
The precursor gas is heated and photoionized by the
shock self-radiation propagating upstream.
The upstream radiation is gradually absorbed in the
radiative precursor, producing a photoabsorption
layer in which the heating and ionization rates
gradually decline with (upstream) distance from
the shock front. 

The ionization parameter in the
radiative precursor is higher than in the post-shock
photoabsorption plateau, because the gas density
in the precursor is four times lower than the
post-shock density (see \S~\ref{flow-eqn}).
Typically, the temperature of the gas approaching
the shock front is a few times $10^4$~K, but it may
be as high as $\sim10^5$~K for $T_s=5\times10^7$~K.
In computing the precursor columns, we integrate from 
the shock front to a distance where the heating rate 
in the unshocked gas is sufficiently low that is reaches
a temperature $T_{\rm low}=1000$~K.

We use Cloudy to construct equilibrium photoionization
models for the radiative precursors. 
For the radiation flux entering the precursor, 
we use the post-shock flux at the position in which the 
post-shocked gas first becomes thick at the Lyman limit, $l_{thick}$,
similarly to the scheme presented in \S~\ref{precursor-phys}.
For each shock model, we construct a precursor model assuming
that a flux $F_\nu(l_{thick})$ enters a gas of density $n_0/4$.
The metal column densities are then $N_i=\int n_{\rm H} Z A_{\rm el} x_i dl$,
where $l$ is the upstream distance from the shock front, and $n_{\rm H}=n_0/4$.
In all cases, the precursors are assumed to have a constant density.

\begin{figure*}
\epsscale{1.0}
\plotone{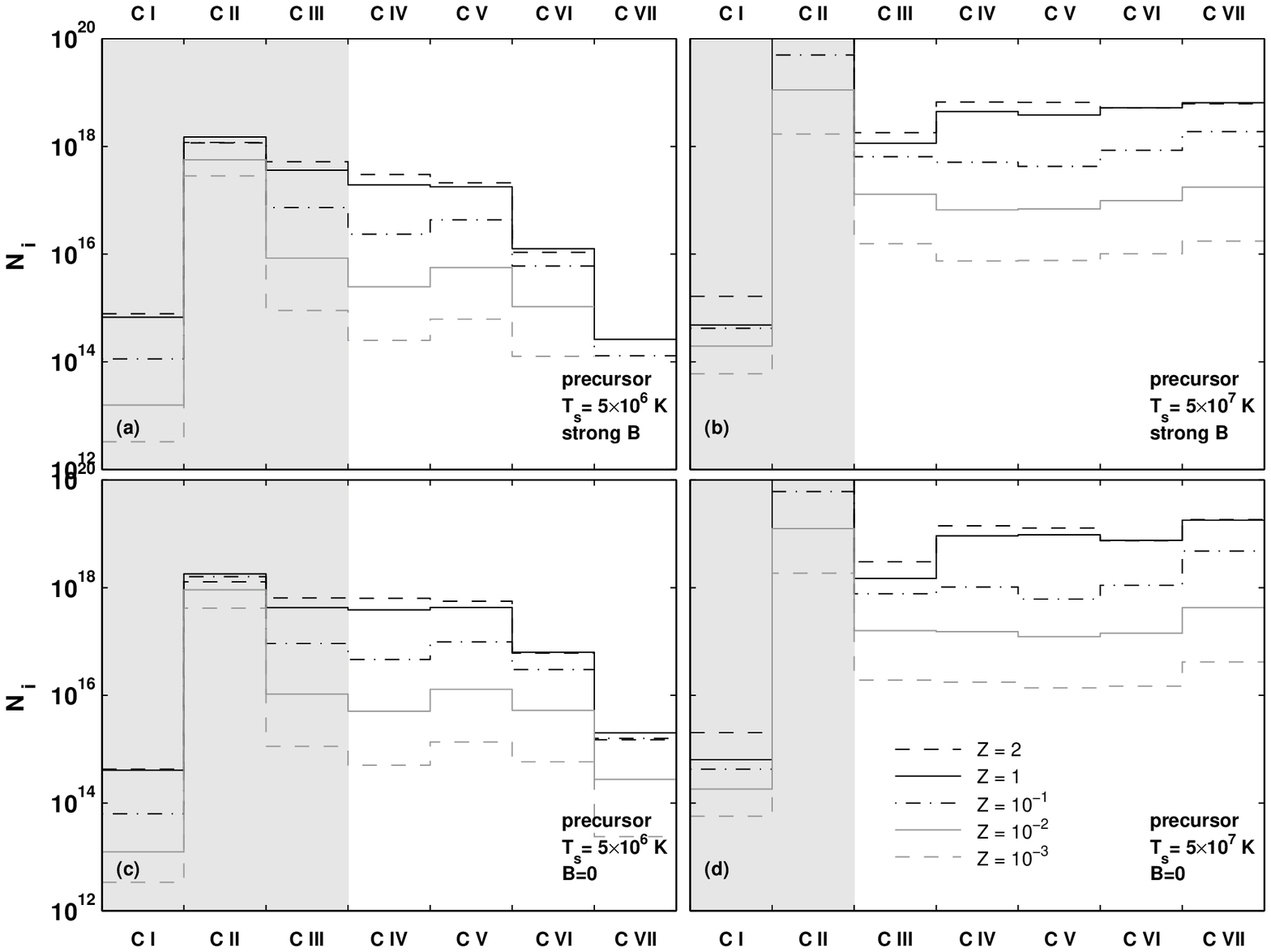}
\caption{ Carbon ion-column densities in radiative precursor,
for $Z$ from $10^{-3}$ to $2$ times solar.
(a)~$T_s = 5\times10^6$~K, for strong-$B$.
(b)~$T_s = 5\times10^7$~K, for strong-$B$.
(c)~$T_s = 5\times10^6$~K, for $B=0$.
(d)~$T_s = 5\times10^7$~K, for $B=0$.
The shaded areas show ions that are still abundant at $T_{\rm low}=1000$~K
where we stop the integration. The column densities of these ions are
sensitive to the choice of $T_{\rm low}$.
}
\label{C-pre}
\end{figure*}

\begin{figure*}
\epsscale{1.0}
\plotone{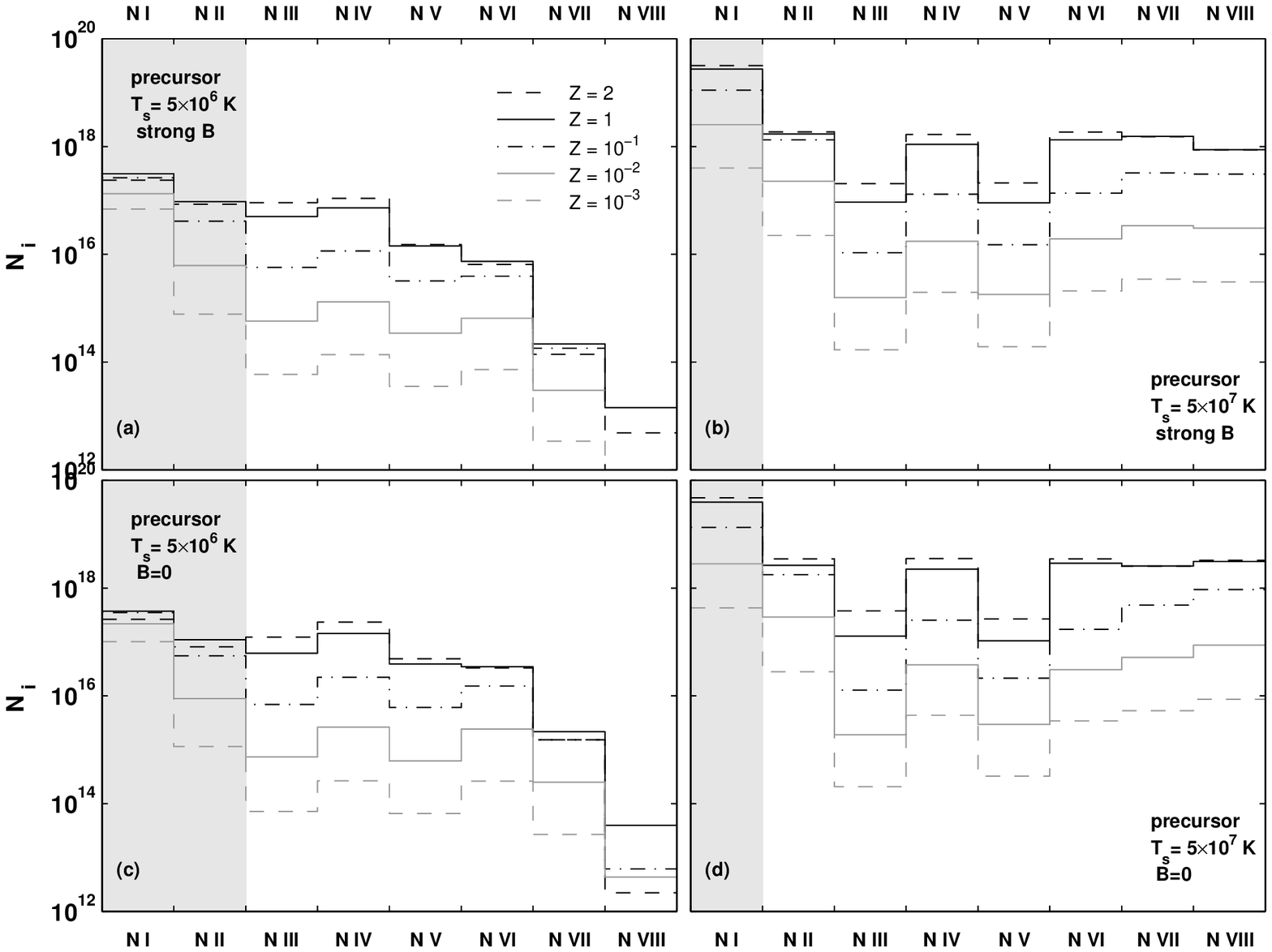}
\caption{ Same as Figure \ref{C-col} but for Nitrogen.}
\label{N-pre}
\end{figure*}

\begin{figure*}
\epsscale{1.0}
\plotone{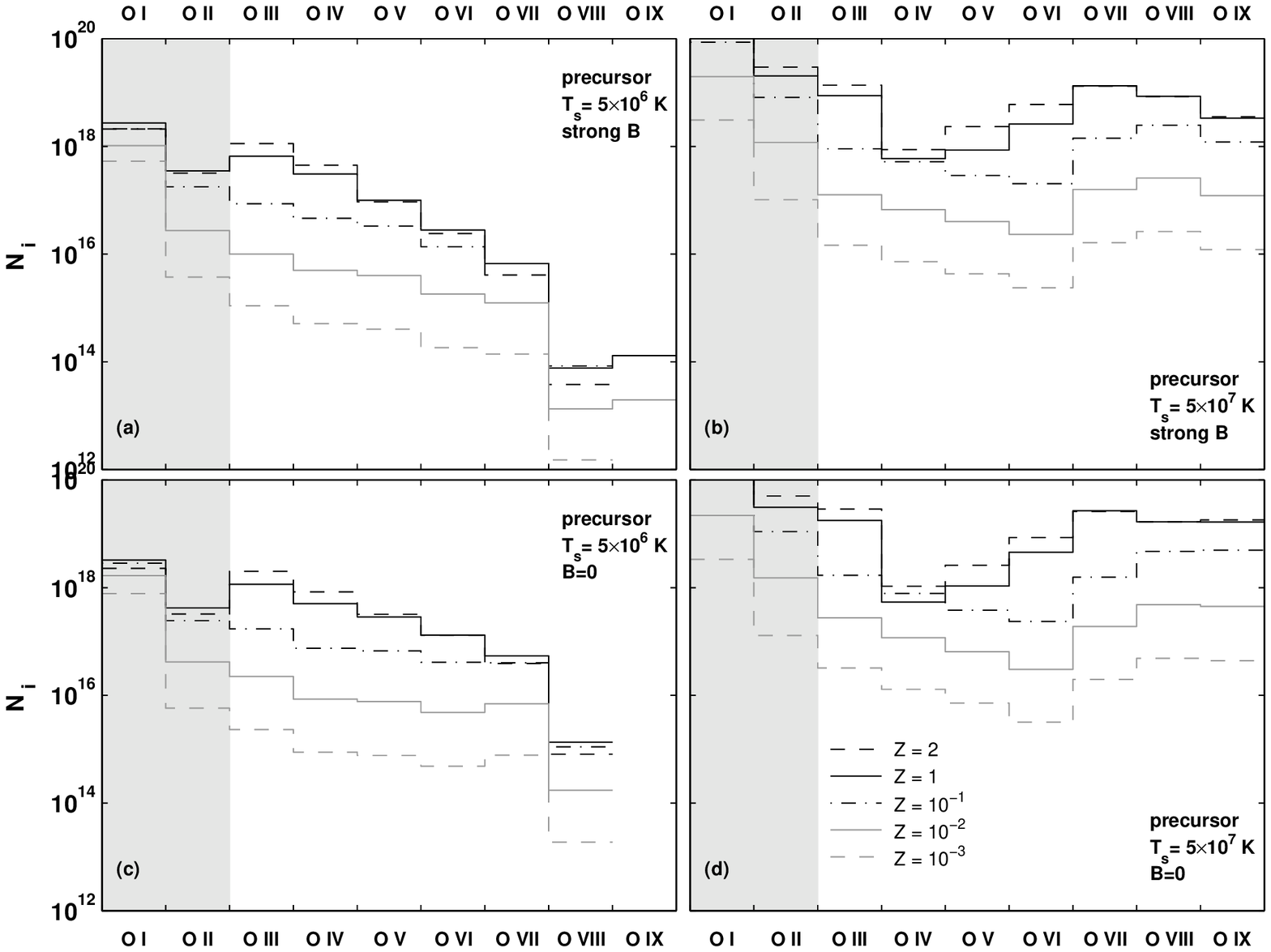}
\caption{ Same as Figure \ref{C-col} but for Oxygen.}
\label{O-pre}
\end{figure*}

\begin{figure*}
\epsscale{1.0}
\plotone{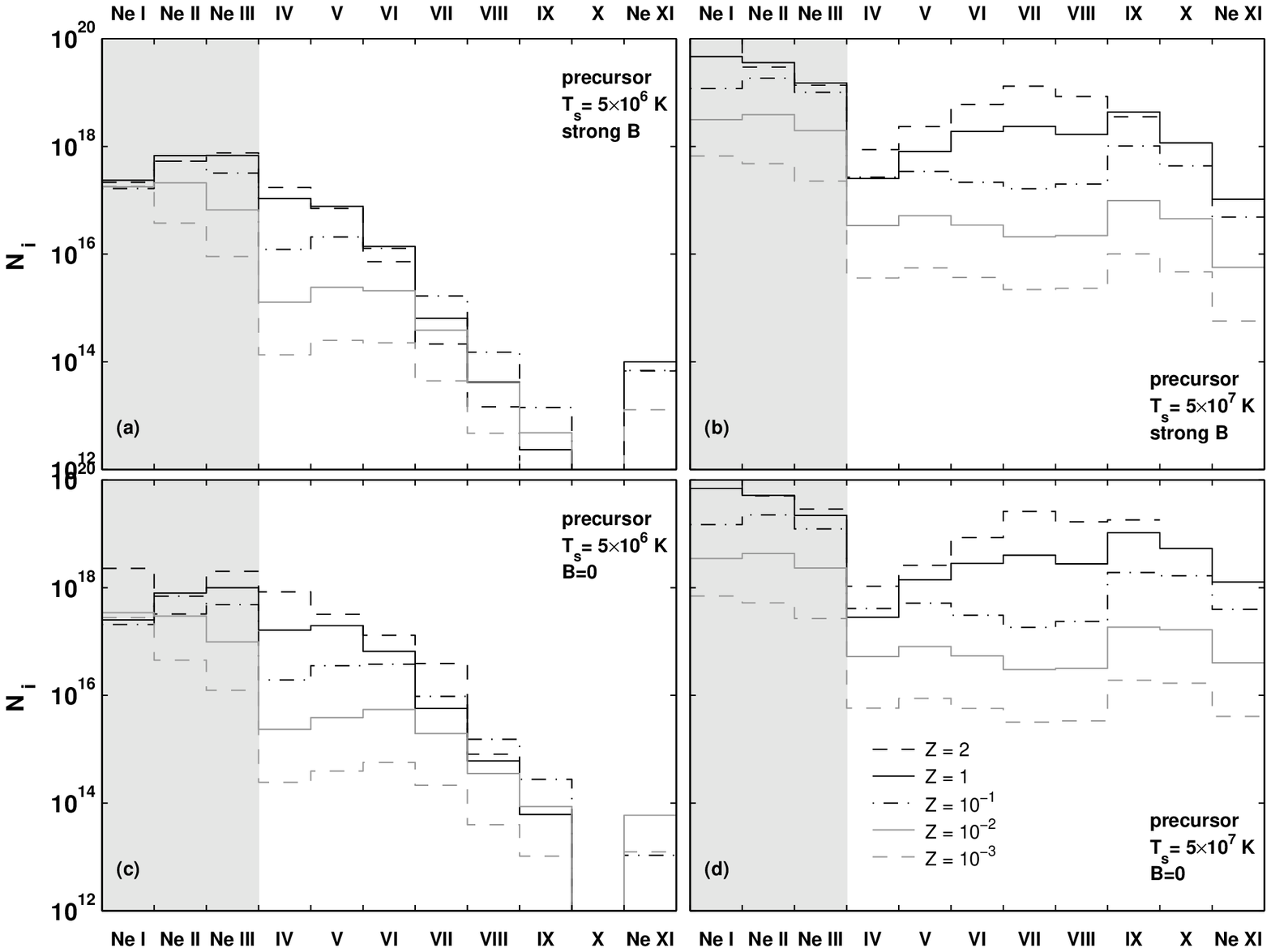}
\caption{ Same as Figure \ref{C-col} but for Neon.}
\label{Ne-pre}
\end{figure*}

Figures~\ref{C-pre}-\ref{Ne-pre} display the full set of
carbon-, nitrogen-, oxygen-, and neon-ion
column densities, for $Z$ between $10^{-3}$ and $2$, for
the radiative precursors of strong-$B$ (upper panels) and
$B=0$ (lower panels) shocks.
Left hand panels are for $T_s=5\times10^6$~K, and right hand
panels are for $T_s=5\times10^7$~K.
The shaded areas show ions that are still abundant at
$T_{\rm low}=1000$~K. 
The column densities of these ions 
(e.g.~O and O$^+$)
are sensitive to the
choice of $T_{\rm low}$. 
The full set of column densities for all the metal ions that we consider
are listed in Table~\ref{prec-columns-table}, 
as described in Table~\ref{guide}.
Our integrated precursor column densities are in good
qualitative agreement with the recent results of Allen et al.~(2008)
for the case of a solar metallicity gas cooling behind
a $600$~km~s$^{-1}$ shock, in which $B\simeq0$ (their model ``J\_n1\_b0'').

\begin{deluxetable*}{llcccc}
\tablewidth{0pt}
\tablecaption{Precursor Column Densities in a $5\times10^6$~K, $2$ times solar, strong-$B$-shock}
\tablehead{
\colhead{ionization} &
\colhead{H} &
\colhead{He} & 
\colhead{C} & 
\colhead{N} & 
\colhead{\ldots} \\
\colhead{}&
\colhead{(cm$^{-2}$)}&
\colhead{(cm$^{-2}$)}&
\colhead{(cm$^{-2}$)}&
\colhead{(cm$^{-2}$)}&
\colhead{\ldots}
}
\startdata
I   & $2.2\times10^{21}$ & $1.7\times10^{20}$ & $7.8\times10^{14}$&$2.4\times10^{17}$&\ldots\\
II  & $2.3\times10^{21}$ & $1.5\times10^{20}$ & $1.2\times10^{18}$&$8.4\times10^{16}$&\ldots\\
III & --                 & $5.5\times10^{19}$ & $5.2\times10^{17}$&$9.1\times10^{16}$&\ldots\\
\enddata
\tablecomments{The complete version of this table is in the electronic
edition of the Journal. The printed edition contains only a sample.
The full table lists precursorn column densities for $B=0$ and in the
strong-$B$ limit,
for shock temperatures of $5\times10^6$~K and $5\times10^7$~K, and for
$Z=10^{-3}$, $10^{-2}$, $10^{-1}$, $1$, and $2$ times solar metallicity
gas (for a guide, see Table~\ref{guide}). }
\label{prec-columns-table}
\end{deluxetable*}

Figures \ref{C-pre}-\ref{Ne-pre} show that significant column densities
are created in the radiative precursors. 
For the high-ions, which collisional abundance peak at $T\sim T_s$, 
the column densities are dominated by the post-shock cooling layers.
However, the higher ionization parameters in the lower density precursors 
allow for a very efficient production of lower ions that are
created below $T_s$.
The precursor columns often exceed the post-shock cooling columns,
especially for ions that are maintained by photoionization in the
post-shock gas.
For example, in solar metallicity gas,
the ratio of post-shock to precursor column density
of the high ion \ion{O}{8}, is $N_{\rm post}/N_{\rm pre} = 1000$
for a $T_s=5\times10^6$~K strong-$B$ shock. The \ion{O}{8} column
is completely dominated by the post-shock gas.
However, for \ion{O}{4} this ratio is  $0.25$, so that most
of the \ion{O}{4} column is created in the precursor.

Higher shocks velocities produce harder and more intense radiation-fields,
resulting in efficient production of high-ions and in deeper,
higher column-densities, radiative precursors. 
For example the upper panels in Figure~\ref{O-pre} show that when $Z=1$,
the \ion{O}{8} column is $8.5\times10^{18}$~cm$^{-2}$ for $T_s=5\times10^7$~K,
but $7.7\times10^{13}$~cm$^{-2}$ for $T_s=5\times10^6$~K.
The \ion{O}{4} column is $5.9\times10^{17}$~cm$^{-2}$ for $T_s=5\times10^7$~K,
but $3.1\times10^{17}$~cm$^{-2}$ for $T_s=5\times10^6$~K.

The column densities produced in the radiative precursors of strong-$B$
shocks are similar to those produced in $B=0$ precursor.
The radiative fluxes in the two cases differ only by a
factor $5/3$ (see Figure~\ref{specs-B}). A comparison of the upper and lower panels in 
Figures~\ref{C-pre}-\ref{Ne-pre} shows that the precursor columns
are indeed similar.

For a given radiation field and absorption depth, the metal-ion column
densities scale linearly with gas metallicity.
However, since both the spectral energy distribution and the gas opacities
are functions of the gas metallicity, some deviations from this linear
correlation are expected.
For example, lower-$Z$ shocks produce harder radiation fields, which
are more efficient in producing high ions.
Figures~\ref{C-pre}-\ref{Ne-pre} shows that the metal-ion column densities
generally scale with gas metallicity, but departures from exact linear
correlation are apparent in some cases.

While the column densities in the radiative precursor may far exceed the
columns produced in the post-shock cooling layers, kinematic differences
may distinguish the two components.
For high-ions created at $\sim T_s$, the kinematic offset between the
radiative precursor and the post-shocked gas is $3v_s/4$, which is
of order the thermal width.  Ions produced at lower temperatures
have narrower widths, making it possible to distinguish the 
precursor and post-shock components.
In $B=0$ shocks, the post-shock low-ion columns are greatly suppressed
(see \S~\ref{columns}),
and are orders of magnitudes lower than the precursor columns.
In strong-$B$ isochoric shocks,  significant post-shock low-ion columns 
are expected to 
form with a velocity offset of $3v_s/4$ relative to the
precursor, and with the same velocity centroid as the post-shock high-ions.

\section{Summary}

In this paper we present new computations
of the metal-ion column-densities produced in post-shock
cooling layers behind fast, radiative shocks.
We have constructed a new 
(one-dimensional, steady)
shock code in which we
explicitly follow the 
non-equilibrium 
ionization states of post-shock gas containing
H, He, C, N, O, Ne, Mg, Si, S, and Fe.
We present results for initial post-shock temperatures, $T_s$,
of $5\times10^6$ and $5\times10^7$~K, corresponding to 
shock velocities, $v_s$, of 
$600$ and $\sim2000$~km~s$^{-1}$. We consider shocks in which
there is no magnetic field ($B=0$) for which the cooling flows
are approximately isobaric ($P_\infty = (4/3)P_0$).
We also consider shocks in which the magnetic field dominates
the pressure everywhere ("strong-$B$") for which the flows
are isochoric ($n_\infty=n_0$).
We assume that the gas is dust free, and we present results for
metallicities $Z$ ranging from $10^{-3}$ to twice the
solar photospheric abundances of the heavy elements.

For the shock temperatures that we consider,
the post-shock gas emits energetic radiation that
may later be absorbed by cooler gas further downstream, 
and by the unperturbed gas that is approaching the shock front.
The shock self-radiation significantly affects the ionization
states and thermal properties of the gas. Its absorption
by the downstream gas creates a photoabsorption
zone, in which the temperature and ionization states
of the gas are set by the shock self-radiation.

In following the time-dependent ion fractions, we rely 
on the code and work presented in Gnat \& Sternberg (2007).
In this paper, in addition
to calculating the non-equilibrium ionization and cooling, 
we also follow the radiative transfer of the shock 
self-radiation through the post-shock cooling layers;  
take into account the resulting photoionization and heating
rates; follow the dynamics of the cooling gas; and self-consistently
compute the initial photoionization states in the precursor gas.
We 
use up-to-date rate coefficients
for all of the atomic ionization and recombination
processes, and for the energy loss and absorption mechanisms.
The equations and our numerical method are
presented in \S~\ref{physics}.

In \S~\ref{struct} we discuss the shock structure and emitted radiation,
and we discuss how these quantities depend on the controlling parameters, 
including the gas metallicity, shock velocity, magnetic field
and gas density (see, e.g.~Fig.~1).
For $B=0$ shocks (nearly isobaric)
the gas is compressed and decelerated 
as it cools. 
Because the cooling time-scale is proportional to $1/n$,
the final evolutionary stages of isobaric shocks occur much
more rapidly than in "strong-$B$" isochoric shocks.
The effects of photoionization by the shock radiation are much more important in
isochoric shocks because the ionization parameter remains large in
the downstream absorbing layers.

The evolution of the post-shock cooling layers is significantly affected
by the gas metallicity in several ways. High metallicity shocks cool more rapidly,
because for $10^4\lesssim T\lesssim10^7$~K the cooling efficiency
is dominated by metal-line cooling. The degree to which 
recombination lags behind cooling, and the resulting
departures from ionization equilibrium therefore depend on 
the gas metallicity. Second,
while the total energy emitted by the cooling gas is independent
of $Z$, and equals the energy flux entering the shock,
the spectral energy distribution of the shock self-radiation
is a function of gas metallicity.
For high-$Z$, a high fraction of the input energy
is radiated as line-emission, while for low-$Z$, the relative
contribution of lines is small, and most of the initial energy
flux is radiated as thermal bremsstrahlung continuum.
At low $Z$ the shock self-radiation is harder. The photoionization
states of the downstream layers depend on the metallicity-dependent
spectral energy distributions.

In \S~\ref{ion-frac} we present our computations of the non-equilibrium
ionization states in the post-shock cooling layers. 
In Table~\ref{ion-frac-table} we list the time-dependent ion fractions
as a function of time and temperature for the various
shock models that we consider.
As the gas enters the shock, its ionization state rapidly adjusts
toward CIE at the shock temperature.
Later the gas gradually cools, recombines, and radiates its
thermal energy.
We describe how photoionization by the shock self-radiation
affects the ion fractions in the cooling gas. We find that
in strong-$B$ (isochoric) shocks, photoionization plays a 
central role in setting the ionization states in the cooling
layers, even when the gas is still optically thin.
Photoionization increases the abundances of
high-ions at low temperatures, above the enhancements  
due to the recombination lags. In those phases where
photoionization dominates, the ionization states remain close to
photoionization equilibrium. For
$B=0$ (isobaric) shocks, the gas compression
significantly suppresses the role of photoionization.

We further demonstrate how departures from ionization equilibrium 
alter
the gas ionization. We find that for strong-$B$ shocks, 
departures from photoionization equilibrium occur at temperatures 
between $\sim2\times10^4$~K and $\sim10^6$~K, but the ion fractions
differ by $\lesssim25\%$.
However, when $B=0$, departures from equilibrium
ionization are larger due to the smaller effect of photoionization,
and the non-equilibrium ion-fractions are mostly due to the
recombination lags in collisionally ionized gas.
We also study the how the ion distributions depend on the gas
metallicity, shock velocity, and magnetic field.

We present our computations of the radiative cooling 
and heating efficiencies in \S~\ref{cooling}. We list our results in
Table~\ref{cool-table} and display them in Figure~\ref{highT-cool}.
We discuss the different cooling and heating processes that
operate in the various cooling ``zones'' (or "phases") and
shown in Figures~\ref{Tt} and~\ref{schem}.

Finally, in \S~\ref{columns} and \S~\ref{prec-cols} 
we present our results for the integrated metal-ion 
column-densities.
In \S~\ref{columns} we discuss the columns
that are produced in the post-shock cooling gas.
We give results for gas metallicities between
$10^{-3}$ and $2$ times the solar abundances.
The computed column densities are listed in Table~\ref{columns-table}.
We show that the cooling column densities created in the post
shock cooling layers are strong functions of the magnetic
field intensity, shock velocity and gas metallicity.
Ionic ratios are useful as diagnostic probes.
The predicted column densities show that UV and X-ray
absorption-line observations of shocked gas may 
be used to probe the intensity of magnetic fields
in interstellar and intergalactic shocks, and to infer the
physical properties of gas in fast radiative shock waves.
In \S~\ref{prec-cols} we compute the integrated equilibrium metal-ion
column densities produced in the
upstream radiative precursors. 
Our precursor column densities are
listed in Table~\ref{prec-columns-table}. 
For ``low-ions'',
the column densities
produced in the radiative precursors are often comparable to
or greater than the post-shock cooling columns. However, kinematic
differences between the precursor and post-shock components
may provide a means to 
distinguish observationally
between the two contributions.

\section*{Acknowledgments}

We thank Gary Ferland for his invaluable assistance in our
non-standard use of Cloudy. We thank Hagai Netzer for generously providing us
with his up-to-date Ion atomic data set.
We thank Chris McKee, Re'em Sari, and Ehud Nakar for many helpful
discussions.
Our research is supported by the US-Israel Binational
Science Foundation (grant 2002317).


\end{document}